\documentclass[12pt,preprint]{aastex}

\usepackage{float}

\shorttitle{MESA}
\shortauthors{Paxton et al.}

\newcommand{\MESA}{\texttt{MESA}}
\newcommand{\alert}{\texttt{alert}}
\newcommand{\utils}{\texttt{utils}}
\newcommand{\const}{\texttt{const}}
\newcommand{\chem}{\texttt{chem}}
\newcommand{\mtx}{\texttt{mtx}}
\newcommand{\MESAstar}{\texttt{MESA star}}
\newcommand{\num}{\texttt{num}}

\newcommand{\kap}{\texttt{kap}}
\newcommand{\eos}{\texttt{eos}}
\newcommand{\intone}{\texttt{interp\_1d}}
\newcommand{\inttwo}{\texttt{interp\_2d}}
\newcommand{\atm}{\texttt{atm}}
\newcommand{\diff}{\texttt{diffusion}}
\newcommand{\mlt}{\texttt{mlt}}
\newcommand{\rates}{\texttt{rates}}
\newcommand{\net}{\texttt{net}}
\newcommand{\neu}{\texttt{neu}}
\newcommand{\weak}{\texttt{weaklib}}
\newcommand{\jina}{\texttt{jina}}
\newcommand{\screen}{\texttt{screen}}
\newcommand{\ioniz}{\texttt{ionization}}
\newcommand{\Msun}{{M_{\odot}}}
\newcommand{\Lsun}{{L_{\odot}}}
\newcommand{\Rsun}{{R_{\odot}}}
\newcommand{\Teff}{{T_{\rm eff}}}
\newcommand{\logg}{{\log(g)}}

\begin{document}

\title{Modules for Experiments in Stellar Astrophysics ($\MESA$)}

\author{Bill Paxton and Lars Bildsten}
\affil{Kavli Institute for Theoretical Physics and Department of Physics, Kohn Hall, University of California, Santa Barbara, CA 93106 USA}
\author{Aaron Dotter\altaffilmark{1} and Falk Herwig} 
\affil{Department of Physics and Astronomy, University of Victoria, PO Box 3055, STN CSC, Victoria, BC, V8W 3P6 Canada}
\altaffiltext{1}{Current address: Space Telescope Science Institute, 3700 San Martin Drive, Baltimore, MD, 21218, USA}
\author{Pierre Lesaffre}\affil{LERMA-LRA, CNRS UMR8112, Observatoire de Paris and Ecole Normale Superieure, 24 Rue Lhomond, 75231 Paris cedex 05, France}
\author{Frank Timmes} \affil{School of Earth and Space Exploration, Arizona State University, PO Box 871404, Tempe, AZ, 85287-1404 USA}

\begin{abstract}
Stellar physics and evolution calculations enable a broad range of
research in astrophysics.  Modules for Experiments in Stellar
Astrophysics ($\MESA$) is a suite of open source, robust, efficient,
thread-safe libraries for a wide range of applications in
computational stellar astrophysics.  A 1-D stellar evolution module,
$\MESAstar$, combines many of the numerical and physics modules for
simulations of a wide range of stellar evolution scenarios ranging
from very-low mass to massive stars, including advanced evolutionary
phases.  $\MESAstar$ solves the fully coupled structure and
composition equations simultaneously.  It uses adaptive mesh
refinement and sophisticated timestep controls, and supports shared
memory parallelism based on OpenMP.  State-of-the-art modules provide
equation of state, opacity, nuclear reaction rates, element
diffusion data, and atmosphere boundary conditions.  Each module is
constructed as a separate Fortran 95 library with its own explicitly
defined public interface to facilitate independent development.
Several detailed examples indicate the extensive verification and testing
that is continuously performed, and demonstrate the wide range of
capabilities that $\MESA$ possesses.  These examples include
evolutionary tracks of very low mass stars, brown dwarfs, and gas
giant planets to very old ages; the complete evolutionary track of a
$1\Msun$ star from the pre-main sequence to a cooling white dwarf; the
Solar sound speed profile; the evolution of intermediate mass stars
through the He-core burning phase and thermal pulses on the He-shell
burning AGB phase; the interior structure of slowly pulsating B Stars
and Beta Cepheids; the complete evolutionary tracks of massive stars
from the pre-main sequence to the onset of core collapse; mass
transfer from stars undergoing Roche lobe overflow; and
the evolution of helium accretion onto a neutron star. $\MESA$ can be
downloaded from the project web site.\footnote{\url{http://mesa.sourceforge.net/}}
\end{abstract}


\tableofcontents

\section{Introduction\label{intro}}
Much of the information that astronomers use to study the universe comes from starlight.
Interpretation of that starlight requires a detailed understanding of stellar astrophysics,
especially as it relates to stellar atmospheres, structure, and evolution. Stellar structure 
and evolution models underpin much of modern astrophysics as they are used to analyze: 
the Sun through helioseismology \citep[e.g.,][]{bah98}, the pulsational properties of many 
nearby stars with asteroseismic data from, e.g., {\it Corot} \citep{deg09} and {\it Kepler} 
\citep{gil10}, the color-magnitude diagrams of resolved stellar and sub-stellar populations in 
the Milky Way and nearby galaxies \citep[e.g.,][]{vdb00,dot10}, the integrated light of distant 
galaxies and star clusters via population synthesis techniques \citep[e.g.,][]{wor94,coe07}, 
stellar yields and galactic chemical evolution \citep[e.g.,][]{tim95}, physics of the first 
stars \citep{fuj00}, and a variety of aspects in time domain astrophysics (e.g., LSST\footnote{\url{http://www.lsst.org/lsst/scibook}}).

Stellar evolution is broadly recognized as the first key problem in computational astrophysics.
The introduction of electronic computers enabled the solution of the highly non-linear, coupled 
differential equations of stellar structure and evolution, and the first detailed reports of computer 
programs for stellar evolution soon
appeared \citep{iben62,hen64,hof64,kip67}. Implicit in the development of these codes
was a sufficiently mature theoretical understanding of stars \citep[][see as well the compilation of 
references later in this section]{chandra38,mar58}, development of a concise
yet sufficiently accurate treatment of convection \citep{bv58}, as well as a better 
understanding of the properties of stellar matter, including nucleosynthesis \citep{b2fh,cam57}. 
Further improvements and alternative implementations became available addressing, for example,
the numerical stability of computations \citep{sug70}, more efficient methods for following shell
burning in low mass stars \citep{egg71}, and the hydrodynamics of advanced burning in  massive stars 
\citep{wea78}. Progress continues on stellar evolution codes, with code developments and comparisons 
often facilitated by the opening of new observational windows. For example, the participants
\citep{astec,franec,yrec,geneva,tgec,cesam,starox,cles,aton,garstec} in the CoRoT Evolution and Seismic
Tools Activity \citep{leb08} are a representative sample of the active community.

Modules for Experiments in Stellar Astrophysics ($\MESA$) began as an effort to improve upon the EZ stellar
evolution code \citep{egg71, pax04}. It employs modern software engineering 
tools and techniques to target modern computer architectures that are 
significantly different from those available to the pioneers half a 
century ago. As the pieces of the new system
started to emerge, it became clear that the parts would be of greater
value than the whole if they were carefully structured for independent
use. $\MESA$ includes a new 1-D stellar evolution code, $\MESAstar$,
but is designed to be useful for a wide range of stellar physics
applications.  The physical inputs to stellar evolution models, like
the equation of state, opacities, and nuclear reaction networks, have
a broader application than stellar evolution calculations
alone. $\MESA$ is designed so that each of the individual components
is usable on its own, with the intention of facilitating verification
test suites amongst different codes and encouraging new computational
experiments in stellar astrophysics.

$\MESAstar$ approaches stellar physics, structure, and evolution with
modern, sophisticated numerical methods and updated physics that give
it a very wide range of applicability. The numerical and computational
methods employed allow $\MESAstar$ to consistently evolve stellar
models through challenging phases, e.g., the He core flash in low
mass stars and advanced nuclear burning in massive stars, that have
posed substantial challenges for stellar evolution codes in the past.

$\MESA$ is open source: anyone can download the source code, compile
it, and run it for their own research or education purposes.  It is 
meant to engage the broader community of astrophysicists in related 
fields and encourage contributions in the form of testing, finding 
and fixing bugs, adding new capabilities, and, generally, sharing 
experience with the $\MESA$ community.  The philosophy and guidelines 
of $\MESA$ are described in more detail in the $\MESA$ manifesto (see
Appendix \ref{manifesto}).

This paper serves as an introduction to $\MESA$ and demonstrates its
current capabilities.  We assume that the reader is familiar with the 
basic stellar physics and numerical methods, both of which are essential to arrive at meaningful 
solutions when using $\MESA$. For background material we refer the reader to
\citet{edd}, \citet{chandra39}, \citet{mar58}, \citet{cox68}, \citet{clay84},
\citet{iben91}, \citet{han95}, \citet{arn96}, and \citet{kip96}.

The $\MESA$ codebase is in constant development, and
future capabilities and applications will be detailed in subsequent papers.
The paper is outlined as follows: \S\ref{mod} explains the design and 
implementation of $\MESA$ modules; \S\ref{num}-\ref{macro} describe the
numerical, microphysics, and macrophysical modules; \S\ref{star} describes the
stellar evolution module $\MESAstar$; \S\ref{vandv} presents 
a series of tests and code comparisons that serve as rudimentary verification
and demonstrates the broader capabilities of of $\MESAstar$; 
and \S\ref{summary} summarizes the material presented.

\begin{deluxetable}{lll}
\tabletypesize{\scriptsize}
\tablecolumns{3}
\tablewidth{0pc}
\tablecaption{Variable Index\label{vardefs}} 
\tablehead{\colhead{Name}&\colhead{Description}&\colhead{First appears}}
\startdata
A                                  & atomic mass number                    & \S \ref{rates} \\
$a$                                & acceleration at the cell face         & \S \ref{s.stareqns} \\
$\alpha$                           & order of convergence                  & \S \ref{solarVandV} \\
$\alpha_{\rm MLT}$                  & mixing length parameter               & \S \ref{mlt} \\
C                                  & ``spacetime'' parameter for convergence study & \S \ref{solarVandV} \\
$C_P$                              & specific heat at constant pressure    & \S \ref{eos} \\
$C_V$                              & specific heat at constant volume      & \S \ref{eos} \\
$c_s$                              & sound speed                           & \S \ref{s.pulsate} \\
$\chi_{\rho}$                        & $\equiv d{\rm ln}P/d{\rm ln}\rho|_T$ & \S \ref{eos} \\
$\chi_T$                           & $\equiv d{\rm ln}P/d{\rm ln}T|_{\rho}$ & \S  \ref{eos} \\
$D$                                & Eulerian diffusion coefficient        & \S \ref{overshoot} \\
$D_{\rm OV}$                        & overshoot diffusion coefficient       & \S \ref{overshoot} \\
$\Delta$                           & grid difference                       & \S \ref{s.mesh} \\ 
$\delta$                           & time difference                       & \S \ref{s.mesh} \\ 
$dm$                               & mass of a cell                        & \S \ref{s.stareqns} \\
$dP_s$                             & $P$ difference between surface and center of first cell    & \S \ref{s.stareqns} \\
$dT_s$                             & $T$ difference in between surface and center of first cell & \S \ref{s.stareqns}\\
$\delta t$                         & timestep                              & \S \ref{s.timestep} \\
$E$                                & energy                                & \S \ref{eos} \\
e$_{\rm nuc}$                       & nuclear energy generation in ergs/g  & \S \ref{net} \\
$\epsilon$                         & power per unit mass (nuclear, thermal neutrino, gravity) & \S \ref{s.stareqns} \\
$\epsilon_F$                       & Fermi energy                          & \S \ref{sec:lowmass} \\
$F$                                & mass flow rate                        & \S \ref{s.stareqns} \\
$f$                                & overshoot mixing parameter            & \S \ref{overshoot} \\
$g$                                & local gravity                         & \S \ref{mlt} \\
$\Gamma_1$                         & $\equiv d{\rm ln}P/d{\rm ln}\rho|_S$  & \S \ref{eos} \\
$\Gamma$                           & Coulomb coupling parameter            & \S \ref{eos} \\
$\Gamma_3$                         & $\equiv d{\rm ln}T/d{\rm ln}\rho|_S+1$ & \S \ref{eos} \\
$\nabla_{\rm ad}$                      & adiabatic temperature gradient        & \S \ref{eos} \\
$\nabla_{\rm rad}$                     & radiative temperature gradient        & \S \ref{mlt} \\
$\nabla_T$                         & actual temperature gradient           & \S \ref{mlt} \\
$\kappa_s$                         & opacity at the surface of the outermost cell & \S \ref{atm} \\
$L$                                & total luminosity                      & \S \ref{mlt} \\
$L_{\rm conv}$                      & convective luminosity                 & \S \ref{mlt} \\
$\Lambda$                          & mixing length ($\alpha_{\rm MLT} \lambda_P$) & \S \ref{mlt} \\
$\lambda_P$                        & pressure scale height                 & \S \ref{mlt} \\
$m$                                & mass interior to cell                 & \S \ref{s.stareqns} \\
$M_c$                              & inner mass (not modeled) for central BC & \S \ref{s.masschanges} \\
$M_m$                              & modeled mass                          & \S \ref{s.masschanges} \\
$\mu$                              & mean molecular weight per gas particle & \S \ref{eos} \\
$\mu_e$                            & mean molecular weight per electron    & \S \ref{eos} \\
$N$                                &  Brunt-V\"ais\"al\"a  frequency          & \S \ref{s.pulsate} \\
$\eta$                             & dimensionless electron degeneracy parameter & \S \ref{s.mesh} \\
$P$                                & total pressure                        & \S \ref{eos} \\
$P_{\rm gas}$                       & gas pressure                          & \S \ref{eos} \\
$P_s$                              & pressure at surface of outermost cell & \S \ref{atm} \\
$q$                                & relative mass coordinate              & \S \ref{s.masschanges} \\
$\rho$                             & density                               & \S \ref{eos} \\
$R$                                & total radius                          & \S \ref{mlt} \\
$R_{\rm CZ}$                        & radius of the base of the solar convective zone & \S \ref{sec:lowmass} \\
$r$                                & radius at the cell face               & \S \ref{s.stareqns} \\
$S$                                & entropy                               & \S \ref{eos} \\
$\sigma$                           & Lagrangian diffusion coefficient      & \S \ref{mlt} \\
$T$                                & temperature                           & \S \ref{eos} \\
$T_s$                              & temperature at surface of outermost cell & \S \ref{atm} \\
$\Teff$                            & effective temperature                 & \S \ref{atm} \\
$\tau_s$                           & optical depth at the surface of the outermost cell & \S  \ref{atm} \\
$\tau$                             & optical depth     & \S \ref{atm} \\
$v$                                & velocity at the cell face             & \S \ref{s.stareqns} \\
$v_c$                              & timestep control target               & \S \ref{s.timestep} \\
$v_{\rm conv}$                      & convective velocity                   & \S \ref{mlt} \\
$v_t$                              & timestep control variable             & \S \ref{s.timestep} \\
$w$                                & diffusion velocity                    & \S \ref{diff} \\
$X$                                & H mass fraction                       & \S \ref{eos} \\
$X_i$                              & mass fraction of the $i^{th}$ isotope  & \S \ref{s.stareqns} \\
$\xi$                                  & relative difference in convergence study & \S \ref{solarVandV} \\
$Y$                                & He mass fraction                      & \S \ref{kap} \\
$Y_e$                              & electrons per baryon ($\bar{\rm Z}$/$\bar{\rm A}$) & \S \ref{s.vv_highmass} \\
$Z$                                & metals mass fraction ($1-X-Y$)       & \S \ref{eos} \\
Z                                  & atomic number                         & \S \ref{diff} \\
$z$                                & distance from convective boundary     & \S \ref{overshoot} \\
\enddata
\end{deluxetable}
\clearpage

\section{Module design and implementation\label{mod}}  
Each $\MESA$ module is responsible for a different aspect of numerics
or physics required to construct computational models for stellar
astrophysics. Each has a similar organization: a public interface, a
private implementation, a makefile to create a library, and a test
suite for verification.  Each module includes an installation script
that builds the library, tests it, and, if the test succeeds, exports
it to the $\MESA$ libraries directory. Comparisons between local and
expected results are carried out with the open source \texttt{ndiff}
utility.\footnote{See
\url{http://www.math.utah.edu/~beebe/software/ndiff/}. $\MESA$ installs
its own copy of \texttt{ndiff} the first time the main installation
process is performed.}  There is a global install script for $\MESA$
that performs the installation of each of the modules in the required
order to satisfy dependencies.  The installation on UNIX-like systems,
including Linux and Mac OS X requires a modern, up-to-date Fortran
compiler.\footnote{Information about supported compilers and
installation is provided on the $\MESA$ project website.}  A template
module, \texttt{package\_template}, exists for initiation of new
modules by the community. All current $\MESA$ modules are listed in Table 
\ref{moddefs}, along with the  function they perform and the section in 
this paper where the description resides. 

\begin{deluxetable}{lllc}
\tabletypesize{\scriptsize}
\tablecolumns{4}
\tablewidth{0pc}
\tablecaption{$\MESA$ Module Definitions and Purposes\label{moddefs}} 
\tablehead{\colhead{Name}&\colhead{Type}&\colhead{Purpose}&\colhead{Section}}
\startdata
$\alert$ &               utility       & error handling  & \ref{num} \\ 
$\atm$ &                microphysics   &     grey and non-grey atmospheres; tables and integration & \ref{atm} \\ 
$\const$ &               utility  &        numerical and physical constants & \ref{const} \\ 
$\chem$	&			microphysics	&		properties of elements and isotopes & \ref{const} \\
$\diff$ &           macrophysics  &       gravitational settling and chemical and thermal diffusion & \ref{diff} \\
$\eos$  &                microphysics &       equation of state & \ref{eos} \\
$\intone$ &          numerics     &  1-D interpolation routines & \ref{num}\\
$\inttwo$ &     numerics  &     2-D interpolation routines & \ref{num} \\
$\ioniz$    &  microphysics  &      average ionic charges for diffusion & \ref{diff} \\
$\jina$      &    macrophysics     &   large nuclear reaction nets using reaclib & \ref{net} \\
$\kap$   &    microphysics &       opacities & \ref{kap} \\
\texttt{karo}  &    microphysics  &      alternative low-T opacities for C and N enhanced material & \ref{kap} \\
$\mlt$   &   macrophysics  &      mixing length theory & \ref{mlt} \\
$\mtx$               &   numerics  &     linear algebra matrix solvers & \ref{num} \\
$\net$                &  macrophysics &       small nuclear reaction nets optimized for performance & \ref{net} \\
$\neu$                & microphysics     &   thermal neutrino rates & \ref{net} \\
$\num$               &   numerics  &     solvers for ordinary differential and differential-algebraic equations & \ref{num} \\
\texttt{package\_template} & utility & template for creating a new $\MESA$ module & \ref{mod} \\
$\rates$               &  microphysics  &      nuclear reaction rates & \ref{rates} \\
$\screen$           &    microphysics  &       nuclear reaction screening & \ref{net} \\
\texttt{star}                &  evolution   &   1-D stellar evolution & \ref{star} \\
$\utils$              &  utility  &      miscellaneous utilities & \ref{num} \\
$\weak$         &     microphysics &       rates for weak nuclear reactions & \ref{net} \\
\enddata
\end{deluxetable}

The $\MESA$ modules are ``thread-safe''---meaning that more than one
process can execute the module routines at the same time---allowing
applications to utilize multicore processors. A module is thread-safe
if all of its shared data is read-only after initialization. This
prohibits the use of common blocks and ``SAVE'' statements.  Working
memory must be allocated as local variables of routines or allocated
dynamically.  To take full advantage of shared memory on multicores,
an operation that is performed in parallel needs to fit in the
processor cache.
Evaluations of local microphysics, such as the equation of state, 
opacity, and nuclear reaction networks can be carried out in parallel
using the OpenMP application programming interface.\footnote{\url{http://openmp.org}}
The capability of $\MESAstar$ to take advantage of multithreading is discussed 
in \S \ref{s.perform}.

\section{Numerical methods\label{num}} 
$\MESA$ includes several modules that provide numerical methods.
The following briefly describes each one presently available and references the 
relevant literature (or web-based resource) where the full description resides. 

The $\mtx$ module provides an interface to linear algebra routines
for matrix manipulation.  A large set of BLAS and LAPACK routines
are included, but the $\mtx$ module can easily be modified to
accept these routines from other linear algebra packages, e.g. 
GotoBLAS\footnote{\url{http://www.tacc.utexas.edu/resources/software}} or the Intel 
Math Kernel Library\footnote{\url{http://software.intel.com/en-us/intel-mkl/}} (MKL). 
Sparse matrix operations are supported, including a subset of the 
SPARSKIT sparse matrix iterative solver\footnote{\url{http://www-users.cs.umn.edu/saad/software/SPARSKIT/sparskit.html}}  and an 
interface to the Intel version of the PARDISO sparse matrix direct solver. The 
routines in $\num$ make use of these matrix routines.

Modules $\intone$ and \texttt{interp\_2d} deal with 1-D and 2-D interpolation, 
respectively. One dimensional interpolation is carried out using
either a piecewise monotonic cubic method \citep{huy93,sur97}
or a monotonicity-preserving method \citep{stef90}. Compared to the
piecewise monotonic method, the monotonicity-preserving
method is stricter and does not allow an interpolated 
value to range outside of the given values \citep{stef90}.
Module $\inttwo$ includes parts of the PSPLINE 
package\footnote{\url{http://w3.pppl.gov/NTCC/PSPLINE}} and routines by both \citet{ak96} 
and \citet{ren99} for bivariate interpolation and surface fitting on a grid or with 
a scattered set of data points. Both single- and double-precision versions of the 
2-D interpolation routines are provided.

Module $\num$ provides a variety of solvers for stiff and non-stiff 
systems of ordinary differential equations (ODEs) and a Newton-Raphson 
solver for multidimensional, nonlinear root-finding. 
The family of ODE solvers is derived from the routines of \citet{hairer96}.  
The non-stiff ODE class are explicit Runge-Kutta integrators of orders 5 and 8
with dense output, automatic stepsize control, and optional monitoring for 
stiffness.  The stiff ODE solvers are linearly implicit Runge-Kutta, with 2nd, 
3rd, and 4th order versions and two implicit extrapolation integrators of
variable order: either midpoint or Euler.  All integrators support dense, 
banded, or sparse matrix routines, analytic or numerical difference Jacobians,
explicit or implicit ODE systems, dense output, and automatic stepsize control.

The Newton-Raphson solver for multidimensional, nonlinear root-finding supports 
square, banded, and sparse matrices and analytic or automatic numerical differencing for 
the Jacobians.  It has the ability to reuse Jacobians and employs a line search 
method to give improved convergence.  The multidimensional Newton-Raphson solver 
is used by $\MESAstar$ to solve highly non-linear systems of 
differential-algebraic equations with tens of thousands of variables (see 
$\S$\ref{star}).  The structure of the Newton-Raphson solver is derived from 
Lesaffre's version of the Eggleton stellar evolution code \citep{egg71,pols95,les06}
and some details of the implementation will be described in \S \ref{star}.

The $\alert$  module provides a framework for reporting messages, including errors, to the 
terminal. The $\utils$ module provides a number of functions for checking if a variable has 
been assigned a bad value (e.g., NaN or Infinity) and tracking Fortran I/O unit 
numbers in use.  It also provides subroutines for basic file I/O and for 
allocating arrays of different types and dimensions, including a Fortran
implementation of a hash tree that is used by the stellar evolution module to
update the model mesh. Programs and scripts that are used for testing that
each module has compiled correctly are stored in $\utils$.

\section{Microphysics\label{micro}}

The $\MESA$ microphysics  modules provide the physical properties of stellar matter,
with each module focusing on a separate aspect of the physics. 

\subsection{Mathematical constants, physical and astronomical data \label{const}}
The $\MESA$ module $\const$ contains mathematical, physical, 
and astronomical constants relevant to stellar astrophysics 
in cgs units. The primary source for physical constants is 
the CODATA Recommended Values of the Fundamental Physical 
Constants \citep{codata}. Values for the Solar age, mass, radius,
and luminosity are taken from \citet{bah05}.

The $\MESA$ module $\chem$ is a collection of data, functions, and subroutines to 
manage the chemical elements and their isotopes. It contains 
basic information about the chemical elements and their isotopes from 
Hydrogen through Uranium. It includes routines for translating between 
atomic weights and numbers and isotope names. It contains full listings
of Solar abundances on several scales \citep{ag89, gn93, gs98, lod03, ags04}.
Module $\chem$ contains a framework for the user to provide an arbitrary set 
of species in a text file.

\subsection{Equation of state \label{eos}}

The equation of state (EOS) is delivered by the $\eos$ module. It
works with density,  $\rho$,  and temperature, $T$, as independent variables. These are the natural variables 
in a Helmholtz free energy formulation of the thermodynamics.
However, as some calculations are more naturally performed using pressure, $P$, 
and $T$ (as in a Gibbs free energy formulation),  a simple root find can provide $\rho$  given the desired $P_{\rm gas}=P-aT^4/3$ and $T$.  While conceptually simple, this can impose a substantial computational overhead if 
done for each $\eos$ call. To alleviate this computational burden, the root finds 
are pre-processed, creating a set of tables indexed by $P_{\rm gas}$ and $T$. As a result, 
the runtime cost of evaluating $\eos$ using $P_{\rm gas}$ and $T$ is the same as for using 
$\rho$ and $T$, as long as the $P_{\rm gas}-T$ requests are within the pre-computed ranges. 
When outside those ranges, the root find is performed during runtime, slowing the computations. 

\begin{figure}[H]
\begin{center}
\includegraphics[width=0.75\textwidth]{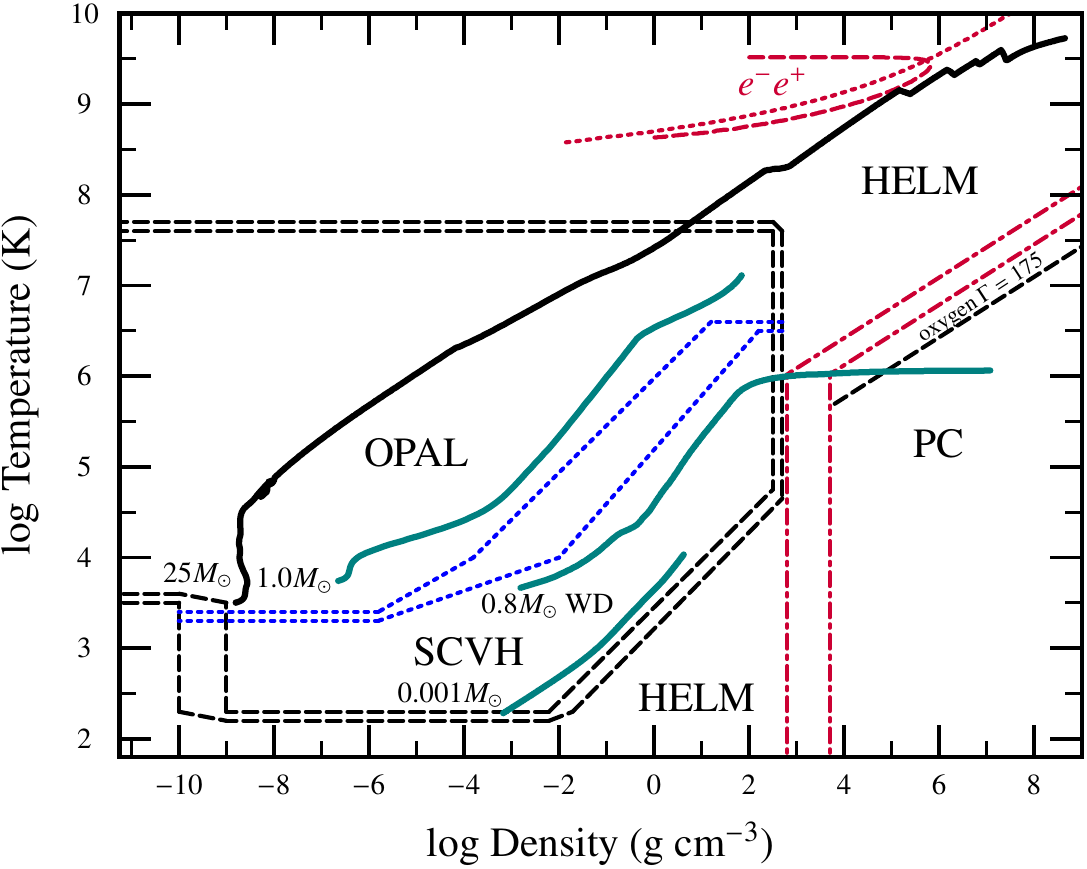}
\end{center}
\caption{The $\rho-T$ coverage of the equations of state used by the $\eos$ module
for $Z\leq 0.04$.  Inside the region bounded by the  black dashed lines we use $\MESA$ 
EOS tables that were constructed from the OPAL and SCVH tables.  The OPAL and SCVH 
tables were blended in the region shown by the blue dotted lines, as described in the 
text. Regions outside of the black dashed lines utilize the HELM and PC EOSs, which, 
respectively,  incorporate  electron-positron pairs at high temperatures and crystallization 
at low temperatures. The blending of the $\MESA$ table and the HELM/PC results occurs between 
the black dashed lines and is described in the text. The dotted red line shows where the number 
of electrons per baryon has doubled due to pair production, and the region to the left of the 
dashed red line has $\Gamma_1<4/3$. The  very low density cold region in the leftmost part of 
the figure is treated as an ideal, neutral gas. The region below the black dashed line labeled 
as $\Gamma=175$  would be in a crystalline state for a plasma of pure oxygen and is fully 
handled by the PC EOS.  The red dot-dashed line shows  where $\MESA$ blends the PC and HELM EOSs. 
The green lines show stellar profiles for a main sequence star ($M=1.0\Msun$), a contracting object 
of $M=0.001\Msun$ and a cooling white dwarf of $M=0.8\Msun$. The heavy dark line is an evolved 
$25\Msun$ star that has a maximum infalling speed of $1000 \ {\rm km \ s^{-1}}$. The jagged behavior 
reflects the distinct burning shells.\label{eosRhoT}}
\end{figure}

The $\MESA$ $\rho-T$  tables are based on the 2005 update of the OPAL EOS tables \citep{rn02}.  To extend to 
lower temperatures and densities, we use the SCVH tables \citep{scvh95}, and construct a smooth transition between these tables in the overlapping region that we define (shown by the blue dotted lines in Figure \ref{eosRhoT}). The limited thermodynamic information available from 
these EOSs restricts our blending to the output quantities  listed in Table  \ref{eosvars}.
The resulting $\MESA$  tables are more finely gridded than the original tables (so that no information is lost) and 
are provided at six $X$ and three $Z$ values:
$X=(0.0, 0.2, 0.4, 0.6, 0.8, 1.0)$ and $Z=(0.0, 0.02, 0.04)$ in keeping with the OPAL
tables, allowing for Helium rich compositions. 
In order to save space, the $\MESA$ tables are not rectangular in the independent 
variables. Instead, the region occupied by usual stellar models is roughly rectangular 
in the stellar modeling motivated variables, $\log T$ and $\log Q=\log \rho-2\log T+12$.
The range in $\log T$  is from 2.1 to 8.2 in steps of 0.02 and the 
range in $\log Q$  is from -10.0 to 5.69 in steps of 0.03. Partials with 
$\log T$ and $\log Q$  are derived from the interpolating polynomials, while 
partials with respect to $\log \rho$ then follow.  
The resulting region of these $\MESA$ tables is that inside of the dashed black lines of Figure \ref{eosRhoT}. 
The $\MESA$ $P_{\rm gas}-T$ tables are rectangular in $\log T$ 
and $\log W=\log P_{\rm gas}-4\log T$ over a range $-17.2\leq \log W\leq -2.9$, and $2.1\leq \log T\leq8.2$. 

Outside the region covered by the $\MESA$ tables, the HELM \citep{ts00} and PC \citep{pot10}  EOSs are employed. Both HELM and PC assume complete ionization and were explicitly constructed from a free energy approach, 
guaranteeing thermodynamic consistency. In nearly all cases, the full 
ionization assumption is appropriate since the OPAL and SCVH tables are used at those cooler
temperatures where partial ionization is significant.\footnote{We discuss the 
ionization states of trace heavy elements in \S \ref{diff}.}
Since the $\MESA$ tables are only constructed  for $Z \leq 0.04$, $\eos$ uses HELM and PC for $Z > 0.04$ in the whole
$\rho-T$ plane.

\begin{deluxetable}{ccc}
\tablecolumns{3}
\tablewidth{0pc}
\tablecaption{$\eos$ output quantities and units\label{eosvars}}
\tablehead{\colhead{Output}&\colhead{Definition}&\colhead{Units}}
\startdata
$P_{\rm gas}$ & gas pressure & ${\rm ergs \ cm^{-3}}$\\
$E$ & internal energy& ${\rm ergs \ g^{-1}}$ \\
$S$ & entropy per gram & ${\rm ergs \ g^{-1}  \ K^{-1}}$\\
$dE/d\rho|_T$ & &${\rm ergs \ cm^{3} \ g^{-2}} $\\\
$C_{\rm V}$ & specific heat at constant $V\equiv 1/\rho$  & ${\rm ergs \ g^{-1}  \ K^{-1}}$\\
$dS/d\rho|_{T}$ & &${\rm ergs \ cm^{3} \ g^{-2}  \ K^{-1}}$\\\
$dS/dT|_{\rho}$ & &${\rm ergs \ g^{-1}  \ K^{-2}}$\\\
$\chi_{\rho}$ & $\equiv d{\rm ln}P/d{\rm ln}\rho|_T$ & none \\
$\chi_T$ & $\equiv d{\rm ln}P/d{\rm ln}T|_{\rho}$ &none \\
$C_{\rm P}$ & specific heat at constant pressure& ${\rm ergs \ g^{-1}  \ K^{-1}}$\\
$\nabla_{\rm ad}$ & adiabatic T gradient with pressure & none \\
$\Gamma_1$ & $\equiv d{\rm ln}P/d{\rm ln}\rho|_S$ & none\\
$\Gamma_3$ & $\equiv d{\rm ln}T/d{\rm ln}\rho|_S+1$ & none \\
$\eta$ & ratio of electron chemical potential to $k_BT$ & none\\
$\mu$ & mean molecular weight per gas particle & none \\
$1/\mu_e$ & mean number of free electrons per nucleon & none\\
\enddata
\end{deluxetable}

HELM was constructed for high temperatures (up to $\log T=13$) and densities (up to $\log \rho=15$), and accounts for the  onset of  electron-positron pair production at high temperatures. 
The dotted red line in Figure \ref{eosRhoT} shows where the number of electrons per baryon has doubled due to pair production.  The domination of pairs in the plasma creates a region where $\Gamma_1<4/3$ (to the left of  the dashed red line). The blending region to HELM  (from the $\MESA$ tables) is shown by the black dashed lines
in Figure \ref{eosRhoT}, and can by modified by the user.  In this transition region, the blend of the two EOSs is performed in a way that preserves thermodynamic consistency. Therefore, if  each separate EOS satisfies Maxwell's relations,  the blend will also satisfy them. 
To accomplish this, we linearly sum the EOS quantities $Q_i$ (i.e. $P,E,S$ and their partial derivatives with respect to $\rho$ and $T$) 
needed to satisfy Maxwell's relations \citep{ts00}.\footnote{The more conventional forms of these nine thermodynamic quantities are displayed in the 
first nine rows in Table \ref{eosvars}.} The blend is calculated by defining the boundary limits,  inside of which we define a fractional ``distance", $F$, from the boundary. As $F$ varies from zero to one, we use the  smoothing function ${\rm S}=(1-\cos({\rm F}\pi))/2$  and for each of the nine quantities we construct $Q_i={\rm S}Q^A_i+(1-{\rm S})Q^B_i$, where $Q^A_i$ and $Q^B_i$ are the outputs from the two EOSs. We then use these to rederive the thermodynamic quantities ($\chi_\rho, \chi_T, C_P, \nabla_{\rm ad}, \Gamma_3, \Gamma_1$) delivered by the $\eos$ routine. 

 In late stages of the cooling of white dwarfs, the ions in the core will  crystallize. 
  For pure oxygen, the crystallization limit corresponds to a value of the Coulomb coupling parameter,
$\Gamma \approx 175$, shown by the black dashed line in Figure \ref{eosRhoT}. In this region, we use the PC EOS, which accounts for the modified thermodynamics of a crystal, as well as carefully handling mixtures (e.g. carbon and oxygen).  The blend between PC and HELM  (as shown by the dot-dashed red lines in Figure \ref{eosRhoT})
is performed in the same manner as described above.  In the dense 
liquid realm, the blending region is defined by  the Coulomb coupling parameter, $\Gamma=\overline{\rm Z}^2e^2/a_i k_BT$, where $a_i$ is the mean ion spacing, and $\overline{\rm Z}$ is the average ion charge. 
The default choice is PC for $\Gamma>80$ and HELM for $\Gamma<40$. The PC EOS is not constructed for arbitrarily low densities, forcing a transition to HELM at $\log \rho<2.8$, with the blend beginning at $\log \rho=3.7$.
These boundaries may be re-defined by the user if needed. 

In addition to the two independent variables,  
the $\eos$ module requires as input $X, Z$, $\overline{{\rm A}}$ (the mass-averaged atomic 
weight of metals), and $\overline{\rm Z}$ (the mass-averaged atomic charge of metals). 
When operating in the regime where the PC EOS is implemented, the mass fractions for all isotopes with mass fractions above a specified minimum are needed (default is $0.01$), allowing PC to correctly handle isotope mixtures.  It returns a total of sixteen 
quantities (listed in Table \ref{eosvars})  as well as the partial derivatives of each quantity with respect to the 
independent variables. The tables are interpolated in the independent variables using 
bicubic splines from $\inttwo$ with partial derivatives determined from the splines. 
Separate quadratic interpolations are performed in $X$ and $Z$. 

The construction of $\eos$ tables as outlined above is the default
option for $\MESA$ but the $\eos$ module has the flexibility to accept
tables from any source so long as the tables conform to the $\MESA$ standard
format. For example, the comparison with the Stellar Code Calibration
project \citep{weiss07} described in $\S$\ref{SCC} utilizes tables 
constructed using FreeEOS.\footnote{\url{http://freeeos.sourceforge.net}}   The FreeEOS code does not
cover the same range of $\rho$ and $T$ as SCVH+OPAL+HELM but the $\eos$
module is designed with this flexibility in mind: the table dimensions
are specified in the table headers and the module dynamically allocates
arrays of the appropriate size to hold them when the tables are read in.

Since not all EOS sources may be in the tabular form desired by $\eos$, we have created a
module, \texttt{other\_eos},  that provides the user an opportunity to incorporate their own EOS and use it 
with the stellar evolution module $\MESAstar$. 

\subsection{Opacities \label{kap}}

The pre-processor \texttt{make\_kap} resides within the $\kap$ module and constructs the $\MESA$ opacity 
tables by combining radiative opacities with the electron conduction opacities from \citet{cas07}. In the 
rare circumstances where $\rho$ or $T$ are outside the region covered by \citet{cas07} 
($-6 \leq \log \rho \leq 9.75 $ and $3 \leq \log T\leq 9$), the \citet{iben75} fit to the \citet{hub69} 
electron conduction opacity is used for non-degenerate cases while the \citet{yak80} fits are used for degenerate cases.
Radiative opacities are taken from \cite{ferg05} for $2.7 \leq \log T \leq 4.5$ and 
OPAL \citep{ir93,ir96} for $3.75 \leq \log T \leq 8.7$. The low $T$ opacities of \citet{ferg05} include the effects of molecules and grains
on the radiative opacity. Tables from OP \citep{op} can be used in place of OPAL as the table format is identical. 
The radiative opacity is dominated by Compton scattering for 
$\log T > 8.7$ and is calculated using the equations of \citet{buch76} up to a density of $10^6 \ {\rm g \ cm^{-3}}$. We use the HELM EOS to calculate the 
increasing number of electrons and positrons per baryon 
when pair production becomes prevalent, an important opacity enhancement. 

The OPAL tables with fixed metal distributions are called Type 1 \citep{ir93,ir96} and 
cover the region  $0.0 \leq X \leq  1-Z $ and $0.0\leq Z \leq 0.1$.
Additionally, there is support for the OPAL Type 2 \citep{ir96} tables that allow for varying 
amounts of C and O beyond that accounted for by $Z$; these are needed during helium burning and beyond. These have a range $0.0 \leq X \leq 0.7 $,  $0.0\leq Z\leq0.1$. 

\begin{figure}[H]
\plotone{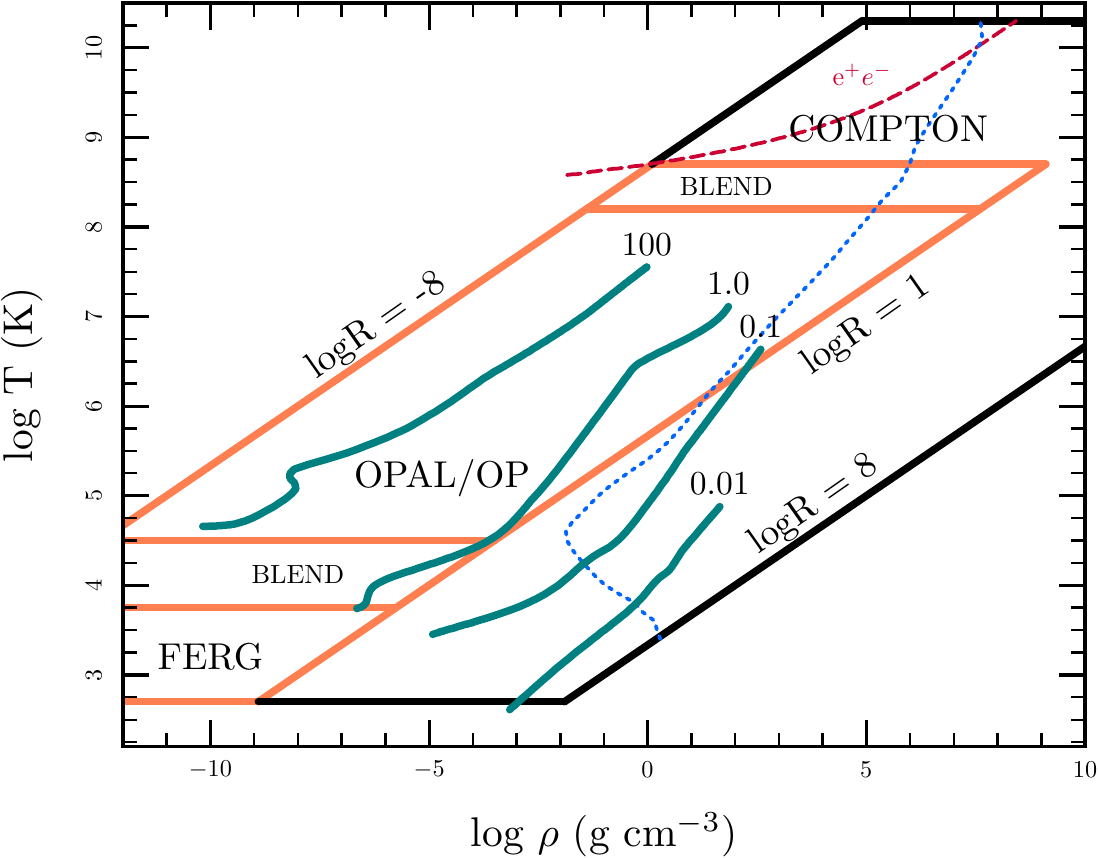}
\caption{The sources of the standard $\MESA$ opacity tables. Construction of opacity tables requires incorporating different sources, denoted by the labels. The heavy orange lines denote regions where input tables exist for radiative opacities, whereas the heavy black lines extend into regions where we use algorithms to derive the total opacities,  described in the text. Above the dashed  red line, the number of electrons and positrons from pair production exceeds the number of electrons from ionization,  and is accounted for in the opacity table. The opacity in the region to the right of the 
dashed blue line is dominated by electron conduction. Also shown are stellar profiles for stars on 
main sequence  ($M=0.1, 1.0,\ \&\ 100\Msun$) or just below (a contracting $M=0.01\Msun$ brown dwarf). 
\label{kapregions}}
\end{figure}

The resulting $\kap$ tables cover the large range  $2.7 \leq \log T \leq 10.3$ and $-8 \leq \log R  \leq 8$ ($R=\rho/T_6^3$, so $\log R=\log \rho-3\log T+18$), as shown by the heavy orange and black lines in Figure \ref{kapregions}. The $\MESA$ release includes $\MESA$ opacity tables derived from Type 1 and 2 OPAL tables, tables from OP, and \citet{ferg05}. The heavy orange lines delineate the 
boundaries where we use existing tables to make the $\MESA$ opacity table. The blended 
regions in Figure \ref{kapregions} are where two distinct sources of radiative opacities exist for the same parameters, requiring a smoothing function that 
blends them  in a manner adequate for derivatives. The blend is 
calculated at a fixed $\log R$ by defining the upper ($\log T_U$) and lower ($\log T_L$) 
boundaries of the blending region in $\log T$ space, where $\kappa_U$ ($\kappa_L$) is the opacity source above (below) the blend. We perform the interpolation by defining ${\rm F}=(\log T-\log T_L)/(\log T_U-\log T_L)$, and using a smooth 
function ${\rm S}=(1-\cos({\rm F}\pi))/2$ for 
\begin{equation} 
\label{blendequation} 
\log \kappa={\rm S} \log \kappa_U(R,T)+(1-{\rm S})\log \kappa_L(R,T).
\end{equation} 
At high temperatures, the blend from Compton to OPAL (or OP) has $\log T_U=8.7$ and $\log T_L=8.2$. At low temperatures, the 
blend between  \citet{ferg05} and OPAL has $\log T_U=4.5$ and $\log T_L=3.75$.

\begin{figure}[H]
 \plotone{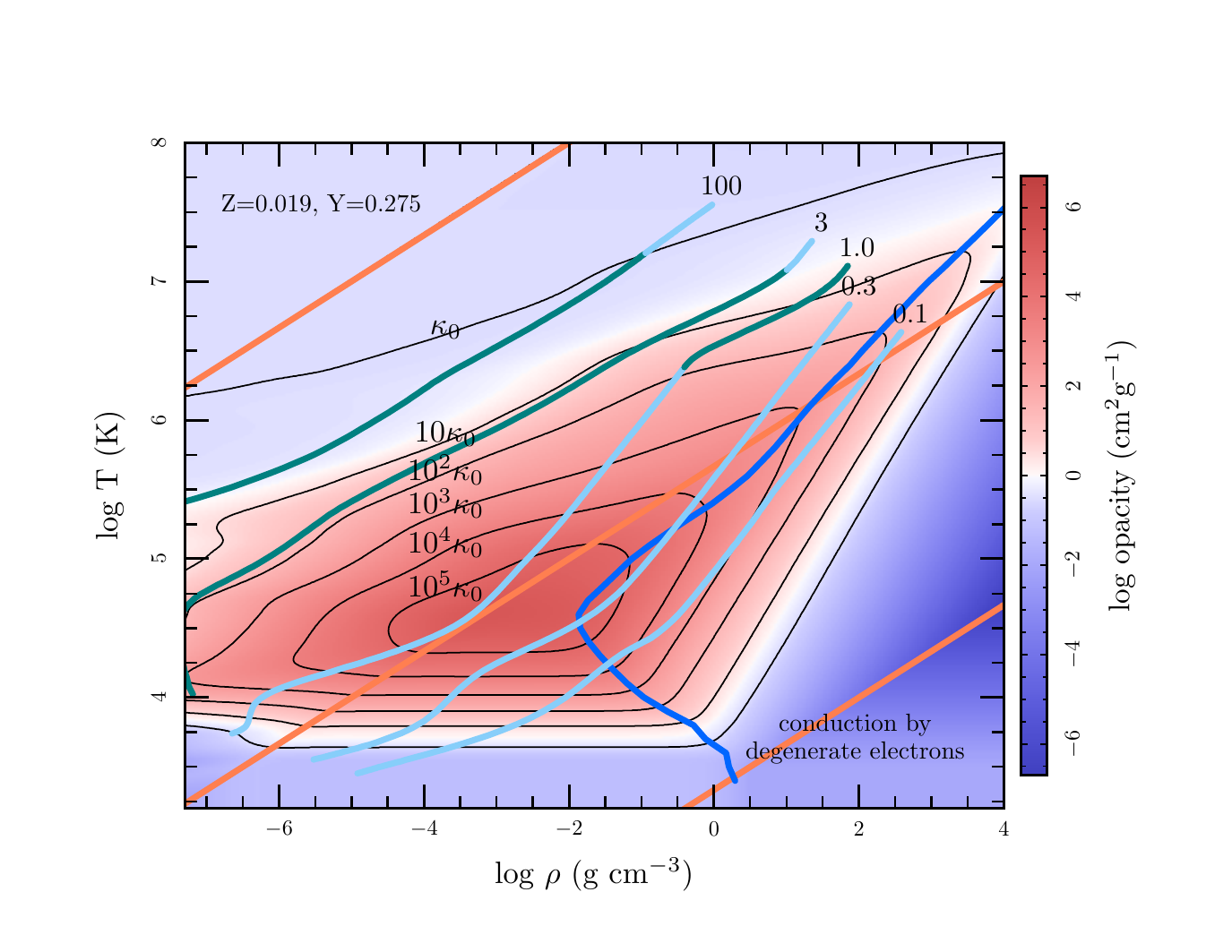}
\caption{The resulting $\MESA$ opacities for $Z=0.019, Y=0.275$. The underlying shades  show 
the value of $\kappa$, whereas the contours are in units of the electron scattering opacity, 
$\kappa_{0}=0.2(1+X) \ {\rm cm^2 \ g^{-1}}$. The orange lines show (top to bottom) 
where  $\log R=-8$, $\log R=1$ and $\log R=8$.
Stellar interior profiles for main sequence stars of mass $M=0.1,0.3,1.0, 3.0 \ \& \ 100\Msun$ are shown by the green(radiative regions )-light blue(convective regions) lines. Electron conduction dominates the opacity to the right of the  dark blue line (which is where  the radiative opacity equals the conductive opacity). 
\label{kapevaluations}}
\end{figure}

The absence of tabulated radiative opacities for $\log R> 1$ and $\log T<8.2$ (the region  below the heavy dashed line 
in Figure \ref{kapregions})  leads us to use the radiative opacity at $\log R = 1$ (for a specific $\log T$) when combining 
with the electron conduction opacities.  This  introduces errors in the $\MESA$ opacity table between $\log R=1$ and the 
region to the right of the dashed blue line in Figure \ref{kapregions} where conductive opacities become dominant.  However, as we show in 
Figure \ref{kapevaluations}, main sequence stars are always efficiently convective in this region of parameter space, alleviating the issue. 

The module $\kap$ gives the user the resulting opacities by interpolating in $\log T $ and $\log R$  with bicubic splines from $\inttwo$. The user has the option of either linear or cubic interpolation in $X$ and $Z$
and can specify whether to use the fixed metal (Type 1) tables or the varying C and O (Type 2) tables.  In the latter case, the user must specify the reference C and O mass fractions, usually corresponding to the C and O in the initial composition. 

\clearpage

For requests outside the  $\log T $ and $\log R$  boundaries, the following is done. The region to the left of $\log R=-8$ and below $\log T=8.7$ is electron scattering dominated, so the cross-section per electron is density independent. However, the increasing importance of the Compton effect as the  temperature increases (which is incorporated in the OPAL/OP tabulated opacities) must be included,  so we use the opacity from the table at $\log R=-8$ at the appropriate value of $\log T$. For higher temperatures ($\log T>8.7$)  electron-positron pairs become prevalent, as exhibited by the red dashed line that shows where the number of positrons and electrons from pair production exceeds the number of electrons from ionization.   $\MESA$ incorporates the enhancement to the opacity from these increasing numbers of leptons per baryon.

At the end of a star's life, low enough entropies can be reached 
that an opacity for $\log R>8$ is needed.  When $\kap$ is called in this region, we simply use the value at $\log R=8$ for the same $\log T$. For regions where $Z>0.1$, the table at $Z=0.1$ is used.

The resulting opacities for $Z=0.019$ and $Y=0.275$ are shown in Figure \ref{kapevaluations}, both
as a color code, and as contours relative to the electron scattering opacity, $\kappa_{0}=0.2(1+X) \ {\rm cm^2 \ g^{-1}}$. The orange lines show (top to bottom) 
where  $\log R=-8$, $\log R=1$ and $\log R=8$. We show a few stellar profiles for main sequence
stars as marked. The green parts of the line are where heat transfer is dominated by heat transport, 
requiring an opacity, whereas the light blue parts of the line are where the model is convective. As is evident, nearly all of the stellar cases of interest (shown by the green-blue lines) are safely within the boundaries or the $\MESA$ tables. The lack of radiative opacities in the higher density
region to the right of $\log R=1$ implies opacity uncertainties until the dark blue line is reached (where the conductive opacity takes over). However, the stellar models are convectively 
efficient in this region, so that the poor value for $\kappa$ does not impact the result as long as the convective 
zone's existence is independent of the opacity (the typical case for these stars, where the ionization zone causes  the convection). 

It is also possible to generate a new set of $\kap$ readable opacity tables using the 
\texttt{make\_kap} pre-processor.  The requirements are high-temperature radiative
opacities in the standard OPAL format  and low-temperature radiative opacities
in the number and format provided by \citet{ferg05}.\footnote{\url{http://webs.wichita.edu/physics/opacity}}
Specific high-temperature radiative opacities can be made by 
using the  OPAL site\footnote{\url{http://opalopacity.llnl.gov/new.html}}  or
the  Opacity Project site\footnote{\url{http://cdsweb.u-strasbg.fr/topbase/op.html}}.    

Since not all opacity sources can be placed in the tabular form desired by $\kap$, we have created a
module, \texttt{other\_kap}, that provides the user an opportunity to incorporate their own opacity source.
A simple flag tells $\MESAstar$ to call \texttt{other\_kap} rather than $\kap$, 
allowing for  experiments with new opacity schemes and  physics updates.
The first example of such an implementation that has now become a $\MESA$ module is
\texttt{karo}. It was developed  to
 study the stellar evolution effects of dust-driven winds in Carbon-rich stars, using the Rosseland opacities of \citet{led09} and the hydro-dynamical wind models of \citet{mat10}.

\subsection{Thermonuclear and weak reactions  \label{rates}}

The $\rates$ module contains thermonuclear reaction rates from
\citet[][CF88]{cau88} and \citet[][NACRE]{nacre}, with preference
given to the NACRE rate when available. The reaction rate library
includes more than 300 rates for elements up to Nickel, and includes
the weak reactions needed for Hydrogen burning (e.g. positron
emissions, electron captures), as well as neutron-proton conversions
and a few other electron and neutron capture reactions.  Significant
updates to the NACRE rates have been included for
$^{14}$N(p,$\gamma$)$^{15}$O \citep{imb04}, triple-$\alpha$
\citep{fyn05}, $^{14}$N($\alpha,\gamma$)$^{18}$F \citep{ga00} and
$^{12}$C($\alpha,\gamma$)$^{16}$O \citep{kunz02}.  In these special
cases, the rate can be selected from CF88, NACRE, or the newer
reference by the user at run time.

The $\weak$ module calculates lepton captures and $\beta$-decay rates
for the high densities and temperatures encountered in late stages of
stellar evolution.  The rates are based on the tabulations of
\citet{full85}, \citet{oda94}, and \citet{lmp00} for isotopes with $45
< {\rm A} < 65$. The most recent
tabulations of \citet{lmp00} take precedence, followed by
\citet{oda94}, then \citet{full85}.  The user can override this to
create tables using any combination of these or other sources.

The $\screen$ module calculates electron screening factors for thermonuclear
reactions in both the weak and strong regimes. The treatment
has two options. One is based on \citet{dew73} and \citet{grab73}.
The other\footnote{\url{http://cococubed.asu.edu/code\_pages/codes.shtml}\label{coco}}  
combines  \citet{grab73} in the weak regime and \citet{aj78} with plasma parameters 
from \citet{itoh79} in the strong regime.

The $\neu$ module calculates energy loss rates and their derivatives
from neutrinos generated by a range of processes including plasmon
decay, pair annihilation, Bremsstrahlung, recombination and
photo-neutrinos (i.e. neutrino pair production in Compton
scattering). It is based on the publicly available routine (see
footnote \ref{coco}) derived from the fitting formulas of
\citet{itoh96}.

\subsection{Nuclear reaction networks  \label{net}}

The $\net$ module implements nuclear reaction networks and is derived
from publicly available code (see footnote \ref{coco}). It includes a
``basic'' network of 8 isotopes: $^1$H, $^3$He, $^4$He, $^{12}$C,
$^{14}$N, $^{16}$O, $^{20}$Ne, and $^{24}$Mg, and extended networks
for more detailed calculations including coverage of hot CNO
reactions, $\alpha$-capture chains, ($\alpha$,p)+(p,$\gamma$)
reactions, and heavy-ion reactions \citep{tim99}. In addition to using
existing networks, the user can create a new network by listing the
desired isotopes and reactions in a data file that is read at run
time. The amount of heat deposited in the plasma by reactions is
derived from the nuclear masses in $\chem$, taken from the JINA
Reaclib database \citep{rau00,sak06,cyb10}, and accounts for positron
annihilations and energy lost to weak neutrinos, using
\citet{bah97,bah02} for the hydrogen burning reactions. The list of
approximately 350 reactions is stored in a data file that catalogs the
reaction name, the input and output species, and their heat release.

\begin{deluxetable}{lcccccc}
\tablecolumns{7}
\tablewidth{0pc}
\tabletypesize{\footnotesize}
\tablecaption{Comparison of 1-zone Solar burn results at 10 Gyr\label{burnH}}
\tablehead{ \colhead{Network} & \colhead{log$_{10}$ $\mathrm{e}_{\rm nuc}$} & \colhead{log$_{10}$ X($^1$H)} &
  \colhead{log$_{10}$ X($^4$He)} & \colhead{log$_{10}$ X($^{12}$C)} & \colhead{log$_{10}$ X($^{14}$N)} & \colhead{log$_{10}$ X($^{16}$O)} 
}
\startdata
$\jina$ 25   &     18.63757961   &    -3.87550319   &    -0.008144854   &   -4.40235799  &  -1.9195882  &   -3.07400339 \\
$\net$ 25    &     18.63685339   &    -3.87550517   &    -0.008145036   &   -4.40235799  &  -1.9195882  &   -3.07400333 \\
$\net$ 8     &     18.63675658   &    -3.93650004   &    -0.008137607   &   -4.39650625  &  -1.9135911  &   -3.04585377 \\
\enddata
\end{deluxetable}

\begin{deluxetable}{lccccc}
\tablecolumns{6}
\tablewidth{0pc}
\tabletypesize{\footnotesize}
\tablecaption{Comparison of 1-zone He-burn results at 10 Gyr\label{burnHe}}
\tablehead{ \colhead{Network} & \colhead{log$_{10}$ $\mathrm{e}_{\rm nuc}$} & \colhead{log$_{10}$ X($^{12}$C)} & \colhead{log$_{10}$ X($^{16}$O)} &  \colhead{log$_{10}$ X($^{22}$Ne)} &  \colhead{log$_{10}$ X($^{26}$Mg)}}
\startdata
jina 200   &    17.9085633    &    -0.721578469  &    -0.108630252  &    -1.50380756  &  -4.01520633 \\
net 11     &    17.9086380    &    -0.721576540  &    -0.108630957  &    -1.50385214  &  -3.99780015 \\
net 8      &    17.9083877    &    -0.718866029  &    -0.107692784  &    \nodata      &  \nodata     \\
\enddata
\end{deluxetable}

\begin{figure}[p]
\plotone{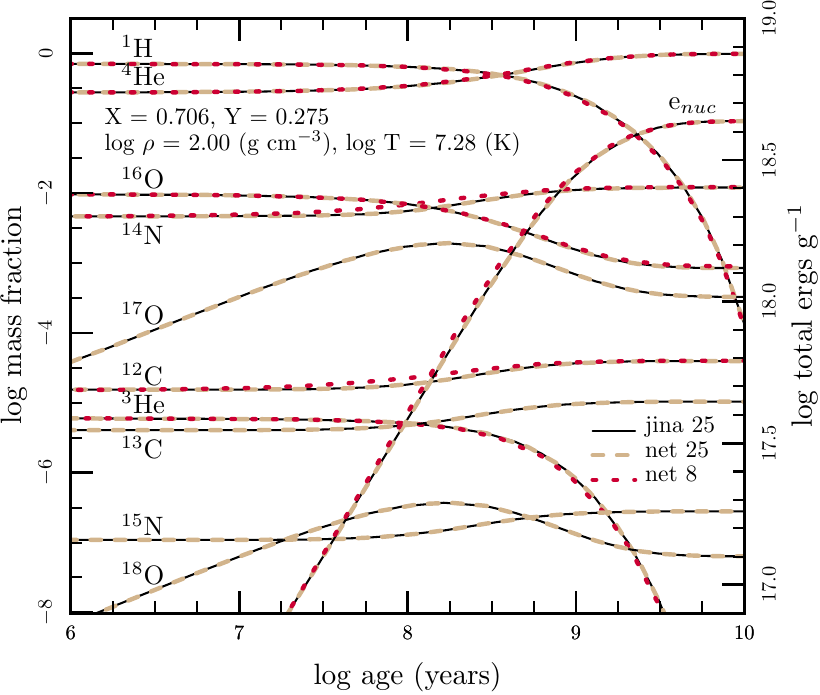}
\caption{A 1-zone hydrogen burn at constant $T= 19\times 10^6$ K and $\rho=100
\ {\rm g \ cm^{-3}}$ by three different networks.  The number
following $\net$ or $\jina$ indicates the number of isotopes
considered in that network. The 25 isotope networks expand on the 8
isotope network by including minor contributors to the pp and CNO
cycles.  The plot shows the evolution of the mass fraction abundances
of the 10 most abundant isotopes and net energy generation per unit
mass, $\mathrm{e}_{\rm nuc}$ (ergs g$^{-1}$), as a function of
time. The left-hand axis shows the mass fraction while the right-hand
axis shows the net energy generation per unit mass.
\label{nethydburn}}
\end{figure}

\begin{figure}[p]
\plotone{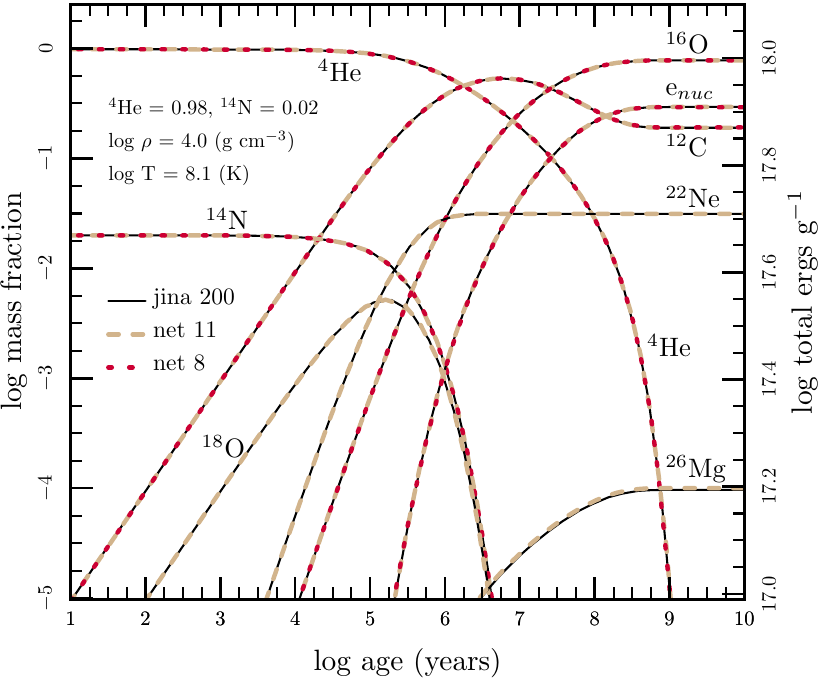}
\caption{Equivalent to Figure \ref{nethydburn} but now showing a 1-zone helium
burn at constant $\log T=8.1$, $\log \rho=4.0$.  The ``$\net$ 11''
network adds $^{18}$O, $^{22}$Ne, and $^{26}$Mg to the 8 isotope
network; ``$\jina$ 200'' includes about 200 isotopes up to $^{71}$Ge.
\label{netheburn}}
\end{figure}

The $\jina$ module is an alternative nuclear network module that
specializes in large networks. It is based on the `netjina' package by
Ed Brown and uses the JINA Reaclib database for thermonuclear reaction
rates \citep{rau00,sak06},\footnote{\url{http://groups.nscl.msu.edu/jina/reaclib/db/index.php}}
the $\rates$ and $\weak$ modules for weak interactions, and $\screen$
for electron screening. Most importantly, it allows the user to create
large nuclear networks by specifying the list of isotopes to consider.
All nuclear reactions (both strong and weak) linking the isotopes in
the set are automatically included in the network. In all, $\jina$
covers more than 76,000 nuclear reactions involving more than 4,500
isotopes.  The $\jina$ module is slower than $\net$ for small networks
but the flexibility and capacity to handle large networks make it
advantageous in some cases.

Both $\net$ and $\jina$ include one-zone burn routines that operate on
a user-defined initial composition, nuclear network, and a trajectory
comprising density and temperature as a function of time. The one-zone
burn routines interface with $\mtx$ and $\num$, enabling the use of
the sparse matrix solver, which substantially improves performance
compared to the dense matrix solver for networks of more than a few
hundred isotopes.  Figures \ref{nethydburn} and \ref{netheburn}
demonstrate these one-zone routines operating on conditions
appropriate for the Sun on the main sequence and for a core He-burning
star, respectively. Both examples were evolved at fixed density and
temperature for 10 Gyr. Each figure compares three networks of varying
size in terms of the mass fractions and net energy generation per unit
mass, $\mathrm{e}_{\rm nuc}$ (ergs g$^{-1}$), produced. Tables \ref{burnH}
and \ref{burnHe} complement Figures \ref{nethydburn} and
\ref{netheburn}, respectively, by listing the final values from each
of the one-zone burn simulations.  These comparisons indicate that the
8 isotope network produces results that agree with larger networks to
4-5 significant figures in net energy generation per unit mass and
generally 2-3 significant figures in the mass fractions of various
isotopes. Hence, the 8 isotope network is sufficiently accurate to
describe the  energy generation for hydrogen and helium burning.  

\section{Macrophysics\label{macro}}

\subsection{The mixing length theory of convection \label{mlt}}

The $\mlt$ module implements the standard mixing length theory (MLT) 
of convection as presented by \citet[][chapter 14]{cox68}. There are 
options both for computing the actual temperature gradient, $\nabla_T$, 
when the total luminosity, $L$,  is specified and for computing the convective 
luminosity, $L_{\rm conv}$, when $\nabla_T$ is specified. The $\mlt$ module calculates diffusion
coefficients for those codes, such as $\MESAstar$, that treat convective mixing of elements as a 
diffusive process. The quantities listed in Table \ref{mltvars} and their partial derivatives
with respect to several physical variables are returned by the $\mlt$ module.

In addition to the standard MLT of Cox \& Giuli, the $\mlt$ module includes the 
option to use the modified MLT of \citet{hvb65}. Whereas the standard MLT assumes 
high optical depths and no radiative losses, the \citet{hvb65} variation allows the convective 
efficiency to vary with the opaqueness of the convective element, an important effect for convective 
zones near the outer layers of stars. If the \citet{hvb65} option is used, the parameter $\nu$ (a mixing 
length velocity multiplier)  and $y$ (a parameter that sets the temperature gradient in a rising bubble) 
may be set by the user. They default to the recommended values of $y=1/3$ and $\nu=8$. 

Towards the center of a star, the commonly used definition of the pressure scale height, 
$\lambda_P = P/g\rho $, diverges as $g\rightarrow 0$. Therefore, we provide the option of using the 
alternate definition of \citet{egg71}, $\lambda_{P}' = (P/G\rho^2)^{1/2}$, when 
$\lambda_{P}' < \lambda_P$. At the center of the star, $\lambda_{P}' \sim R$.

\begin{deluxetable}{ccc}
\tablecolumns{3}
\tablewidth{0pc}
\tablecaption{$\MESA$ $\mlt$  output quantities and units\label{mltvars}}
\tablehead{\colhead{Output}&\colhead{Definition}&\colhead{Units}}
\startdata
$\nabla_T$ & actual temperature gradient\tablenotemark{a} & Dimensionless \\
$\nabla_{\rm rad}$ & radiative temperature gradient\tablenotemark{a} & Dimensionless\\
$L_{\rm conv}$ & convective luminosity\tablenotemark{b}  & ${\rm ergs \ s^{-1}}$\\
$L $ & total luminosity\tablenotemark{c} & ${\rm ergs \ s^{-1}}$\\
$\lambda_P$ & pressure scale height & ${\rm cm}$ \\
$\Lambda$ & mixing length ($\equiv \alpha_{MLT} \lambda_P$) & ${\rm cm}$\\
$v_{\rm conv}$ & convective velocity & ${\rm cm \ s^{-1}}$\\
$D$ & Eulerian diffusion coefficient & ${\rm cm^2 \ s^{-1}}$ \\
$\sigma$ & Lagrangian diffusion coefficient ($\equiv D(4\pi r^2 \rho)^2$) & ${\rm g^2 \ s^{-1}}$\\
\enddata
\tablenotetext{a}{Only when $L$ is specified.}
\tablenotetext{b}{Only when $\nabla_T$ is specified.}
\tablenotetext{c}{Only when $\nabla_T$ is specified and $L_{\rm conv} > 0$.}
\end{deluxetable}
\clearpage

\subsection{Convective overshoot mixing  \label{overshoot}}

As described in \S\ref{s.stareqns}, $\MESAstar$ treats convective mixing
as a time-dependent, diffusive process with a diffusion coefficient, $D$, 
determined by the $\mlt$ module. In the absence of a 3-D
hydrodynamical treatment of convection it is necessary to account for
the hydrodynamical mixing instabilities at convective boundaries,
termed overshoot mixing, via a parametric model. After the MLT
calculations have been performed, $\MESAstar$ sets the overshoot
mixing diffusion coefficient 
\begin{equation}
D_{\rm OV} = D_{\rm conv,0} \exp\left(-{2z\over f\lambda_{P,0}}\right), 
\end{equation}
where $D_{\rm conv,0}$ is the MLT derived diffusion coefficient at a user-defined location near the 
Schwarzschild boundary, $\lambda_{P,0}$ is the pressure scale height at that location, 
$z$ is the distance in the radiative layer away from that location, and 
$f$ is an adjustable parameter \citep{her00}. In $\MESAstar$ the adjustable parameter, $f$,
may have different values at the upper and lower convective boundaries for non-burning, H-burning, 
He-burning, and metal-burning convection zones.

Parameters are provided to allow the user to set a lower limit on
$D_{\rm OV}$ below which overshoot mixing is neglected and to limit the
region of the star over which overshoot mixing will be considered. So as to model 
the $^{13}$C pocket needed for s-process nucleosynthesis,
$\MESAstar$ also allows an increase in the overshooting parameter at the
bottom of the convective envelope during the third dredge-up compared
to the inter-pulse value \citep{lug03}. There is also an option to
change the value of overshoot mixing at the bottom of the AGB thermal
pulse-driven convection zone compared to the standard value chosen for
the bottom of the He-burning convection zone.

\subsection{Atmosphere boundary conditions  \label{atm}}

As described in \S \ref{s.stareqns}, the pressure, $P_s$, and temperature, $T_s$, at the top of the outermost cell in $\MESAstar$ must be set by an atmospheric model. This is done 
by the $\atm$ module, which uses $M$, $R$, and $L$ to 
provide $P_s$ and $T_s$. It also gives partial derivatives of $T_s$ and $P_s$ with respect to the input variables. The $\atm$ module assumes the plane parallel limit, so that the relevant variables 
are $g=GM/R^2$ and $\Teff^4=L/4\pi \sigma_{\rm SB} R^2$. With some options, the user 
must specify the optical depth $\tau_s$ to the base of the atmosphere, whereas in other cases, the $\atm$ module has an implicit value. Three methods are supplied by $\atm$: direct integrations, interpolations in model atmosphere tables, and a ``simple" recipe. 

The integrations of the hydrostatic balance equation,  $dP_{\rm gas}/d\tau=g/\kappa - (a/3) dT^4/d\tau$, with 
$d\tau=-\kappa \rho dr$ are performed using either the  
relation $T^4(\tau)=3T_{\rm eff}^4(\tau+2/3)/4 $ \citep{edd}, 
or the specific $T-\tau$ relation of \citet{ks66}. These integrations start at
$\tau = 10^{-5}$ and end at a user specified stopping point, $\tau_s$, 
which defaults to $\tau_s=2/3$ (0.312) for Eddington (Krishna Swamy).\footnote{If the first attempt to integrate fails, the code 
makes two further attempts, each time increasing the initial $\tau$ by a factor of 
10. The integration is carried out with the Dormand-Price integrator from the $\num$ 
module.} The routine integrates the gas pressure and then adds the radiation pressure 
at the stopping point to get $P_s$. 

\begin{figure}[H]
\plotone{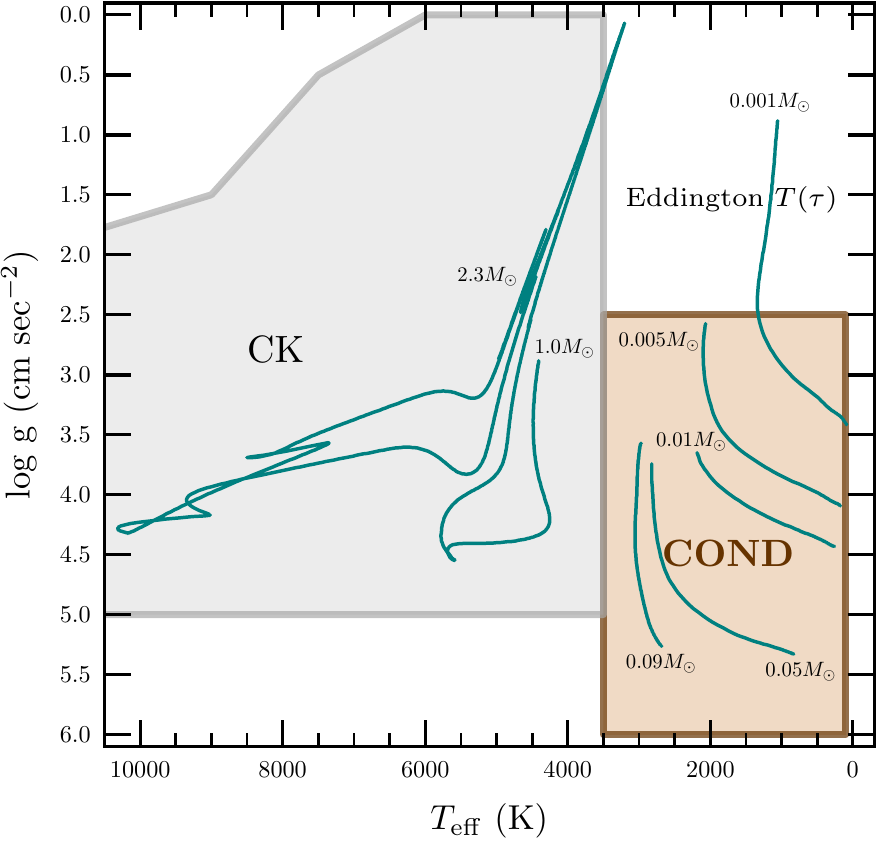}
\caption{The range of $\Teff$ and $\logg$ covered by the $\MESA$ $\atm$ tables for $\tau_s=100$ and Solar metallicity. The CK region uses the tables of \citet{cas03}, whereas the COND region uses 
\citet{all01}. At lower $\logg$ and cold regions, we use direct integrations of the Eddington $T-\tau$ relation. The green lines show evolutionary tracks of stars, brown dwarfs and giant planets of the 
noted masses. 
\label{tau100}}
\end{figure}

The $\MESA$ model atmosphere tables come in two forms. 
The $\MESA$ photospheric tables (which return $T_s\equiv\Teff$ and assume that $\tau_s\approx 1$)
cover $\log Z/Z_{\odot}=-4$ to $+0.5$ assuming the \citet{gn93} 
Solar abundance mixture. They span $\logg = -0.5$ to $5.5$ at 0.5 dex intervals
and $\Teff = $2,000-50,000K at 250K intervals. They are constructed, in precedence order, 
with, first, the PHOENIX \citep{phxa,phxb} 
model atmospheres (which span $\logg = -0.5$ to $5.5$ and $\Teff = 2,000$ to $10,000$ K); 
and second, the \citet{cas03} model atmospheres (which span $\logg=0$ to 
$5$ and $\Teff = 3500$ to $50,000$ K). In regions where neither  table is available, we 
generate the $\MESA$ table entry using the integrations described above with the Eddington T-$\tau$ relation. The second $\MESA$ table is for Solar metallicity and 
gives $P_s$ and $T_s$ at $\tau_s=100$. It is primarily for the evolution of low mass stars, 
brown dwarfs, and giant planets. It is constructed from \citet{cas03}, and for $\Teff<3000$K, the COND model atmospheres 
\citep{all01} which assume gravitational settling of those elements that form dust, depleting  those elements from the photosphere. 
Figure \ref{tau100} shows the regions where the different sources are used, and in those 
regions where there are no published results, we use the integration of the  Eddington T-$\tau$ relation. 

Finally, there is a simple option where the user specifies $\tau_s$ and we use the constant opacity, $\kappa_s$, solution of radiative diffusion, 
\begin{equation}
P_s = \frac{\tau_s g}{\kappa_s} \left[1 + 1.6\times10^{-4} \kappa_s\left(\frac{L/\Lsun}{M/\Msun}\right)\right],\label{atm:P}
\end{equation}
where the factor in square brackets accounts for the nonzero radiation pressure
\citep[see, e.g.,][Section 20.1]{cox68}.
The temperature is simply given by the Eddington relation. The user can either specify $\kappa_s$ or
it will be calculated in an iterative manner using the initial value of $P_s$ from an initial guess at $\kappa_s$ (usually given by $\MESAstar$ as the value in the outermost cell; see \S \ref{s.stareqns}). 
In addition, the $\atm$ module has the option to revert to Equation (\ref{atm:P}) if a model wanders outside the range of the currently used model atmosphere tables or if the atmosphere integration fails for any reason.

\subsection{Diffusion and gravitational settling \label{diff}}

$\MESA$ $\diff$ calculates particle diffusion and gravitational settling by solving 
Burger's equations using the method and diffusion coefficients of \citet{tho94}. 
The transport of material is computed using the semi-implicit, finite difference 
scheme described by \citet{iben85}. Radiative levitation is not presently included.  The $\diff$ module treats the elements present in the stellar model as belonging to ``classes" defined by the user in terms of ranges of atomic masses. For each class, the user specifies a representative
isotope, and all members of that class are treated identically with their 
diffusion velocities determined by the representative isotope, and the diffusion equation solved with the mass fraction in that class. 
The caller can either specify the ionic charge for each class at each
cell in the model or have the charge calculated by the $\ioniz$ module, which 
estimates the typical ionic charge as a function of 
$T$, $\rho$, and free electrons per nucleon from  \citet{paq86}. 

\begin{figure}[H]
\plotone{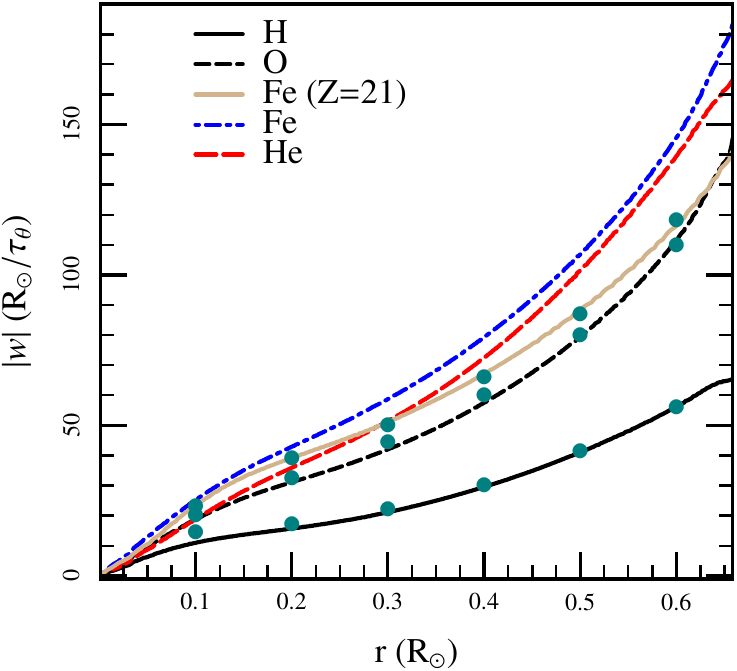}
\caption{
The absolute values of the diffusion velocities from $\diff$ (lines) and those published by \citet{tho94}. All results are plotted in units 
of $\Rsun/\tau_\theta$, where $\tau_\theta=6\times 10^{13} \ {\rm yr}$ is the 
characteristic diffusion timescale for the Sun \citep{tho94}. The dark solid and dashed lines are the $\diff$ results for H and O. The filled green circles show the results of \citet{tho94} for H, O and Fe (${\rm Z}=21$).  The $\diff$ results for Helium are shown as the dashed red line. The $\diff$ results for Fe  include one for ${\rm Z}=21$ (dotted blue line) and one for ionization states determined by $\ioniz$ (the dot-dashed blue line). 
 \label{thoulcompare}}
\end{figure}
\clearpage

The lines in Figure \ref{thoulcompare} plot four classes (H, He, O, and Fe) with a solar model from $\MESAstar$
and compares where possible to the results from Figure 9 of \citet{tho94}, shown by the filled green circles. The agreement is excellent for H, O and Fe (when we fix Fe to have the ${\rm Z}=21$ ionization state chosen by \citet{tho94}). \citet{tho94}  did not exhibit the He velocity, so we have no comparison. For Fe, we also show the diffusion velocity when $\ioniz$ finds a changing ionization state in the ${\rm Z}=16, 17, 18$ region (shown by the upper dot-dashed blue line), highlighting the 
need to better determine the Fe ionization state \citep{gb08}. We also compared the $\diff$ output to the recent calculations of \citet{gb08}, finding  agreement at better than 5\% for the Fe case at ${\rm Z}=26$ and for O. 

The diffusion calculation can be restricted to areas where the  temperature is above some minimum value, 
or where the mass fraction of a diffusing element is above some minimum value, aiding the 
convergence of solutions in a variety of environments. The physics implementation is presently limited to regions where the Coulomb coupling parameter, $\Gamma$, is less than unity. At present, this inhibits an accurate calculation for segregation and settling of the remaining envelope H and He envelope on a cooling white dwarf.

\subsection{Testing $\MESA$ modules in an existing stellar evolution code \label{dsep}}

The complex, nonlinear behavior of stellar structure and evolution models 
makes it difficult to disentangle the effects of model components (e.g., EOS, 
opacities, boundary conditions, etc.) when comparing results of separate codes. By design, 
the modularity of $\MESA$ allows individual physics modules to be incorporated into
an existing stellar evolution code, tested, and then compared against the prior implementation
of comparable physics in the same code. 

During the  development of $\MESA$, several $\MESA$ modules were integrated 
into the Dartmouth Stellar Evolution Program \citep[DSEP,][]{dot08}. 
This section reports the results of using four $\MESA$ modules, $\eos$, $\kap$, $\atm$, and 
$\mlt$, in DSEP to compute the evolution of a $1.0 \Msun$ star with initial values of $X=0.70$ and $Z=0.02$. The star 
was evolved from the fully convective pre-main sequence to the onset of the core He flash. This 
was done six times: once, as the control case, using only DSEP routines and no $\MESA$ modules; 
next, using each of four $\MESA$ modules individually; and, finally, using the four $\MESA$ modules 
at the same time in DSEP.

DSEP employs a $\rho(P,T)$ EOS and so the $\MESA$ $P_{\rm gas}-T$ tables 
were used during the $\eos$ test. Though DSEP and $\kap$ use the same sources for radiative opacities, 
they differ in interpolation methods and the treatment of electron conduction opacities \citep[see][for a 
thorough list of the physics in DSEP]{bc06}. When $\atm$ was tested, we used the Eddington grey atmosphere 
model integrated to $\tau=2/3$. DSEP uses the \citet{hvb65} modification of the mixing length 
theory, which is available in $\mlt$, and assumes that convective regions are instantaneously mixed to a uniform composition.

\begin{figure}[H]
\plotone{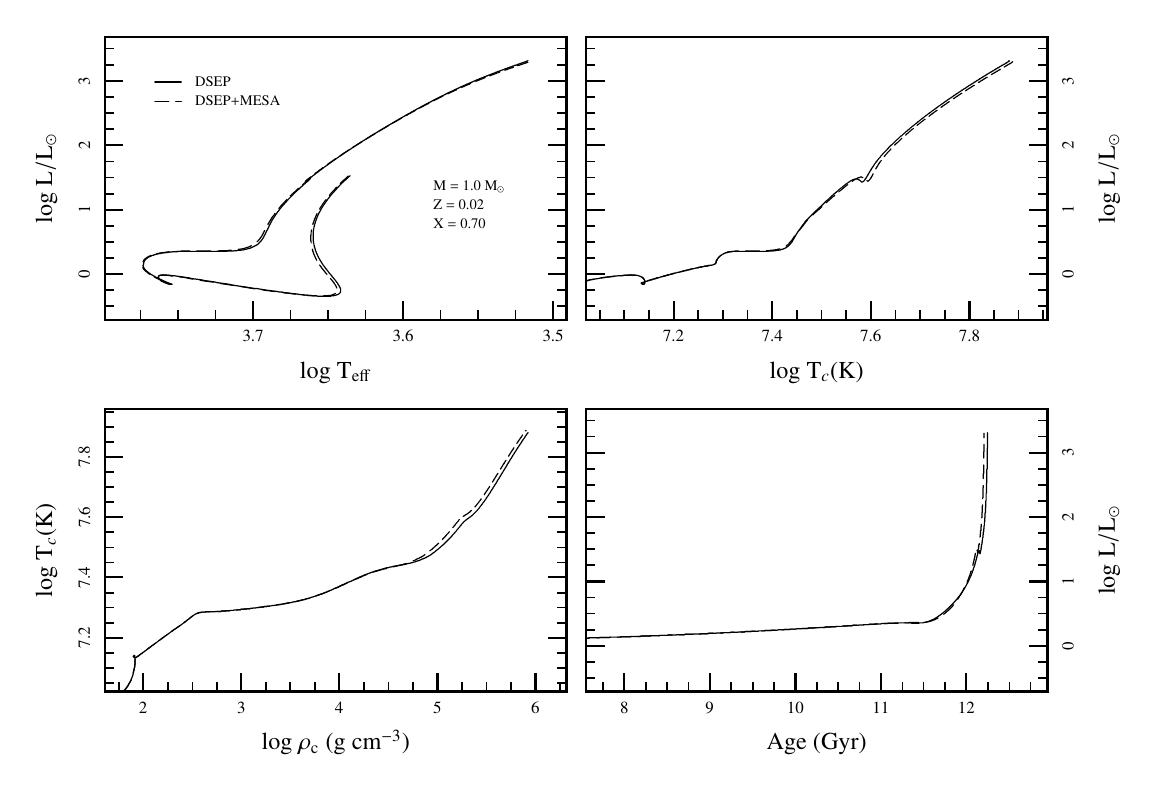}
\caption{Comparison of DSEP tracks using built-in physics modules and
$\MESA$ modules for opacities, EOS, mixing length theory,
and the atmospheric boundary condition. These tracks are for a $1.0\Msun$ 
star with initial $X=0.70$ and $Z=0.02$ evolved from the fully convective pre-main 
sequence to the onset of the He core flash.
Only the H-R diagram shows the full evolutionary track.  The $T_c$ panels omit the pre-main sequence in order to highlight the regions where the differences are most pronounced; the lifetime panel focuses on the end of the main sequence and red giant phase for the same reason.
\label{dsep2x2}}
\end{figure}

DSEP tracks employing either the $\atm$ or the $\mlt$ modules produce results that agree with the 
DSEP-only track to about 1 part in $10^4$. DSEP tracks employing the $\kap$ and $\eos$ modules
exhibit some difference when compared to the DSEP-only track but, even in these cases, the main
sequence lifetime differs by less than $0.3$\% and $\Teff$ differs by less than 10K along the
main sequence. As shown in Figure \ref{dsep2x2}, the largest discrepancy between the DSEP-only track and the one that employs all
four $\MESA$ modules appears in the $T_c-\rho_c$ diagram when $\rho_c > 3\times 10^4 {\rm g \ cm^{-3}}$, 
corresponding to the growing helium core in the center of the red giant. Above $\log \rho_c=4$, the 
track employing $\MESA$ modules is hotter than the DSEP-only track by $\sim0.02$ in $\log T_c$ at constant
$\log \rho_c$. The center of the model has entered the region of electron degeneracy and electron conduction 
has become an important source of opacity. The majority of the difference is due to the EOS whereas
the opacity difference amounts to about $-0.005$ in $\log T_c$, in the opposite direction to the EOS.
The hotter conditions produced by the $\eos$ module is likely the cause for the slightly shorter RGB lifetime
that can be seen in Figure \ref{dsep2x2}.

\section{Stellar structure and evolution\label{star}}
$\MESAstar$ is a full-featured stellar structure and evolution library
that utilizes the numerics and physics modules described in
\S's \ref{num}-\ref{macro}.  It provides a clean-sheet implementation
of a Henyey style code \citep{hen59} with automatic mesh refinement,
analytic Jacobians, and coupled solution of the structure and
composition equations. The design and implementation of $\MESAstar$
was influenced by a number stellar evolution and hydrodynamic codes
that were made available to us: EV \citep{egg71}, EVOL \citep{her04},
EZ \citep{pax04}, FLASH-the-tortoise \citep{les06},
GARSTEC \citep{garstec}, NOVA \citep{star00}, TITAN \citep{geh94}, and
TYCHO \citep{you05}.  

We now briefly describe the primary components of $\MESAstar$.
$\MESAstar$ first reads the input files and initializes the physics
modules (see \S \ref{s.startmodel}) to create a nuclear reaction
network and access the EOS and opacity data. The specified starting
model or pre-main sequence model is then loaded into memory
(see \S \ref{s.startmodel}), and the evolution loop is entered.  The
procedure for one timestep has four basic elements. First, it prepares to take a
new timestep by remeshing the model if necessary
(\S \ref{s.mesh} and \ref{s.timestep}).  Second, it adjusts
the model to reflect mass loss by winds or mass gain from accretion (\S \ref{s.masschanges}) ,
adjusts abundances for element diffusion (\S \ref{diff}), determines the convective 
diffusion coefficients (\S \ref{mlt} and \ref{overshoot}), and solves for the new structure and composition 
(\S \ref{s.stareqns} and \ref{s.converge}) using the Newton-Raphson solver (\S \ref{num}).  
Third, the next timestep is estimated (\S \ref{s.timestep}).  Fourth, output 
files are generated (\S \ref{s.startmodel}).

\subsection{Starting models and basic input/output \label{s.startmodel}}

$\MESAstar$ receives basic input from two Fortran namelist files.  One
file specifies the type of evolutionary calculation to be performed,
the type of input model to use, the source of EOS and opacity data,
the chemical composition and nuclear network, and other properties of the
input model. The second file specifies the controls and options to be
applied during the evolution.

There are two ways to start a new evolutionary sequence with
$\MESAstar$.  The first is to use a saved model from a previous run.
A variety of saved models are distributed with $\MESA$ as
a convenience.  These saved models fall into three general categories:
(1) Zero Age Main Sequence (ZAMS) models for $Z=0.02$ with 32 masses
between 0.08 and 100$\Msun$ ($\MESAstar$ will automatically
interpolate any mass within this range); (2) very low mass, pre-main
sequence models for $Z=0.02$ and masses from 0.001 to 0.025$\Msun$; and
(3) white dwarf models for $Z=0.02$ with He cores of $0.15-0.45\Msun$,
C/O cores of $0.496-1.025\Msun$, and O/Ne cores of $1.259-1.376\Msun$.
The user can also create saved models for essentially any purpose
through available controls.

The second way to start a new evolution is to create a pre-main
sequence (PMS) model by specifying the mass, $M$, a uniform
composition, a luminosity, and a central temperature, $T_c$ low enough
that nuclear burning is inconsequential ($T_c=9\times10^5$ K by
default).  For a fixed $T_c$ and composition, the total mass depends
only on the central density, $\rho_c$. An initial guess for $\rho_c$
is made by using the $n=1.5$ polytrope, which is appropriate for a
fully convective star, but we do not assume the star is fully
convective during the subsequent search for a converged PMS model.
Instead, $\MESAstar$ uses the $\mlt$, $\eos$, and Newton solver from
$\num$ to search for a $\rho_c$ that gives a model of the desired mass.
The PMS routine presently creates starting models for $0.02\le M/\Msun \le 50$.  
Beyond these limits we find challenges converging the generated PMS model within
the $\MESAstar$ evolutionary loop. For such cases it is currently
better to generate a starting model within the acceptable mass range,
save it, relax it to a new mass with a specified mass gain or loss
(see \S \ref{s.masschanges}), and save that model.

$\MESAstar$ has the ability to create a binary file of its complete
current state, called a photo, at user-specified timestep intervals.
Restarting from a photo ensures no differences in the ensuing
evolution.  When restarting from a photo, many controls and options
can be changed.  A photo is different than a saved model in that a saved
model is a text file containing a minimal description of the structure
and composition but does not have enough information to allow a
perfect restart. However, saved models are not tied to a particular 
version of the code and therefore are suitable for long term use 
or sharing with other users.

There are two additional types of output files, logs and profiles. A
log records evolutionary properties over time such as stellar age, 
current mass, and a wide array of other quantities.  A profile
records model properties at a specified timestep at each zone from surface 
to center. $\MESAstar$ can also output models in the FGONG
format\footnote{ \url{http://owww.phys.au.dk/~jcd/solar\_models/}} for
use with stellar pulsation codes and \texttt{se} output for nucleosynthesis 
post-processing with NuGrid codes.\footnote{\url{http://forum.astro.keele.ac.uk:8080/nugrid}}
Finally, a few simple lines of
user-supplied code allows for saving variables or combinations of
variables that are not in the list of predefined options.

\subsection{Structure and composition equations \label{s.stareqns}}

$\MESAstar$ builds 1-D, spherically-symmetric models by dividing the
structure into cells, anywhere from hundreds to thousands depending on
the complexity of nuclear burning, gradients of state variables, composition, 
and various tolerances.  Cells are numbered starting with one at the
surface and increasing inward.  $\MESAstar$ does not require the
structure equations to be solved separately from the composition
equations (operator splitting).  Instead, it simultaneously solves the
full set of coupled equations for all cells from the surface to the
center.  The solution of the equations is done by the Newton solver
from $\num$ using either banded or sparse matrix routines from $\mtx$.
The partial derivatives for use by the solver are calculated
analytically using the partials returned by modules such as $\eos$,
$\kap$, and $\net$.

\begin{figure}[H]
\plotone{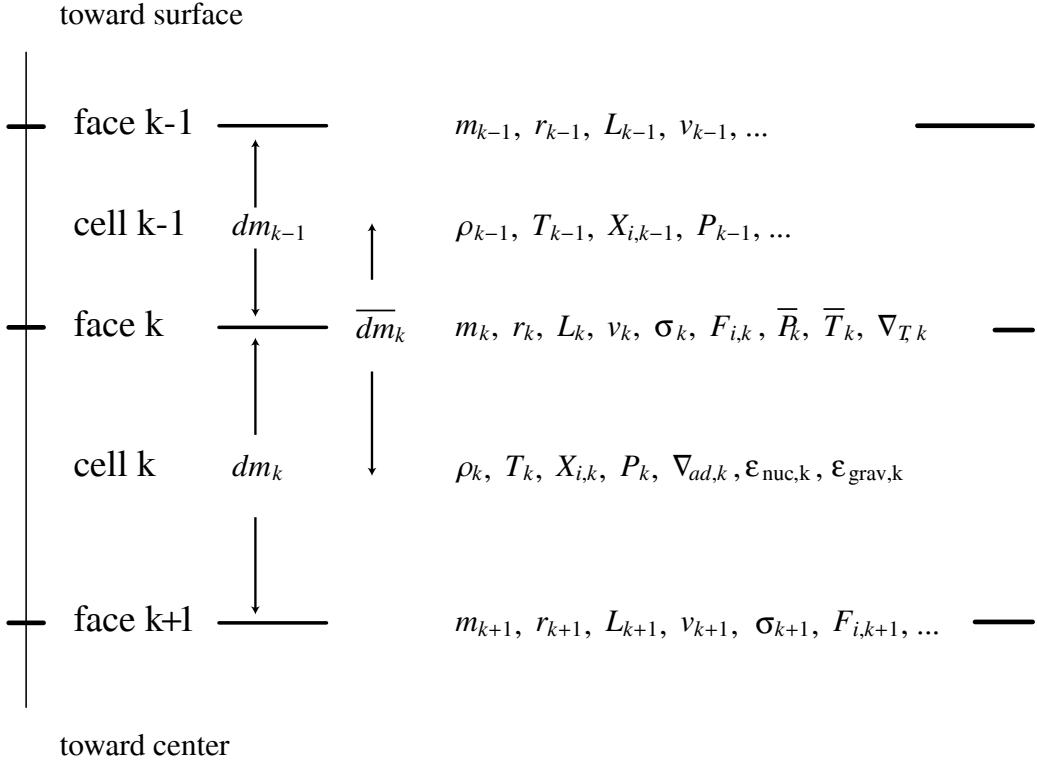}
\caption{Schematic of some cell and face variables for $\MESAstar$.
\label{f.structure}}
\end{figure}

Each cell has some variables that are mass-averaged and others that are defined at
the outer face, as shown in Figure \ref{f.structure}. This way of defining the variables
is a consequence of the finite volume, flux conservation formulation of the equations
and improves stability and efficiency \citep{sug81}.
The inner boundary of the innermost cell is usually the center of the star and,
therefore, has radius, luminosity, and velocity equal to zero. Nonzero center
values can be used for applications that remove the underlying star (e.g., the
envelope of a neutron star), in which case the user must define the values of
$M_c$ and $L_c$ at the inner radius $R_c$. The cell mass-averaged variables
are density $\rho_k$,
temperature $T_k$, and mass fraction vector $X_{i,k}$. The boundary variables
are mass interior to
the face $m_k$, radius $r_k$, luminosity $L_k$, and velocity $v_k$. In addition to
these basic variables, composite variables are calculated for every cell and face, such as
$\epsilon_{\rm nuc}$, $\kappa$,  $\sigma_k$, and $F_k$ (see
Table \ref{vardefs} for variable definitions).
All variables are evaluated at time $t + \delta~t$ unless otherwise specified.

The density evolution of cell $k$ is determined by a finite volume form of the 
mass conservation equation
\begin{equation}
\rho_k = \frac{dm_k}{(4/3) \pi (r_k^3 - r_{k+1}^3)}
\enskip .
\label{e.radius1}
\end{equation}
For the innermost cell, $r_{k+1}$ is replaced by the inner boundary
condition which is typically zero but can be nonzero for some applications.  
We reformulate many of our equations to improve numerical stability of the linear algebra
and minimize round-off errors. We thus rewrite
equation (\ref{e.radius1}) as
\begin{equation}
\log r_k = \frac{1}{3} \log \left [ r_{k+1}^3 + \frac{3}{4\pi}\frac{dm_k}{\rho_k} \right ]
\enskip .
\label{e.radius2}
\end{equation}

The velocity of face $k$ is zero unless the hydrodynamics option is
activated, in which case
\begin{equation}
v_k = r_k \frac{d(\log r_k)}{dt},
\label{e.velocity}
\end{equation}
is the Lagrangian time derivative of the radius at face $k$.
For enhanced numerical stability, we rescale this equation by dividing
by the local sound speed.

The pressure $P_k$ is set by momentum conservation at interior cell boundaries,
\begin{eqnarray}
P_{k-1} - P_k 
&=& {\overline{dm}_k} \left [ \left(\frac{dP}{dm}\right)_{\rm hydrostatic}
+ \left(\frac{dP}{dm}\right)_{\rm hydrodynamic} \right ] \nonumber \\
&=& {\overline{dm}_k} \left [ -\frac{Gm_k}{4 \pi r_k^4} 
- \frac{a_k}{4 \pi r_k^2}
\right ]
\enskip ,
\label{e.pressure}
\end{eqnarray}
where $\overline{dm}_k = 0.5 (dm_{k-1} + dm_k)$, and $a_k$ is the Lagrangian 
acceleration at face $k$,  evaluated by the change in $v_k$ over the timestep 
$\delta t$. The acceleration is set to zero if the hydrodynamic option is not used.  
Similarly, the temperature of interior cells $T_k$ is set by energy transport across 
interior cell boundaries, 
\begin{equation}
T_{k-1} - T_k = {\overline{dm}_k} \left [ \nabla_{T,k} \left(\frac{dP}{dm}\right)_{\rm hydrostatic} 
\frac{\overline{T}_k}{\overline{P}_k}
\right ]
\enskip ,
\label{e.temperature}
\end{equation}
where $\nabla_{T,k} = d\log T/d\log P$ at face $k$ from 
the $\MESA$ module $\mlt$ (see \S\ref{mlt}), $\overline{T}_k = (T_{k-1} dm_k
+ T_k dm_{k-1})/(dm_k + dm_{k-1})$ is the temperature interpolated by
mass at face $k$, and $\overline{P}_k = (P_{k-1} dm_k + P_k dm_{k-1})/(dm_k
+ dm_{k-1})$ is the pressure interpolated by mass at face $k$.  For
enhanced numerical stability, we rescale equation (\ref{e.pressure}) by
dividing by $\overline{P}_k$ and equation (\ref{e.temperature}) by dividing by
$\overline{T}_k$.

The pressure and temperature boundary conditions are constructed by
using $P_s$ and $T_s$ from the $\MESA$ module $\atm$
(see \S\ref{atm}). The difference in pressure and temperature from the
surface to the center of the first cell is found from hydrostatic equilibrium and
$\nabla_T$ by 
\begin{eqnarray}
dP_s &=& \frac{G m_1 dm_1/2}{4 \pi r_1^4} \nonumber \\
dT_s &=& dP_s \nabla_{T,1} \frac{T_1}{P_1}
\enskip .
\label{e.ptdiff}
\end{eqnarray}
The boundary conditions are then
\begin{eqnarray}
\log T_1 &=& \log ( T_s+ dT_s) \nonumber \\
\log P_1 &=& \log ( P_s+ dP_s)
\enskip .
\label{e.ptboundary}
\end{eqnarray}
These implicit equations for $P_1$ and $T_1$ are solved together with
the regular structure and composition equations.

Our finite volume form of energy conservation for cell $k$ is
\begin{equation}
L_k - L_{k+1} = dm_k ( \epsilon_{\rm nuc} - \epsilon_{\nu,{\rm thermal}} + \epsilon_{\rm grav} )
\enskip ,
\label{e.luminosity}
\end{equation}
where $\epsilon_{\rm nuc}$ (from module $\net$ or $\jina$) is the
total nuclear reaction specific energy generation rate minus the
nuclear reaction neutrino loss rate, and $\epsilon_{\nu,{\rm
thermal}}$ (from module $\neu$) is the specific thermal neutrino loss rate.  
The $\epsilon_{\rm grav}$ term is the specific rate of change
of gravitational energy due to contraction or expansion,
\begin{equation}
\epsilon_{\rm grav} = -T \frac{ds}{dt} = 
- T C_{\rm P} \left [ ( 1 - \nabla_{ad} \chi_T) \frac{d\log T}{dt} 
- \nabla_{ad} \chi_{\rho} \frac{d\log \rho}{dt} \right]
\enskip ,
\label{e.epsgrav}
\end{equation}
where $d\log T/dt$ and $d\log \rho/dt$ are Lagrangian time derivatives
at cell center by mass, and the other symbols are defined in
Tables \ref{eosvars} and \ref{mltvars}.  For the innermost cell,
$L_{k+1}$ is replaced by the inner boundary condition which is
typically zero but can be nonzero, $L_c$, in specific applications.  For
additional numerical stability, we rescale equation (\ref{e.luminosity})
by dividing by a scale factor that is typically the surface luminosity
of the previous model.

The equation for mass fraction $X_{i,k}$ of species $i$ in cell $k$ is
\begin{eqnarray}
X_{i,k}(t + \delta t) - X_{i,k}(t) 
&=& dX_{\rm burn} + dX_{\rm mix} \nonumber \\
&=&  \frac{dX_{i,k}}{dt} \delta t + (F_{i,k+1} - F_{i,k}) \frac{\delta t}{dm_k},
\end{eqnarray}
where $dX_{i,k}/dt$ is the rate of change
from nuclear reactions reported by $\net$ or $\jina$,
$F_{i,k}$ is the mass of species $i$ flowing across face $k$
\begin{equation}
F_{i,k} = \left ( X_{i,k} - X_{i,k-1} \right ) \frac{\sigma_k}{\overline{dm}_k}
\enskip,
\label{e.massfractions}
\end{equation}
where $\sigma_k$ is the Lagrangian diffusion coefficient from the combined effects 
of convection (\S \ref{mlt}) and overshoot mixing (\S \ref{overshoot}).
For numerical stability, $\sigma_k$ is calculated at the beginning of the
timestep and held constant during the implicit solver iterations.  This
assumption accommodates the non-local overshooting algorithm and significantly 
improves the numerical convergence. It leads to a small inconsistency between 
the mixing boundary and the convection boundary as calculated at the end of the timestep. 

Equations (\ref{e.radius2}), (\ref{e.pressure}), (\ref{e.temperature}), 
(\ref{e.luminosity}), (\ref{e.massfractions}), and, optionally equation 
(\ref{e.velocity}), are by default solved fully coupled and simultaneously 
with a 1$^{\rm st}$ order backwards differencing time integration.

\subsection{Convergence to a solution \label{s.converge}}

The generalized Newton-Raphson scheme is represented by
\begin{equation}
0 = \vec{F}(\vec{y}) = \vec{F}(\vec{y}_i + \delta \vec{y}_i) = \vec{F}(\vec{y}_i) + \left[ \frac{d\vec{F}}{d\vec{y}} \right]_i \delta \vec{y}_i + O(\delta \vec{y}_i^{\,2})
\end{equation}
where $y_i$ is a trial solution, $\vec{F}(\vec{y}_i)$ is the residual, $\delta \vec{y}_i$ is the 
correction, and $[d\vec{F}/d\vec{y}]_i$ is the Jacobian matrix.

$\MESAstar$ uses the previous model, modified by remeshing, mass change, and element 
diffusion, as the initial trial solution for the Newton-Raphson solver. 
This is generally successful because we use analytic Jacobians and have sophisticated
timestep controls (see \S \ref{s.timestep}). The use of analytic Jacobians in $\MESAstar$
requires that each of the $\MESA$ modules provides not just the required output quantities 
but also quality, preferentially analytic, partial derivatives with respect to the input 
quantities. At each timestep, $\MESAstar$ converges on a final solution by iteratively 
improving upon the trial solution. We calculate the residuals, construct a Jacobian matrix,
and solve the resulting system of linear equations with the solvers in $\mtx$ to find 
the corrections to the variables.

The trial solution is accepted when the corrections and residuals meet a specifiable set of comprehensive convergence criteria.
In most cases, the solver is able to satisfy these limits in 2 or 3
iterations. However, under difficult circumstances like the He core
flash or advanced nuclear burning in massive stars, $\MESAstar$ can 
automatically adjust the convergence criteria.
The corrections to the variables will, generally, not produce zero
residuals because the system of equations is nonlinear.  In some
cases, the corrections might make the residuals larger rather than
smaller. In such cases, the length of the correction vector is reduced by a 
line search scheme\footnote{This is a globally convergent method and 
is similar to what is described in \S 9.7 of \citet{nr}.}  until they 
improve the residuals.  In principle, the residuals can be made arbitrarily 
small, but this may take a prohibitively large number of iterations. In
practice, the use of the line search scheme helps the convergence rate
in many cases, but cannot ensure convergence in all cases.

If convergence cannot be achieved with the current timestep, then
$\MESAstar$ will first try again with a reduced timestep (a ``retry'')
anticipating that a smaller timestep will reduce the non-linearity.  If
the retry fails, $\MESAstar$ will return to the previous model and
with a smaller timestep than it used to get to the current model (a
``backup''). If the backup fails, $\MESAstar$ will continue to reduce
the timestep until either the model converges or the timestep reaches
some pre-defined minimum, in which case the evolutionary sequence is
terminated.

\subsection{Timestep selection}
\label{s.timestep}

Timestep selection is a crucial part of stellar evolution.  The
timesteps should be small enough to allow convergence in relatively
few iterations, but large enough to allow efficient evolutions.  Changes
to the timestep should also provide for rapid responses to varying
structure or composition conditions, but need to be carefully
controlled to avoid over-corrections that can reduce the convergence
rate.

$\MESAstar$ does timestep selection as a two stage process.  The first
stage proposes a new timestep using a scheme based on digital control
theory \citep{sod06}.  The second stage implements a wide range of tests that can
reduce the proposed timestep if certain selected properties of the
model are changing faster than specified.  For the first stage, we use
a low-pass filter.  The control variable
$v_{c}$ is the unweighted average over all cells of the relative
changes in $\log\rho$, $\log T$, and $\log R$.  The target value $v_t$
is 10$^{-4}$ by default.
For improved stability and response, the low-pass filter method uses
the previous two results.  Let $\delta t_{i-1}$, $\delta t_{i}$, and
$\delta t_{i+1}$ be the previous, current, and next timestep,
respectively, while $v_{c,i-1}$ and $v_{c,i}$ are the previous and
current values of $v_{c}$.  The timestep for model $i+1$ is then
determined by
\begin{equation}
\delta t_{i+1} = 
\delta t_i f \left[ 
\frac{f(v_{t}/v_{c,i}) f(v_{t}/v_{c,i-1})}{f(dt_i/dt_{i-1})} \right]^{1/4},
\label{filterA}
\end{equation}
where $f(x) = 1 + 2\tan^{-1}[0.5(x-1)]$.
The control scheme implemented by equation (\ref{filterA}) allows rapid
changes in timestep without undesirable fluctuations.

The timestep proposed by this low-pass filtering scheme can be reduced
according to a variety of special tests that have hard and soft
limits.  If a change exceeds its specified hard limit, the current
solution is rejected, and the code is forced to do a retry or a
backup.  If a change exceeds its specified soft limit, the next
timestep is reduced proportionally.  Examples of special tests include
limits on the maximum absolute or relative changes in mesh structure,
composition variables, nuclear burning rate, $\Teff$, $L$, $M$, $T_c$, $\rho_c$, 
and integrated luminosity from various types of nuclear burning.

\subsection{Mesh adjustment}
\label{s.mesh}

$\MESAstar$ checks the structure and composition profiles of the model at the beginning of each
timestep and, if necessary, adjusts the mesh. Cells may be split into two or more
pieces, or they may be made larger by merging two or more
adjacent cells.  The overall remeshing algorithm is designed such that
most cells are not changed during a typical remesh.  This minimizes
numerical diffusion and tends to help convergence.  Remeshing is
divided into a planning stage and an adjustment stage.

The planning stage determines which cells to split or merge based on
allowed changes between adjacent cells.  Mesh revisions minimize the
number of splits and maximize the number of merges while ensuring that
the magnitudes, $\Delta$, of differences between any two adjacent cells
are below specific thresholds: $\Delta \log P < \theta_P$,
$\Delta \log T < \theta_T$, and $\Delta \log[X(^4\mathrm{He}) +
X(^4\mathrm{He_0})] < \theta_{\rm He}$ where $X(^4\mathrm{He})$ is the
helium mass fraction and $X(^4\mathrm{He_0})$ sets an effective lower
lower limit on the sensitivity to the helium abundance.  The default thresholds
are $\theta_P = 1/30$, $\theta_T = 1/80$, $\theta_{\rm He} = 1/20$,
and X($^4\mathrm{He_0}) = 0.01$.  Options are available for specifying
allowed changes between cells for other mass fractions,
$\Delta \nabla_{ad}$ and $\Delta \log(T / (T + T_0))$ for arbitrary
$T_0$.

Local reductions in the magnitude of allowed changes will place higher resolution in 
desired regions of the star. For example,
the default is to increase resolution in regions of nuclear burning
having $\Delta\log\epsilon_{\rm nuc}$ large compared to $\Delta \log P$. 
This increase 
takes effect at a minimum $\log\epsilon_{\rm nuc} =-2$ and increases to a
maximum factor of 4 in resolution for $\log\epsilon_{\rm nuc} \ge 4$. The size
and range of enhancement can also be set for various specific types of
burning. Similarly, it is possible to increase resolution near the
boundaries of convection zones over a distance measured in units of
the pressure scale height.  Different enhancements and distances can
be specified for above and below the upper and lower boundaries of
zones with or without burning.  There are also options to increase
spatial resolution in regions having $\Delta\log X_i$ large compared
to $\Delta \log P$, or near locations where there are spatial
gradients in the most abundant species.  Finally, further splitting is
done as necessary to limit the relative sizes of adjacent cells.

The adjustment stage executes the remesh plan.  Cells to be split are
constructed by first performing a monotonicity preserving cubic
interpolation \citep{stef90} in mass to obtain the luminosities and
enclosed volumes at the new cell boundaries.  The new densities are
then calculated from the new cell masses and volumes, as shown 
in equation (\ref{e.radius1}). 
Next, new composition mass fraction vectors are calculated.  For cells
being merged, this is straightforward.  For cells being split,
neighboring cells are used to form a linear approximation of mass
fraction for each species as a function of mass coordinate within the
cell.  The slopes are adjusted so that the mass fractions sum to
one everywhere, and the functions are integrated over the new cell mass to
determine the abundances.

Finally, the method for calculating the new temperature varies
according to electron degeneracy.  As the electrons become 
degenerate (i.e. $\eta>0$), split cells simply inherit their temperature while merged cells take 
on the mass-average of their constituent temperatures.  If the
electrons are not degenerate (i.e. $\eta<0$), then a reconstruction parabola is
created for the specific internal energy profile of the parent and its
neighbor cells \citep{stir03}. The parabola is integrated over the new
cell to find its total internal energy. The new cell temperature is
determined by repeatedly calling the $\eos$ module using the new
composition and density with trial temperatures until the desired
internal energy is found.

\subsection{Mass loss and accretion}
\label{s.masschanges}

Mass adjustment for mass loss or accretion is done at each timestep
before solving the equations for stellar structure and composition.
$\MESAstar$ offers a variety of ways to set the rate of mass change
$\dot{M}$.  A constant mass accretion or mass loss rate may be
specified in the input files (see \S \ref{s.startmodel}).
Implementations of \citet{rei75} for red giants, 
\citet{block95} for AGB stars, \citet{jag88} for a range of stars in the 
H-R diagram, mass loss for massive stars by \citep{gle09,vin01,nug00,nie90}, 
supersonic mass loss inspired by \citet{pri95}, and super-Eddington mass loss \citep{pac86}
are available options.  An arbitrary mass accretion or mass loss
scheme may be implemented by writing a new module. An example of such
a routine provided with $\MESAstar$ is \citet{mat10} mass loss for
carbon stars.  Finally, one may write a wrapper program that
calculates $\dot{M}$ for each timestep and then calls the $\MESAstar$
module.

Since $\MESAstar$ allows for simulations with a fixed (and unmodeled) 
inner mass, $M_c$, the total mass is $M=M_c+M_m$, where 
$M_m$ is the modeled mass. For cell $k$, $\MESAstar$ stores the relative cell mass $dq_k =
dm_k/M_m$ and the relative mass interior to a cell face $q_k= m_k/M_m= 1
- \sum_{i=1}^{i=k-1} dq_i$ (see Figure \ref{f.structure}). Rather than evaluate $dq_k$ as 
$q_k-q_{k+1}$, it is essential to define $q$ in terms of $dq$ to maintain accuracy 
\citep{les06}.  For example, in the outer envelope of a star 
where the $q_k$ approach 1, the $dq_k$ can be $10^{-12}$ or smaller.
Subtraction of two adjacent $q_k$ to find a $dq_k$ leads to a
intolerable loss of precision.

After a change in mass, $\delta M$, has been determined, the mass structure of the stellar model is modified.  
This procedure changes the mass location of some cells and revises the composition of those cells to match their new location.
It does not add or remove cells, nor does it change the initial trial solution for the structure variables such as $\rho$, $T$, $r$, or $L$.
The mass structure is divided into an inner (usually the central regions of the star), an intermediate, 
and an outer region (usually the stellar envelope).  
The boundaries of the inner and outer regions are initially set
according to temperature, with defaults of $\log T=6$ for the inner
boundary and $\log T=5$ for the outer boundary.  This range is
automatically expanded, for enhanced numerical stability, if the mass
in the intermediate region is not significantly larger than $\delta M$.  The
range is first enlarged by moving the outer boundary to the
surface. If the enclosed mass in the intermediate region is still too
small, then the inner boundary can be moved inward subject to certain
limits. One limit is that the inner boundary does not cross a region
of the model where the composition changes rapidly.  Another limit is
that the fractional mass of the intermediate region cannot change by
more than a factor of two from its previous value nor exceed 10\% of
the total mass.

Once the regions have been defined, the $dq_k$ are
updated.  In the inner region the $dq_k$ are rescaled by $M/(M + \delta M$). 
Thus, $dm_k$, $m_k$, and $X_k$  have the sames value before and after a
change in mass to eliminate the possibility of unwanted numerical mixing in the center.
In the outer region, cells retain the same value of $dq_k$ to improve convergence in the high entropy regions of the star \citep{sug81}.
The $dq_k$ in the intermediate region are scaled so that $\sum dq_k = 1$.
The composition of cells in the intermediate and outer regions are then updated.  In the
case of mass accretion, the composition of the outermost cells whose enclosed mass totals 
$\delta M$ is set to match the specified accretion abundances.  Cells that were part of the old 
structure have their compositions set to match the previous composition.

\subsection{Resolution sensitivity\label{solarVandV}}

We examined the resolution convergence properties of a $1\Msun$ model by 
varying the parameters for mesh refinement and timestepping.  The mesh
refinement parameter multiplies the limits for variable changes across
mesh cells and is closely correlated with the cell size. 
The timestepping parameter controls the tolerance of
the cell average of the relative changes between time steps in $\log
\rho$, $\log T$, and $\log R$ (see \S\ref{s.timestep}) and is 
closely correlated with the timestep.  We varied the mesh refinement and
timestepping controls in tandem through a parameter C, which is a
multiplicative factor on their default values of $1$ and $10^{-4}$,
respectively. Therefore, C is anti-correlated with the time and space
resolution.

\begin{figure}[H]
\plotone{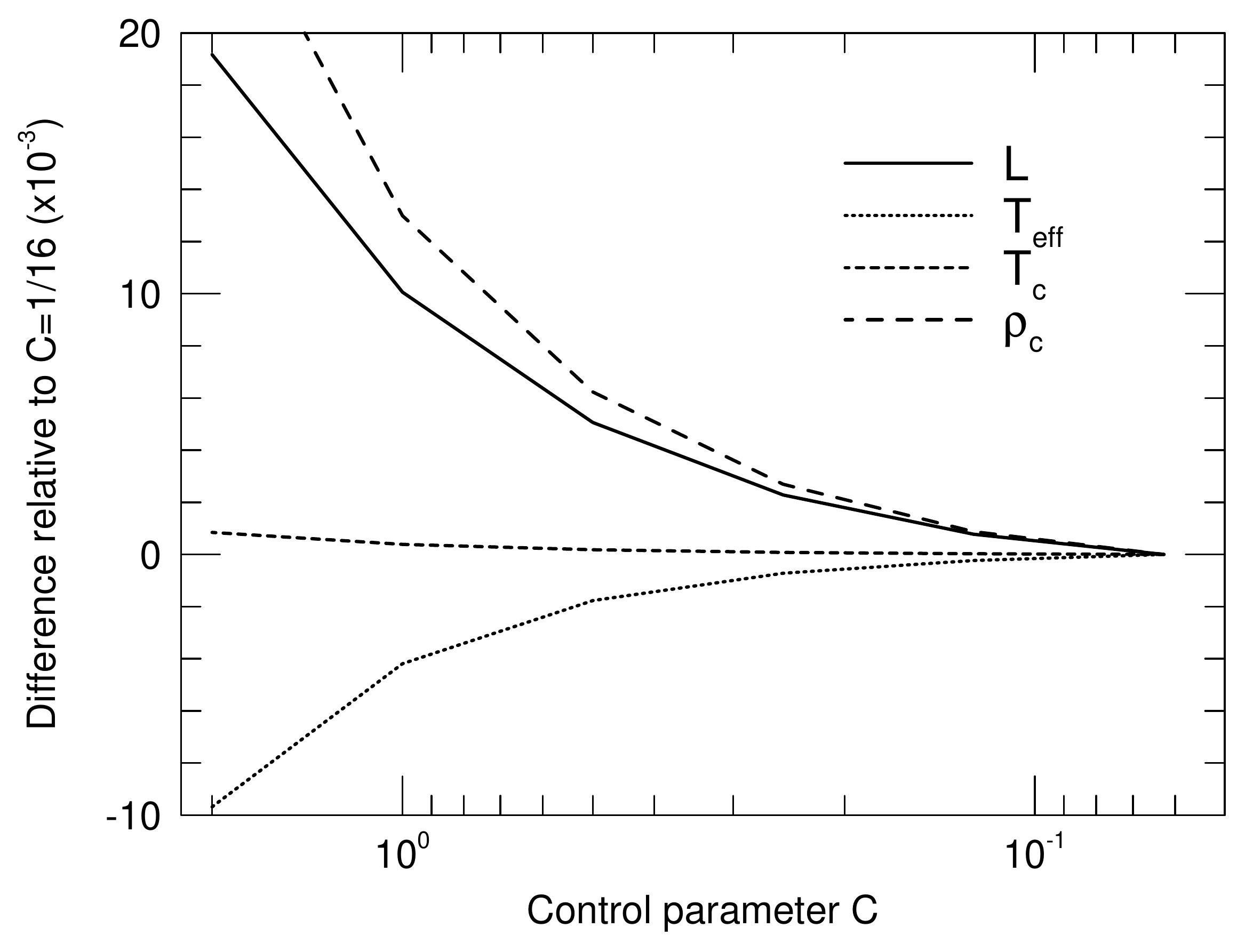}
\caption{
Convergence, $\xi$, for $L$, $T_{\rm eff}$, $T_c$, and $\rho_c$ for a 1$\Msun$ model at
11.0 Gyr as a function of the control parameter C. These differences are all with respect to
 the C=1/16 model.
\label{fig.1msun_converge01}}
\end{figure}

Table \ref{convergence01} and Figure \ref{fig.1msun_converge01} detail
the convergence properties with C of a solar metallicity, 1.0 $\Msun$
model with an $\eta_R$=0.5 Reimers mass loss model \citep[][see
\S\ref{s.masschanges}]{rei75}.  These calculations begin at the ZAMS
and are terminated at 11.0 Gyr, when the model stars are turning off the
main sequence. As a measure of convergence, we use the difference, $\xi$, between
a quantity at a given resolution and the quantity at the highest resolution
considered (C=1/16). In order to determine how convergence depends on 
resolution ($|\xi|\propto {\rm C}^\alpha$), we determine the order of convergence, $\alpha$, 
for increasingly resolved pairs in Table \ref{convergence01}: 
\begin{equation} 
\label{errormodel} 
\alpha = \log \left (\frac{\xi_{\rm fine}}{\xi_{\rm coarse}} \right ) \Big / \log \left (\frac{\rm C_{fine}}{\rm C_{coarse}}\right )
\enskip .
\end{equation}
The convergence orders show that all values converge linearly at large 
values of C and display super-linear convergence ($\alpha \sim$ 1.6) at smaller values of C.
These convergence orders are plausible given that we use a first order
time integration scheme and a finite volume differencing scheme that
is second order accurate on uniform grids.

\begin{deluxetable}{lcccccc}
\tabletypesize{\footnotesize}
\tablecolumns{7}
\tablewidth{0pc}
\tablecaption{$1\Msun$ Model Convergence Properties at 11.00 Gyr \label{convergence01}}
\tablehead{
\colhead{Control parameter C} &
\colhead{2}   & 
\colhead{1}   &
\colhead{1/2} &
\colhead{1/4} &
\colhead{1/8} &
\colhead{1/16} 
           }
\startdata
Number of cells   & 457      & 732      & 1385     & 2740     & 5426     & 10777 \\
Number of timesteps     & 93       & 135      & 225      & 418      & 813      & 1608 \\
& & & & & & \\
L ($\Lsun$)     &  2.06094  & 2.04251  & 2.03241  & 2.02678  & 2.023737    & 2.02217 \\
$\xi$ ($\times 10^{-3}$) &  19.17    & 10.06   & 5.06      & 2.28     & 0.775       &  0.0 \\
$\alpha$          &           & 0.93    & 0.99      & 1.15     & 1.56       &           \\
& & & & & & \\
$T_{\rm eff}$ (K)                & 5543.195 & 5573.935 & 5587.434 & 5593.334 & 5596.064      &  5597.361 \\
$\xi$ ($\times 10^{-3}$) & -9.68   &  -4.19   &  -1.77   &  -0.72   & -0.232        &  0.0   \\
$\alpha$          &         & 1.21  & 1.24  & 1.30  & 1.63 & \\
& & & & & & \\
Log $T_{c}$                  &  7.301645 & 7.298305 & 7.296797 & 7.296052 & 7.295676     &  7.295486 \\
$\xi$ ($\times 10^{-3}$) &  0.8442  & 0.3864   & 0.1797   & 0.0776   & 0.0260       &  0.0    \\
$\alpha$          &          & 1.13  & 1.10  & 1.21  & 1.57  & \\
& & & & & & \\
Log $\rho_{c}$               &  3.393658 & 3.348505 & 3.32615  & 3.314372 & 3.308441     & 3.305552 \\
$\xi$ ($\times 10^{-3}$) & 26.65    & 12.99   & 6.23     & 2.69    & 0.874          & 0.0    \\
$\alpha$          &          & 1.04  & 1.06  & 1.22  & 1.61  & \\
\enddata
\end{deluxetable}

\begin{figure}[H]
\plotone{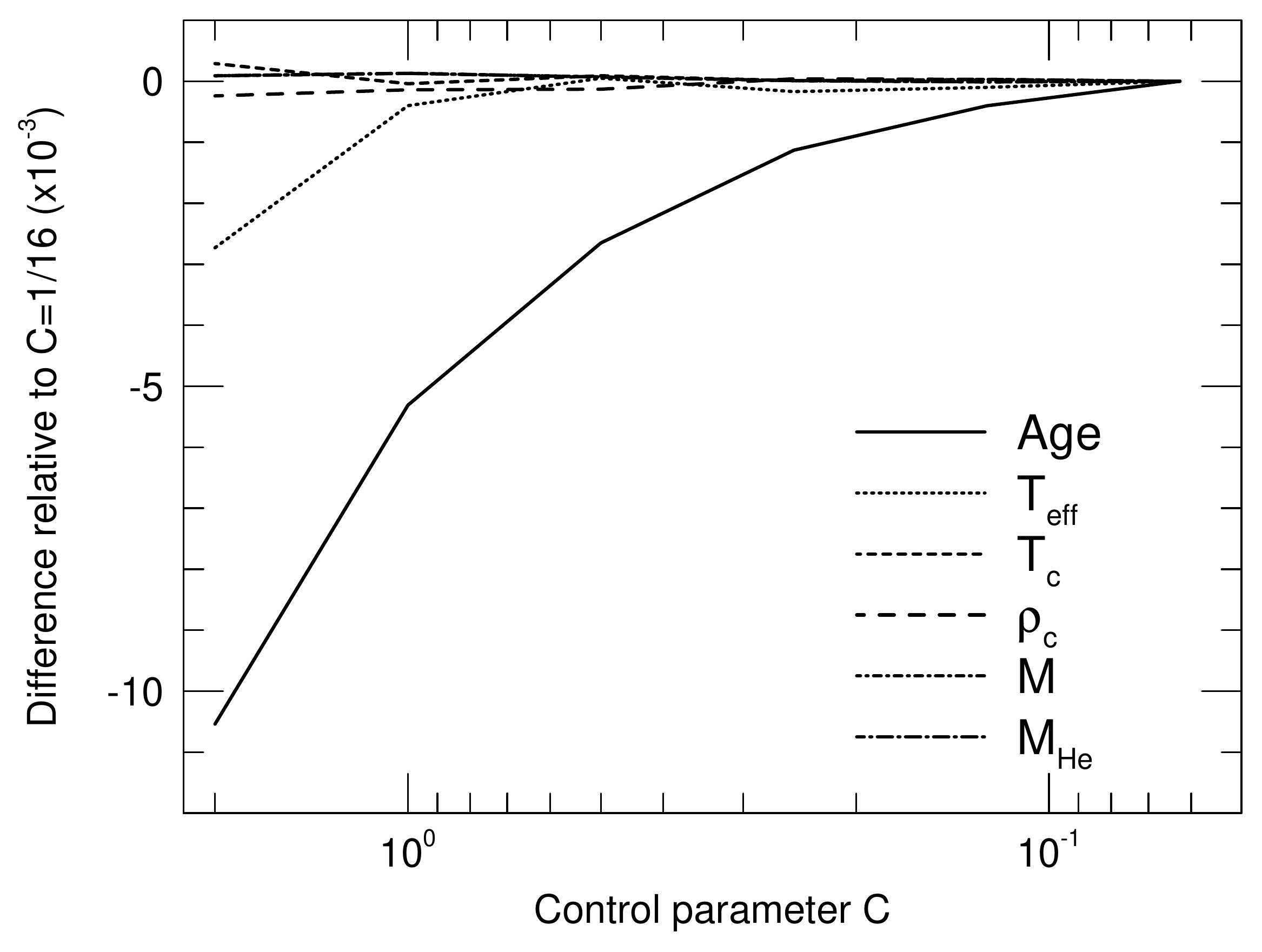}
\caption{
Convergence properties in stellar age, $T_{\rm eff}$, $T_c$, $\rho_c$, $M$, 
and $M_{\rm He}$ for a $M_i=1 \Msun$ model at $100\Lsun$ as a
function of the control parameter C.  These differences are all
relative to the C=1/16 model.  Factors other than the spacetime
resolution are dominating the differences for quantities other than
the stellar age.
\label{fig.1msun_converge_100lsun}}
\end{figure}

Table \ref{convergence02} and Figure \ref{fig.1msun_converge_100lsun}
detail the same stellar models as a function of C except the
calculations are stopped at $L=100\Lsun$, when the stars are on
the RGB. Table \ref{convergence02} suggests the age of the star converges
linearly at larger values of C and super-linearly
($\alpha \sim$ 1.5) at smaller values of C.  However, $T_{\rm eff}$,
$T_c$, $\rho_c$, $M$, and $M_{\rm He}$ all display oscillatory
behavior about the C=1/16 solution, suggesting factors other than
spacetime resolution are dominating the error at this stage of the
evolution. Such factors could be limits in the precision attained by
interpolation in the various tables (e.g., 4 significant figures for
opacities, $\sim$ 6 significant figures for the OPAL and SCVH EOS),
or small changes in boundary conditions.

\begin{figure}[H]
\plotone{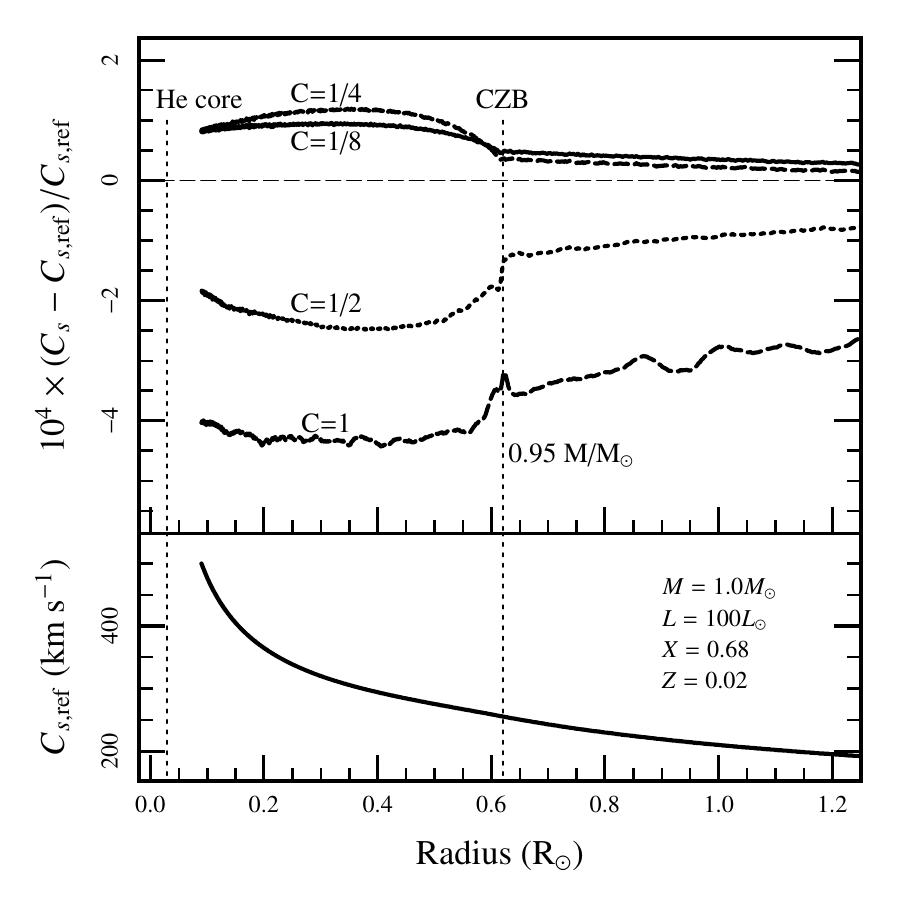}
\caption{
Sound speed profile (lower panel) and convergence properties
(upper panel) for the 1$\Msun$ model at $100\Lsun$ with respect to the reference case, C=1/16.
The helium core boundary and convective zone boundary are labeled.
\label{fig.duo_convergence}}
\end{figure}

The lower panel of Figure \ref{fig.duo_convergence} shows the 
sound speed profile for the $100\Lsun$ model with
C=1/16, our highest resolution case.  The helium core and
convective zone boundaries are labeled. The mass interior to the
convective zone boundary is 0.95 $\Msun$.  The upper panel of Figure
\ref{fig.duo_convergence} shows the convergence properties of the
sound speed profile with resolution in the hydrogen layer.
Both the C=1 and C=1/2 profiles have sound speeds that
are smaller than the C=1/16 profile, while the C=1/4 and C=1/8
profiles have larger sound speeds. Note that the difference between
the various profiles becomes less as the parameter C is made smaller,
indicating that the convergence rate is becoming smaller. This suggests 
factors other than spacetime resolution are dominating the convergence 
rates at this stage of the evolution.  Again, this could be due to the 
precision attained by table interpolations, small changes in boundary
conditions, or differencing errors.

\begin{deluxetable}{lcccccc}
\tabletypesize{\footnotesize}
\tablecolumns{7}
\tablewidth{0pc}
\tablecaption{$1\Msun$ Model Convergence Properties at $100\Lsun$\label{convergence02}}
\tablehead{
\colhead{Control parameter C} &
\colhead{2}   & 
\colhead{1}   &
\colhead{1/2} &
\colhead{1/4} &
\colhead{1/8} &
\colhead{1/16} 
           }
\startdata
Number of cells   & 763      & 1616     & 3262     & 6550     & 13146    & 26248 \\
Number of timesteps     & 689      & 1181     & 2291     & 4547     & 8992     & 17812 \\
& & & & & & \\
Age (Gyr)                     & 12.302   & 12.367   & 12.400   & 12.419   & 12.428   & 12.433 \\
$\xi$ ($\times 10^{-3}$) & -10.54   & -5.31    &  -2.65   &  -1.13   &  -0.402  & 0.0 \\
$\alpha$          &          & 0.99  & 1.00  & 1.24  & 1.49 & \\
& & & & & & \\
$T_{\rm eff}$ (K)                 &  4173.850 & 4183.587 & 4185.450 & 4184.537  & 4184.823   & 4185.260 \\
$\xi$ ($\times 10^{-3}$) & -2.73     & -0.40     & 0.05    & -0.17     &  -0.10 & 0.0 \\
$\alpha$          &           & 2.77      & 3.14    & -1.93     &   0.73  & \\
& & & & & & \\
Log $T_{c}$                   & 7.61573  & 7.61327  & 7.61421  & 7.613652  & 7.613793   & 7.613556 \\
$\xi$ ($\times 10^{-3}$) & 0.29     & -0.04    & 0.09     & 0.01      & 0.03  &  0.0 \\
$\alpha$          &          & 2.93     & -1.19    & 2.77      & -1.30 & \\
& & & & & & \\
Log $\rho_{c}$                & 5.42340  & 5.42393  & 5.42402  & 5.424929  & 5.424868   & 5.424713 \\
$\xi$ ($\times 10^{-3}$) & -0.24    & -0.14    & -0.13    & 0.04      & 0.03 & 0.0 \\
$\alpha$          &          & 0.75     & 0.18     & 1.68      & 0.48 & \\
& & & & & & \\
$M$ ($\Msun$)         &  0.984444 & 0.984482 & 0.984416 & 0.984356  & 0.984345   & 0.984351 \\
$\xi$ ($\times 10^{-3}$) & 0.09      & 0.13     & 0.07  & 0.01 & -0.01 & 0.0 \\
$\alpha$          &           & -0.49    & 1.01  & 3.70 & -0.26 & \\
& & & & & & \\
$M_{\rm He}$ ($\Msun$)       &  0.28080  & 0.28062  & 0.28085  & 0.281016  & 0.281039   & 0.281001 \\
$\xi$ ($\times 10^{-3}$) &  0.09  & 0.13  & 0.07  & 0.01 & -0.01 &  0.0 \\
$\alpha$          &          & -0.49  & 1.01  & 3.70 & -0.26 & \\
\enddata
\end{deluxetable}

\subsection{Multithreading}
\label{s.perform}
$\MESA$ modules are designed to be thread-safe (see \S \ref{mod}), thereby enabling parallel execution.
Table \ref{thread} lists the execution times in seconds of several specific tasks from 4 identical evolutionary 
calculations, each with a different number of threads (one thread per core). The essential difference  
between modules that scale as the inverse of the number of threads ($\eos$ and $\net$) and those that don't 
(e.g., $\kap$ and $\neu$) is the ratio of the overhead associated with parallel execution to the actual time 
required for each module to perform its calculation. The time required to complete serial tasks sets a lower 
limit on the total execution time of $\MESAstar$. 

\begin{deluxetable}{lrrrr}
\tablecolumns{5}
\tablewidth{0pc}
\tablecaption{Execution times (s) with multiple threads\label{thread}}
\tablehead{ \colhead{}        &           \multicolumn{4}{c}{Number of Threads} \\
            \colhead{}&\colhead{1}&\colhead{2}&\colhead{4}&\colhead{8}}
\startdata
total\tablenotemark{a} &  12.2146 &    8.0099 &    5.8963 &   5.0634 \\
          $\ \ $ratio  & \nodata  &    1.53 &    1.36  &   1.16 \\
\cutinhead{Threaded tasks}
               $\net$  &   6.2602 &    3.1721 &    1.6047 &   0.8182 \\
               $\eos$  &   1.7185 &    0.8897 &    0.4539 &   0.2399 \\
               $\mlt$  &   0.2479 &    0.1384 &    0.0704 &   0.0357 \\
               $\kap$  &   0.2285 &    0.1386 &    0.1044 &   0.0875 \\
               $\neu$  &   0.0240 &    0.0209 &    0.0139 &   0.0098 \\
              subtotal &   8.4791 &    4.3597 &    2.2473 &   1.1911 \\
          $\ \ $ratio  &  \nodata &      1.95 &      1.94 &     1.89 \\
$\ \ $fraction of total&     0.69 &      0.54 &      0.38 &     0.24 \\
\cutinhead{Serial tasks}
          file output  &   1.1848 &    1.0301 &    1.0036 &  1.1679 \\
matrix linear algebra  &   0.6569 &    0.7654 &    0.7885 &  0.7926 \\
        miscellaneous  &   1.8938 &    1.8547 &    1.8569 &  1.9118 \\
             subtotal  &   3.7355 &    3.6502 &    3.6490 &  3.8723 \\
$\ \ $fraction of total&     0.31 &      0.46 &      0.62 &    0.76 \\
\enddata
\tablenotetext{a}{These numbers do not include initialization, e.g., 
loading of data tables.}
\end{deluxetable}

\subsection{Visualization with \texttt{PGstar} \label{pgstar}}

By default, $\MESAstar$ provides alpha-numeric output at regular intervals.
In addition, it provides the option for concurrent graphical output. \texttt{PGstar} 
uses the PGPLOT\footnote{We thank Philip Pinto for initiating  this $\MESA$ capability, see \url{http://www.astro.caltech.edu/~tjp/pgplot/}.} 
library to create on-screen plots or images in PNG format that
can be post-processed into animations of an evolutionary sequence. A wide variety of visualization options are provided
and these are all configurable through the \texttt{PGstar} inlist. For example,
the \texttt{PGstar} window can simultaneously hold: an H-R diagram, a
$T_c-\rho_c$ diagram, and interior profiles of physical 
variables, such as nuclear energy generation, and composition.
Animation is very useful for visualizing complex, time dependent processes. 
For example, view the short selection from the $\MESA$ website that shows 
the He core flash in a $1 \Msun$  star.\footnote{ $\ $ \url{http://mesa.sourceforge.net/pdfs/1MHeflash.mov}}

\section{$\MESAstar$ results: comparisons and capabilities\label{vandv}}
 As with any modeling approach, $\MESAstar$ must be verified \citep[``Is it 
solving the equations correctly?"][]{roa98}
and validated (``Does it solve the right equations?") to demonstrate its 
accuracy and predictive credibility \citep[e.g.,][]{ober98}. V\&V is a maturing discipline \citep[e.g.,][]{roa98,cal02}, with the goal of 
assessing the error and uncertainty in a numerical simulation, which also includes addressing sources of error in theory, experiment, 
observation, and computation \citep{cal04}. The results of V\&V testing are historical statements of reproducible evidence that a simulation 
demonstrates a quantified level of accuracy in the solution of a specific problem. 

V\&V is an ongoing activity for $\MESA$ via the $\MESA$ test suite (see Appendix \ref{test}), where code modules are tested individually, and, 
where possible, the integrated code $\MESAstar$ is verified and validated. Verification for $\MESA$ includes a systematic study of the 
effect of mesh and time-step refinement on simulation accuracy (\S \ref{solarVandV}), specific module comparisons (\S \ref{dsep}), and stellar 
evolution code comparisons presented in this section.

This section shows $\MESAstar$ evolution calculations of single stellar and substellar objects with  $10^{-3}\Msun < M < 1000\Msun$ 
(in \S\S \ref{sec:lowmass}, \ref{s.midmass}, \ref{s.vv_highmass}) as well as verification results  (\S\S \ref{s.baraffe}, \ref{SCC},  
\ref{sec:2Msun}, \ref{s.pulsate}, \ref{s.25msun}, and \ref{s.highmasscompare}).  In \S \ref{SSM} we compare the $\MESAstar$ Solar 
model with helioseismic data. As examples of the many other experiments that are possible with $\MESA$, we model prolonged accretion 
of He onto a neutron star and a mass-transfer scenario relevant to cataclysmic variables in \S \ref{s.accr}.

\subsection{ Low mass stellar structure and evolution\label{sec:lowmass}}

\begin{figure}[H]
\begin{center}
\includegraphics[width=0.8\textwidth]{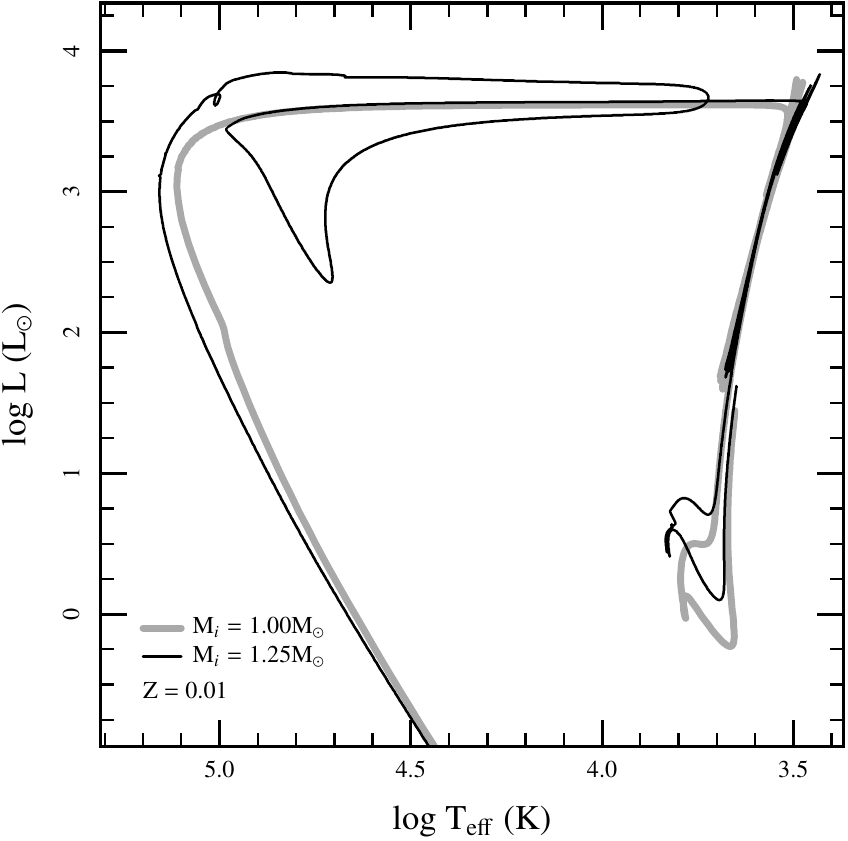}
\end{center}
\caption{The $\MESAstar$ evolution of a $1\Msun$ and a $1.25\Msun$, $Z=0.01$  star from the pre-main sequence to cooling white dwarfs. 
\label{1MHRTRho}}
\end{figure}

$\MESAstar$ has sufficiently broad input physics to compute the evolution of low mass
stars and substellar objects down to Jupiter's mass ($\approx 10^{-3}\Msun$), as well as 
complete evolutionary sequences of low mass stars ($M\la 2\Msun$) from the PMS to the 
white dwarf cooling curve without any intervention. Figure \ref{1MHRTRho} shows evolutionary 
tracks in the H-R diagram for 1 and $1.25\Msun$ models with $Z=0.01$.\footnote{Each calculation
takes a few hours on a laptop computer.} The $1.25\Msun$ model exhibits a late He-shell flash
during the pre-white dwarf phase.

\begin{figure}[H]
\begin{center}
\includegraphics[width=0.7\textwidth]{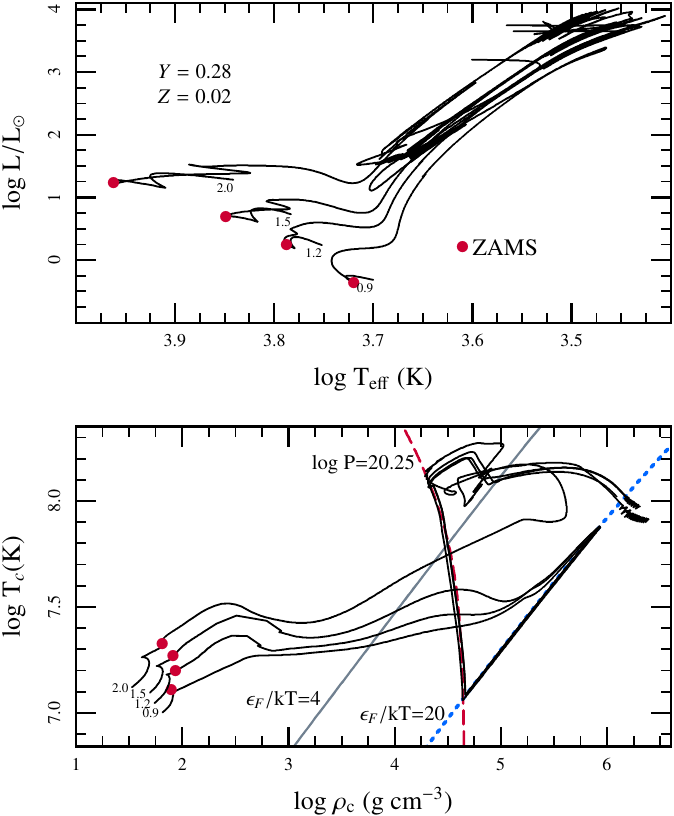}
\end{center}
\caption{Evolution from $\MESAstar$ of 0.9, 1.2, 1.5, and 2
$\Msun$ stars with $Z=0.02$ up to the end of the TP-AGB.  The top panel shows their evolution in the H-R diagram, where the solid red point is the ZAMS. The bottom panel shows the evolution in the $T_c-\rho_c$ plane, exhibiting the He core flash and later evolution of the C/O core during the thermal pulses. The dashed blue (heavy grey) line shows a 
constant electron degeneracy of $\epsilon_F/k_BT=20(4)$. The dashed red line is for a constant pressure of $\log P=20.25$; relevant to the He core flash. 
\label{HRTRho08to2}}
\end{figure}

 Figure \ref{HRTRho08to2} provides further
examples, spanning 0.9-2$\Msun$ at $Z=0.02$; for clarity, the pre-main sequence portion of the tracks were removed and the runs were terminated after the models left the 
thermally-pulsating asymptotic giant branch (TP-AGB). The bottom panel 
shows the evolution in the  $T_c-\rho_c$ plane, exhibiting the convergence of the 0.9, 1.2 and 1.5 $\Msun$ models
 to nearly identical, degenerate, Helium cores when on the RGB. The $2\Msun$ model 
ignites He at a lower level of degeneracy. 

\clearpage

Though the required $\MESAstar$ timesteps get short ($\approx$ hours) during the off-center He core flash in the 0.9, 1.2 and 1.5 $\Msun$ models, 
the stellar model does not become dynamic; the entropy change timescales are always longer than the local dynamical time 
\citep{tho67,serweis05,shen09,moc09}. The reduction of hydrostatic pressure in the core at the onset of flashes leads to the adiabatic 
expansion of the core, visible as the drop in $T_c$ at constant degeneracy. Successive He flashes \citep{tho67,serweis05} work their way into 
the core over a $2\times 10^6$ year timescale, eventually heating it (at nearly constant pressure; see the dashed red line in the bottom panel 
of Figure \ref{HRTRho08to2}) to ignition and arrival onto the horizontal branch. The further evolution during He burning and the thermal 
pulses is seen in the bottom panel, where the small changes in the C/O core during thermal pulses on the AGB are resolved in $\MESAstar$.

We start the more detailed calculations and comparisons to previous work in \S \ref{s.baraffe} by displaying the $\MESAstar$ PMS evolution of low 
mass stars, brown dwarfs and giant planets, and comparing to prior results of \citep{bar03}.

\subsubsection{Low mass pre-main sequence stars, contracting brown dwarfs and giant planets\label{s.baraffe}}

The PMS evolution of low-mass stars \citep{bur89, dant94, bar98} and gravitationally contracting brown dwarfs and giant planets 
\citep{bur97, cha00, chab00, bur01} has been studied extensively. 
The lasting importance of these problems motivates us to ensure that $\MESAstar$ can successfully perform these evolutions. 

\begin{figure}[H]
\begin{center}
\includegraphics[width=0.75\textwidth]{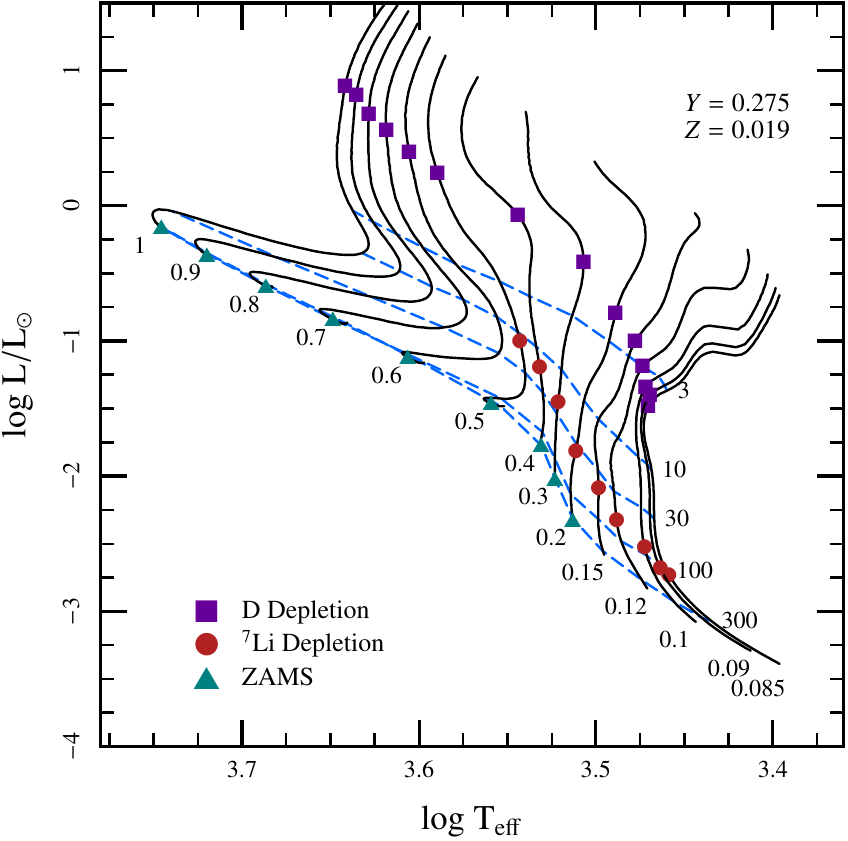}
\end{center}
\caption{ Location in the Hertzsprung-Russell (H-R) diagram for $0.085\Msun < M <1\Msun$ stars as they arrive at the main sequence
for $Y=0.275$ and $Z=0.019$.  The mass of the star is noted by the values at the bottom of the line.
The dashed blue lines are isochrones for ages of 3, 10, 30, 100 and 300 Myr, as noted to the right.
The purple squares (red circles)   show where D ($^7$Li) is depleted by a factor of 100.  The green triangles show  the ZAMS.
\label{vlmhr}}
\end{figure}

Figure \ref{vlmhr} shows the evolution of PMS stars with masses of $0.08\Msun <M<1\Msun$. 
Each solid line starts on the PMS Hayashi track for a fixed mass, $M$, and ends at an age of 1 Gyr.
All stars with $M\ge0.2\Msun$ have reached the ZAMS ($L=L_{\rm nuc}$) by this time. The red circles show the location 
where the $^{7}$Li is depleted by a factor of 100, but only for $M<0.5\Msun$. Stars with $M>0.5\Msun$ deplete their 
$^7$Li after the core becomes radiative \citep{dant94,cha96,bil97}, adding an uncertain dependence on convective overshoot 
that we do not investigate here.

\begin{figure}[H]
\begin{center}
\includegraphics[width=0.75\textwidth]{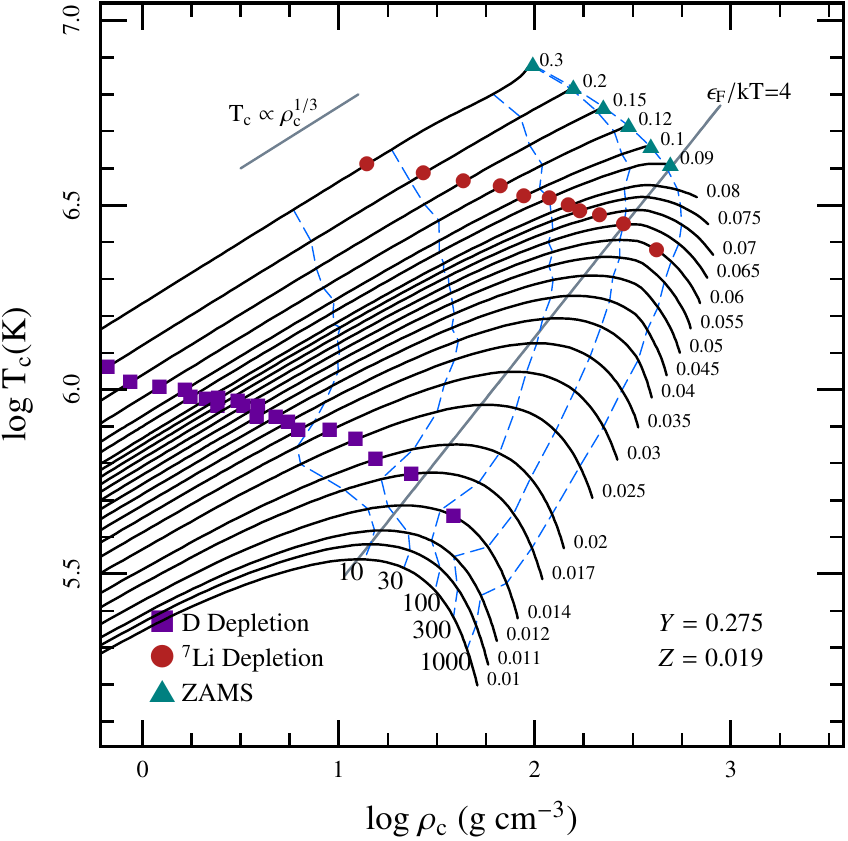}
\end{center}
\caption{ Trajectories of central conditions for fully convective $M<0.3\Msun$ stars as they approach the main sequence ($M>0.08\Msun$) or
become brown dwarfs for $Y=0.275$ and $Z=0.019$. Each solid line shows $T_c$ and $\rho_c$  
for a fixed mass, $M$, noted at the end of the line (when the age is 3 Gyr).
The dashed blue lines are isochrones for ages of 10, 30,100, 300 Myr and 1 Gyr.  The purple squares  and red circles show where D and $^{7}$Li is  depleted by a factor of 100. The green triangles show the ZAMS.
\label{vlmtrho}}
\end{figure}

\begin{figure}[H]
\plotone{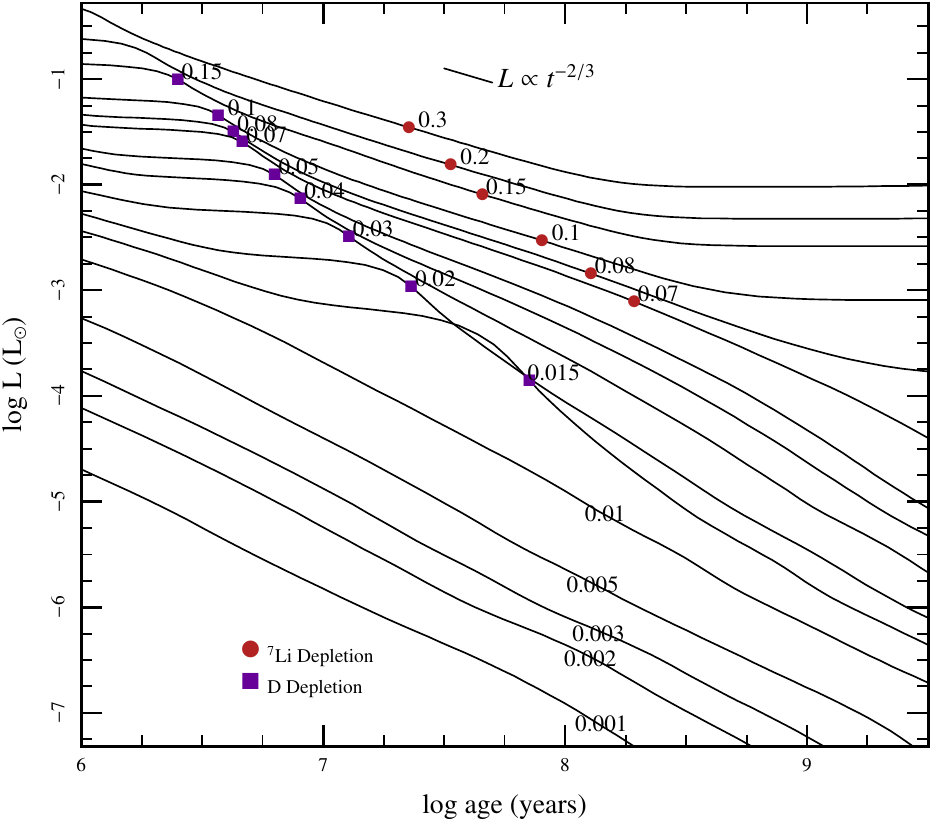}
\caption{Luminosity evolution for fully convective  $M<0.3\Msun$ stars as they approach the main sequence ($M>0.08\Msun$) or become brown dwarfs for $Y=0.275$ and $Z=0.019$. 
From top to bottom, the lines are for $M=0.3$, 0.2, 0.15, 0.1, 0.08, 0.07, 0.05, 0.04, 0.03, 0.02, 0.015, 0.010, 0.005, 0.003, 0.002, \& $0.001\Msun$. The purple squares (red circles) denote where D ($^7$Li) is depleted by a factor of 100.  \label{vlmL}}
\end{figure}

Lower-mass stars ($M<0.3\Msun$) remain fully convective throughout their PMS and arrival on the 
main sequence. In Figure \ref{vlmtrho} we show the evolution in the $T_c-\rho_c$ plane for 
these objects. The light solid line in the upper left denotes the $T_c\propto \rho_c^{1/3}$  relation expected during the Kelvin-Helmholtz contraction for  a fixed mass non-degenerate star. 
Deviations from this relation occur when electron degeneracy occurs, which is 
shown by the grey line at  $\eta\approx \epsilon_F/k_BT=4$, roughly where 
the electron degeneracy has increased the electron pressure to
twice that of an ideal electron gas. We extend the mass range down to 
$M=0.01\Msun$ to reveal the distinction between main sequence stars and brown dwarfs. 
That distinction becomes clearer in Figure \ref{vlmL} which shows the $L$ evolution for a range of stars with $M<0.3\Msun$. Only the $M>0.08\Msun$ stars asymptote at late times to a constant $L$, whereas the others continue to fade. We also exhibit the 
expected scaling for a contracting fully convective star with a constant $T_{\rm eff}$, $L\propto t^{-2/3}$. At the D mass fraction of this calculation, 
$X_{\rm D}=3\times 10^{-5}$, the onset of D burning provides some luminosity for a finite time, causing the evident kink at early times. 

\begin{figure}[H]
\begin{center}
\includegraphics[width=0.8\textwidth]{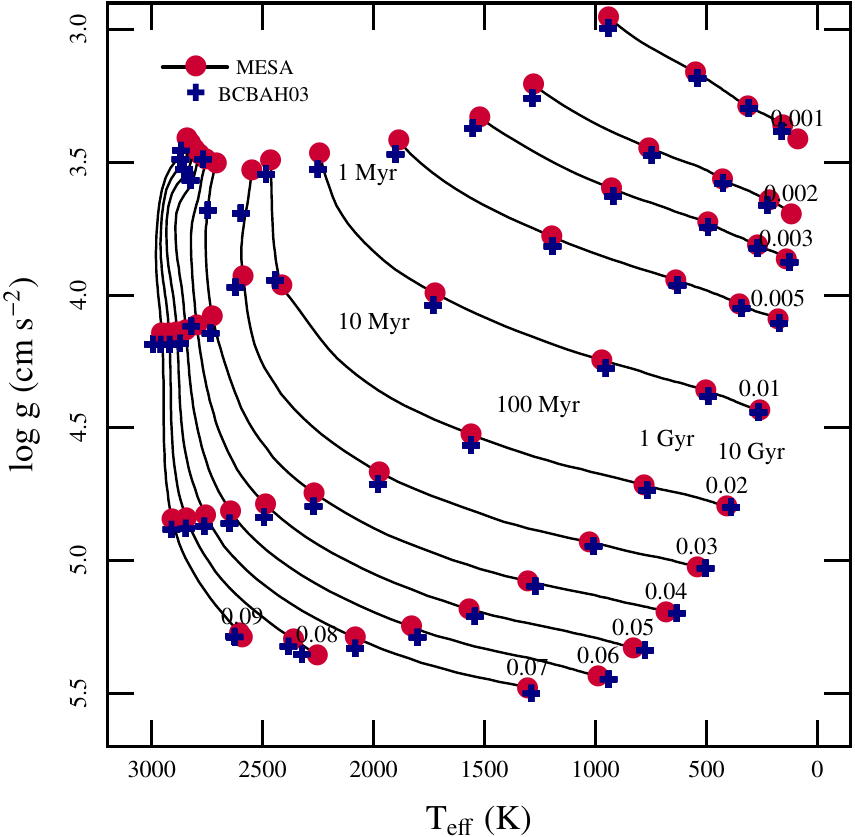}
\end{center}
\caption{Evolution of very low mass stars and substellar objects from 0.09 to 0.001 $\Msun$ for $Z=0.02, Y=0.28$ 
in the $\logg$-$\Teff$ plane. The solid lines are the $\MESAstar$ tracks, with labeled masses (in $\Msun$) at the 
bottom of each. The filled circles denote the location of each track at a given age. The plus symbols (+) mark the 
locations of the BCBAH tracks for the same masses and ages. The two lowest mass tracks from BCBAH do not extend to 
10 Gyr.\label{VLMS}} 
\end{figure}

$\MESAstar$ models evolved from the PMS Hayashi line to 
an age of 10 Gyr with masses ranging from 0.09 to 0.001$\Msun$ are
compared with the models of \citet[][BCBAH]{bar03} in Figure \ref{VLMS}.
Ages in increasing powers of 10 are marked by filled
circles along each track from 1 Myr to 10 Gyr. For comparison,
separate points from the BCBAH evolutionary models are plotted as plus
symbols (``+"). So as to match the choice of 
BCBAH, we set the D mass fraction at $X_D=2\times 10^{-5}$ \citep{cha00}. 
Evolution at the youngest ages is uncertain due
to different assumptions regarding D  burning but beyond 10 Myr the
$\MESAstar$ and the BCBAH models overlap at almost every point. Note
that the BCBAH models were only evolved to 5 Gyr for the two lowest
masses shown in Figure \ref{VLMS}.

\vfill\eject
\subsubsection{Code comparisons of $0.8\Msun$ and $1\Msun$ models \label{SCC}}

The Stellar Code Calibration Project \citep{weiss07} was created to 
provide insight into the consistency of results obtained from 
different state-of-the-art stellar evolution codes. The
contributors performed a series of stellar evolution calculations with
the physics choices held constant to the greatest extent possible.  This
section compares $\MESAstar$ models with published results from that
project for two specific cases. The comparison codes are 
BaSTI/FRANEC \citep{basti}, DSEP \citep{dot08}, and GARSTEC \citep{garstec}.
$\MESAstar$ models lie within the range exhibited by 
BaSTI/FRANEC, DSEP, and GARSTEC in these comparisons.

Two examples are shown here, a $0.8\Msun, Z=10^{-4}$ star and a $1\Msun, Z=0.02$, 
both modeled from the pre-MS to the onset of the He core flash.  The models assume, 
as much as possible, the same nuclear reaction rates (NACRE), opacities \citep[OPAL and ][]{af94}, 
equation of state (FreeEOS), and mixing length ($\alpha_{\rm MLT}=1.6$). These tests do not represent 
the best models for the various codes. Instead, the goal of the comparisons was to see how consistent 
the codes would be when using simple assumptions and comparable input physics \citep{weiss07}.  While 
the agreement is good in most respects, in temporal resolution there is a discrepancy.

\begin{figure}[H]
\plottwo{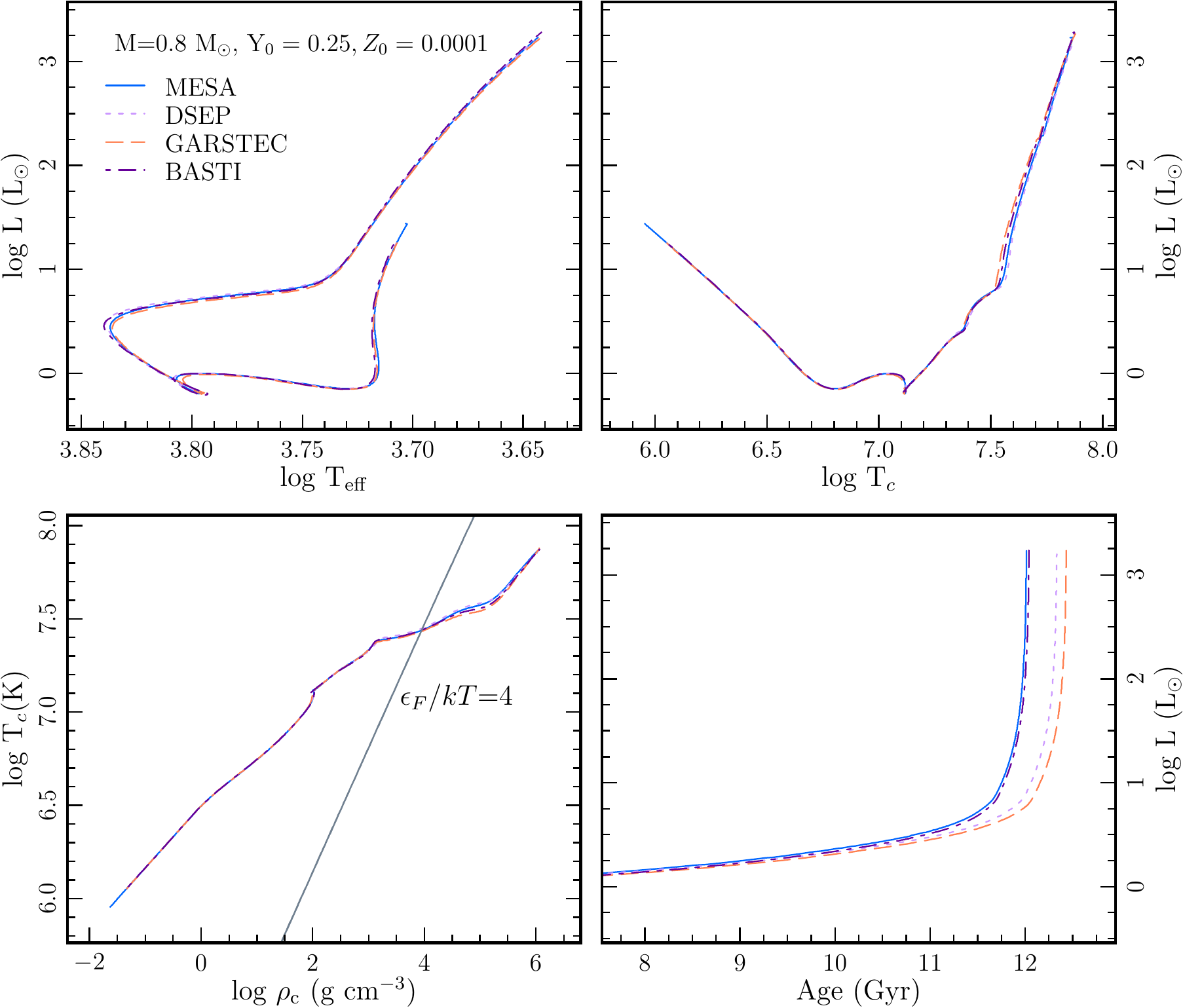}{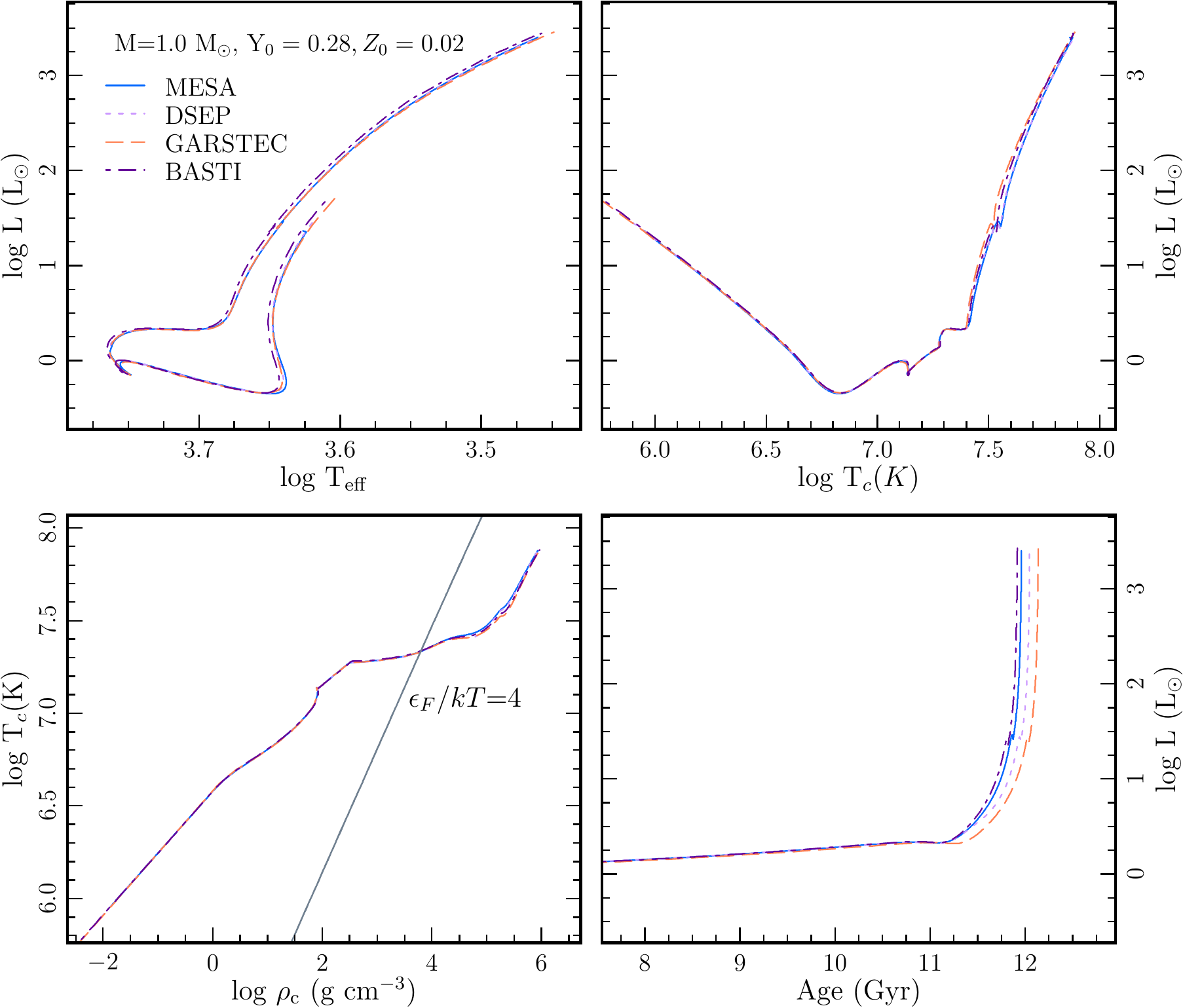}
\caption{
Stellar Code Calibration project models for the $0.8\Msun, Z=10^{-4}$ (left) 
and the $1\Msun, Z=0.02$ (right) cases. The upper-left panels show the H-R diagram;
upper-right panels show luminosity versus central temperature; lower-left 
panels show central T-$\rho$; lower-right shows luminosity versus age. 
\label{SCC:plot}}
\end{figure}

The H-R diagram of the $0.8\Msun, Z=10^{-4}$ case is nearly identical for all
four tracks except near the main sequence turnoff, where DSEP and
BaSTI/FRANEC are hotter than $\MESAstar$ and GARSTEC. These models have
essentially no convection during the main sequence and there is
remarkably little scatter during this phase. In the
$T_c$-$L$ plane, the models are almost indistinguishable until they
enter the red giant phase at $L\approx 10\Lsun$, where the
central temperatures differ slightly only to re-converge at maximum
luminosity on the red giant branch.  Finally, the
lifetime-luminosity plane indicates that the four codes split into two
pairs with one pair shorter lived by $\approx 5\%$  than the other pair. It
is beyond our scope here  to explain the reasons for
these differences; the purpose of the present comparison is to indicate 
that $\MESAstar$ produces results that are consistent with the range exhibited 
among the other three codes.

Convection plays a more prominent role in the $1\Msun, Z=0.02$ case and the 
scatter is greater than in the $Z=10^{-4}$ case.  The BaSTI/FRANEC model is hotter 
than the other three models.
Treatment of convection and, in particular, the resolution of the surface 
convection zone is primarily responsible for the spread seen in the main 
sequence portion of the tracks.

In both cases, the central conditions are very similar until the models 
become red giants. In the $Z=0.02$ case, the range of lifetimes is somewhat 
reduced compared to the $Z=10^{-4}$ case with BaSTI/FRANEC and $\MESAstar$ 
shortest (though the order is reversed with respect to the $Z=10^{-4}$ case) 
followed by DSEP and then GARSTEC.

\subsubsection{ The $\MESAstar$ Solar model \label{SSM}}

$\MESAstar$ performs a Solar model calibration by iterating on
the difference between the final model and the adopted Solar parameters of $\Lsun$ and $\Rsun$ \citep[from][]{bah05}
and the surface value of $Z_s/X_s$ from \citet{gs98} at $4.57$ Gyr. 
This is done by iteratively varying $\alpha_{\rm MLT}$ and the initial $Y_i$ and $Z_i$ values \citep[all for 
the abundance ratios of][]{gs98}, while including diffusion.

\begin{figure}[H]
\plotone{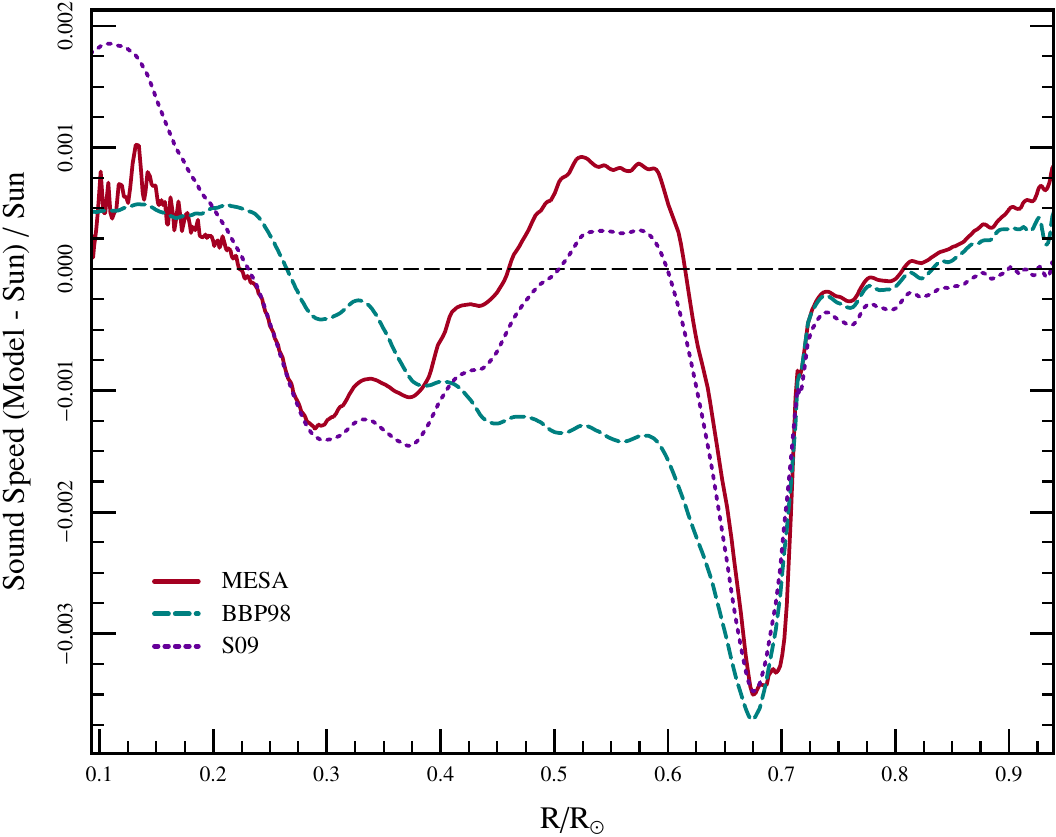}
\caption{Comparisons of the sound speed profiles within the sun. The red solid line shows the 
relative difference in the sound speed between $\MESAstar$ predictions and the inferred sound speed 
profile from helioseismic data (taken from \citet{bah98}). The green-dashed and blue-dotted lines show the same for 
the standard Solar models of \citet[][BBP98]{bah98} and \citet[][S09]{ser09}, respectively. 
\label{sound}}
\end{figure}

The properties of the converged model (which reaches the desired parameters to better than one part in $10^5$) are shown
in Table \ref{t.SSM}, and match the measured depth, $R_{\rm CZ}$, of the surface convection zone 
within 1-$\sigma$  and the surface Helium abundance, $Y_s$, within 2-$\sigma$ \citep{bah05}.
The difference between the model and the helioseismologically inferred Solar
sound speed profile is compared with similar results from \citet[][BBP98]{bah98} and \citet[][S09]{ser09} in
Figure \ref{sound} demonstrating  that $\MESAstar$ is capable of stellar evolution calculations 
at the level of 1 part in $10^3$ demonstrated by others \citep{bas08}.

\begin{deluxetable}{lc}
\tablecolumns{2}
\tablewidth{0pc}
\tablecaption{$\MESAstar$ Standard Solar Model at 4.57 Gyr\label{t.SSM}}
\tablehead{\colhead{Quantity}&\colhead{Value}}
\startdata
\cutinhead{Converged Input Parameters}
$\alpha_{\rm MLT}$   & 1.9179113764 \\
$Y_i$       & 0.2744267987 \\
$Z_i$       & 0.0191292323 \\
\cutinhead{Properties of Converged Model} 
$(Z/X)_s$   &  $0.02293$ \\
$X_s$       &  $0.73973$ \\
$Y_s$ & $0.24331$\\
$Z_s$       &  $0.01696$ \\
$X_c$        &  $0.33550$ \\
$Z_c$        &  $0.02125$ \\
$R_{CZ}/\Rsun$   &  $0.71398$ \\
$\log \rho_c $   &  $2.18644$ \\
$\log P_c$  &  $17.3695$ \\
$\log T_c$  &  $7.19518$ \\
RMS[($c_{\rm Model}-c_{\odot}$)/$c_{\odot}$]   &  $0.00093$
\enddata
\end{deluxetable}

\subsection{Intermediate Mass Structure and Evolution}
\label{s.midmass} 

$\MESAstar$ can calculate the evolution of intermediate mass stars
( $ 2 \la M/\Msun \la 10$) through the He-core burning phase and the advanced He-shell burning Asymptotic Giant Branch (AGB) phase.  $\MESAstar$ produces results
compatible with published results from existing stellar evolution
codes.

\begin{figure}[H] 
\begin{center}
\includegraphics[width=0.7\textwidth]{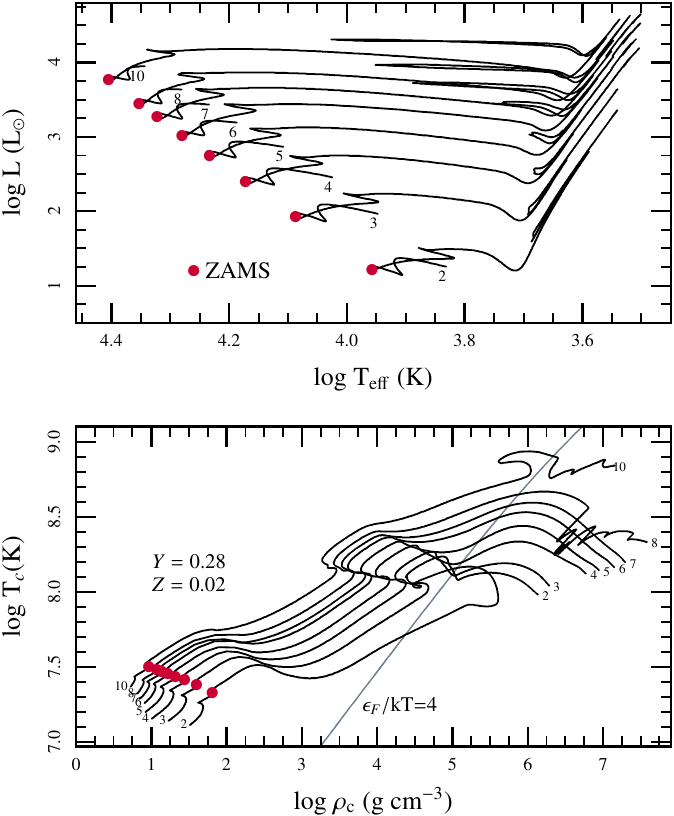}
\end{center}
\caption{Top: $\MESAstar$ H-R diagram for 2-10 $\Msun$ models from the pre-main
sequence to the end of the thermally pulsating AGB. Bottom: trajectories of the central conditions. The filled red points show the ZAMS. 
\label{midduo}}
\end{figure}

We start by showing in Figure \ref{midduo} a grid of $\MESAstar$ evolutionary tracks with masses ranging from 2 to 10$\Msun$ with $Z=0.02$. The top panel shows the evolution in the H-R diagram while the bottom panel shows the evolution in the 
$T_c-\rho_c$ diagram.
 The 8 and $10\Msun$ models start to ignite carbon burning off center, whereas the $2-7\Msun$ models produce C/O white dwarfs. The lack of a complete treatment in $\MESAstar$ of liquid diffusion inhibits our ability to 
verify the resulting white dwarf cooling sequences from $\MESAstar$ at this time.

\subsubsection{Comparison of EVOL and $\MESAstar$  }
\label{sec:2Msun}

We compare $M_i=2\Msun, Z=0.01$ stellar models from $\MESAstar$ and EVOL
\citep{block95,her04,her04b} starting from the pre-main sequence to
the tip of the thermal pulse AGB (TP-AGB).  Both codes employed the
exponentially-decaying overshoot mixing treatment described by
\citet[][see \S \ref{overshoot}]{her00} at all convective boundaries
with $f=0.014$, except during the third dredge-up where we adopt
$f=0.126$ at the bottom of the convective envelope to
account for the formation of a $^{13}$C pocket, and at the
bottom of the He-shell flash convection zone we use
$f=0.008$ \citep{her05}.

In both codes we use the mass loss formula of \citet[][see \S
\ref{s.masschanges}]{block95}.  Thermal pulses start at a slightly lower
core mass, and hence luminosity, in the EVOL model. In order to maintain similar envelope 
mass evolution
through the TP-AGB, the parameter
$\eta_\mathrm{Bl}$ in the mass loss formula was set to 0.05 in
$\MESAstar$ and 0.1 in EVOL. Every effort has been made to tailor the
$\MESAstar$ model to the EVOL model. However, the AGB evolution is
very sensitive to the initial core mass, which depends
on the mixing assumptions and their numerical implementation
during the preceding He-core burning
phase. Consequently,  small differences on the TP-AGB are unavoidable
when comparing tracks from two codes.

\begin{figure}[H]
\plotone{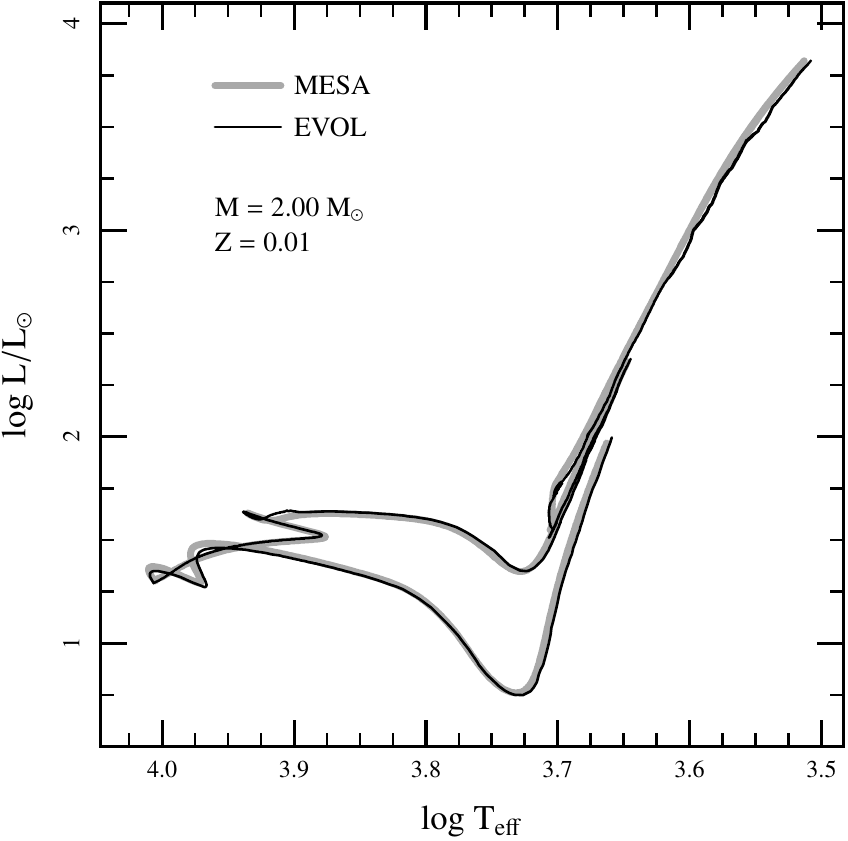}
\caption{ The $2\Msun, Z=0.01$ tracks up to the first thermal pulse from EVOL (solid black line) and $\MESAstar$ (thick grey line)  in the H-R diagram.
 \label{EVOL_MESA_HR}}
\end{figure}

As shown in Figure \ref{EVOL_MESA_HR}, the EVOL and $\MESAstar$ tracks compare well in the H-R diagram. Table \ref{tab:2mz0.01} shows
that key properties differ by less than 5\%.
$\MESAstar$ has the ability to impose a minimum size on convection
zones below which overshoot mixing is ignored.  EVOL does not have
such limits, leading to more mixing of He into the
core and, hence, the $\approx 4$\% larger age of the EVOL sequence at the
first thermal pulse.

\begin{figure}[H] 
\plotone{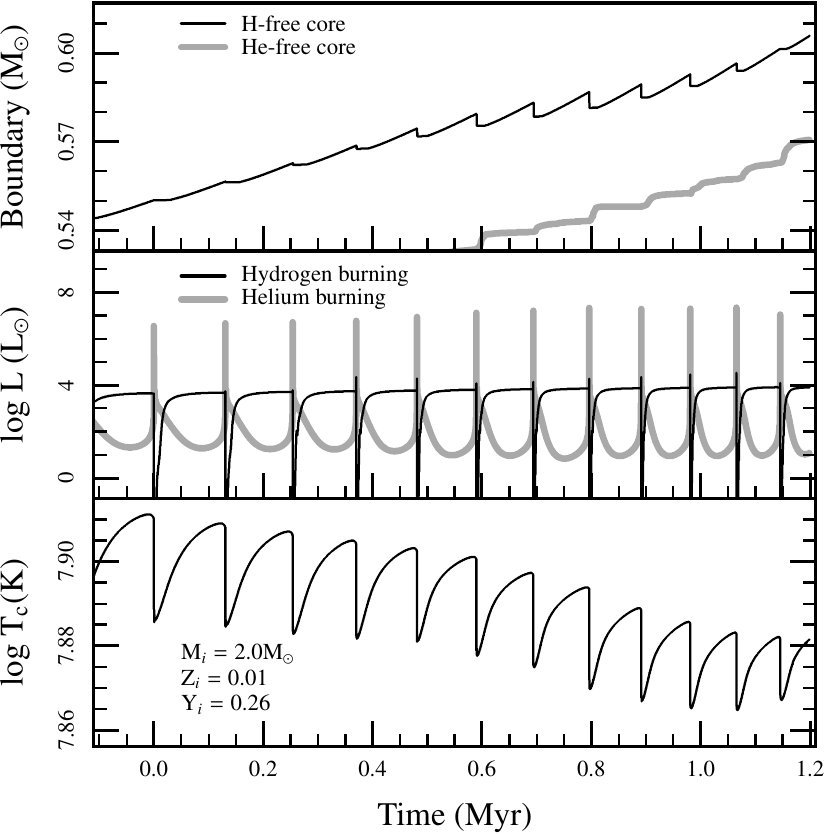}
\caption{Properties of a $M_i=2\Msun$ star from $\MESAstar$ as it approaches the 
end of the AGB. Top: the boundaries of the C/O core and the He layer. Middle: 
luminosities from hydrogen and helium burning. Bottom: central temperature evolution.  
\label{TP-AGB-trio}}
\end{figure}

\begin{figure}[H] 
\begin{center}
\includegraphics[width=0.75\textwidth]{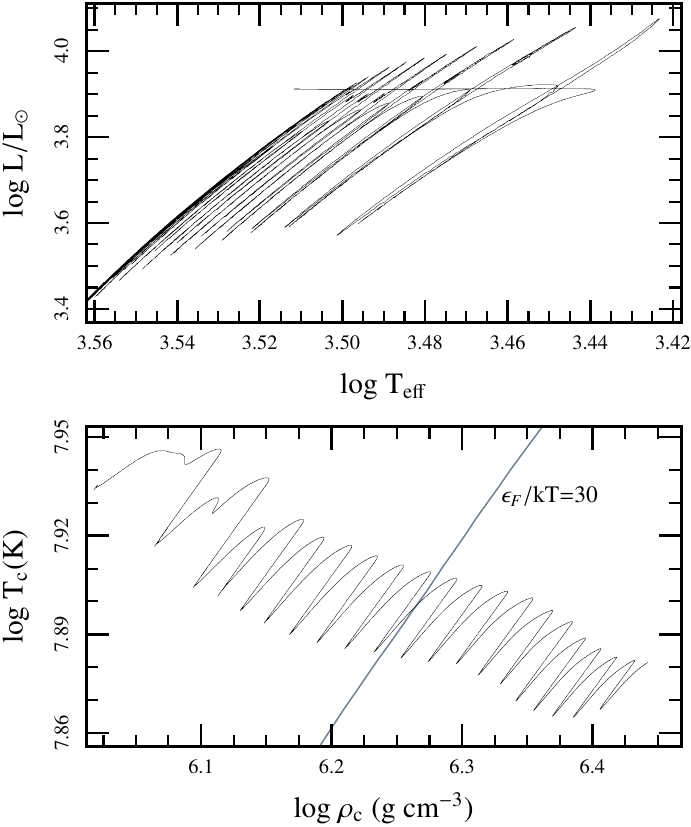}
\end{center}
\caption{Top: H-R diagram for the $2\Msun$ $\MESAstar$ model during the thermal 
pulses on the AGB.  Bottom: trajectories of the central conditions in the C/O core  during the 
thermal pulses. The line showing constant degeneracy is marked. 
\label{midpulse}}
\end{figure}

The thermal-pulse AGB (TP-AGB)  is characterized by recurrent thermonuclear
instabilities of the He-shell, leading to  complex mixing and
nucleosynthesis. These processes are properly represented in $\MESAstar$
calculations, as revealed in Figure \ref{TP-AGB-trio}. The ability of $\MESAstar$ to
calculate the evolution of stellar parameters in a smooth and continuous
manner even during the advanced thermal pulse phases
 and beyond is demonstrated in Figure \ref{midpulse}.  The top panel shows the evolution in the H-R diagram, whereas the bottom panel shows the evolution of the conditions in the C/O core. The adiabatic cooling in the C/O core that occurs 
 during the He flash (due to the pressure dropping at the surface of the C/O core) 
 is evident in the downturns that are parallel to the line of constant degeneracy (which is also the adiabatic slope). The overall trend of increasing $\rho_c$ reflects the growing C/O core mass, which for this model is shown in the top panel of 
 Figure \ref{TP-AGB-trio}.

\begin{deluxetable}{lll}
\tablecolumns{3}
\tablewidth{0pc}
\tablecaption{Comparison of $\MESAstar$ and EVOL models with $M_i=2\Msun, Z=0.01$}
\tablehead{\colhead{}&\colhead{$\MESAstar$}&\colhead{EVOL}}
\startdata
Main sequence lifetime                               (Gyr) & 0.939 & 0.962 \\
Deepest penetration of first dredge-up           ($\Msun$) & 0.328 & 0.327 \\
H-free core mass at the end of He-core burning   ($\Msun$) & 0.466 & 0.454 \\
Core mass at first thermal pulse                 ($\Msun$) & 0.504 & 0.481 \\
Age at first thermal pulse                           (Gyr) & 1.269 & 1.328 \\
Core mass at 2nd thermal pulse with DUP          ($\Msun$) & 0.563 & 0.563 \\
$\ \ $following interpulse time                        (1000 yr) & 116   & 106 \\
$\ \ $following pulse-to-pulse core growth       ($10^{-3}\Msun$) & 6.4   & 6.9 \\
$\ \ $dredge-up mass at following pulse          ($10^{-3}\Msun$) & 1.1   & 1.3 \\
\enddata
\label{tab:2mz0.01}
\end{deluxetable}
\clearpage

\begin{figure}[H]
\plotone{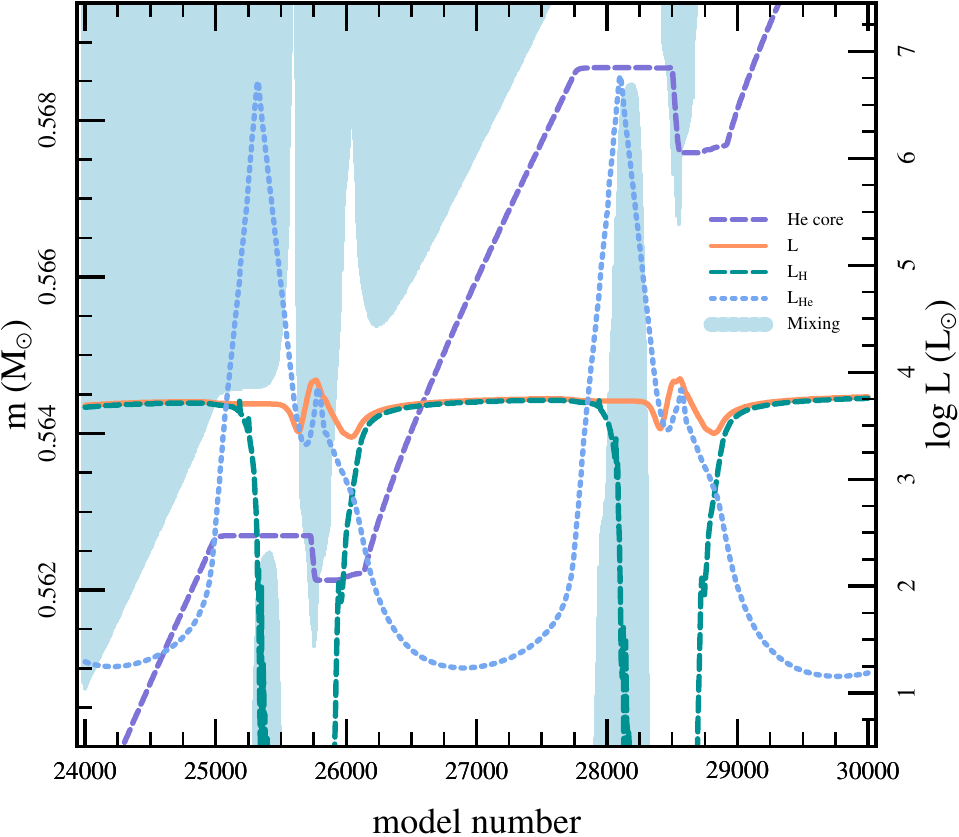}
\caption{
 Kippenhahn diagram with luminosities for the $2^\mathrm{nd}$ and $3^\mathrm{rd}$ thermal 
pulses with third dredge-up of the $2\Msun, Z=0.01$ $\MESAstar$ track shown in Figure
\ref{TP-AGB-trio} .
 \label{Kippenhahn_fromBill_M0024}}
\end{figure}

An example of the evolution of convection zones, shell burning and
total luminosities as well as core boundaries for two subsequent
thermal pulses is shown in Figure \ref{Kippenhahn_fromBill_M0024} as a
function of model number; compare to Figure 3 in \citet{her05}. Quantitative comparison of interpulse time, core growth
and dredge-up amount (see Table \ref{tab:2mz0.01}) shows excellent agreement between the $\MESAstar$ thermal pulses and the
equivalent pulses in the EVOL sequence.

\begin{figure}[H]
\plotone{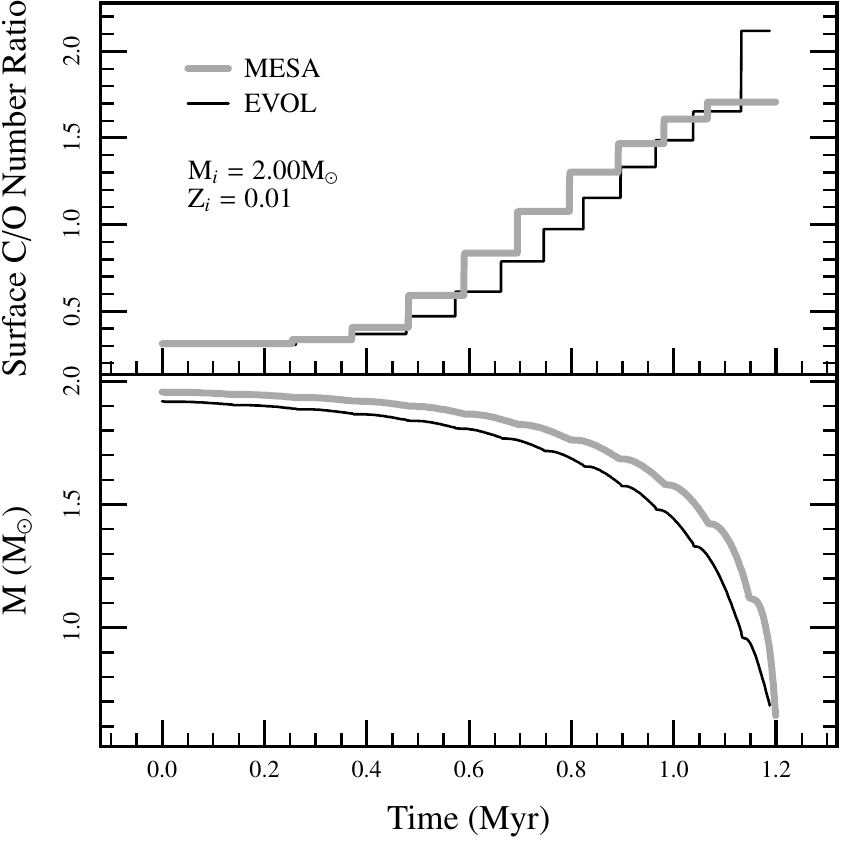}
\caption{
 The C/O number ratio (top panel) and stellar mass, $M$ (bottom panel) 
 as a function of time from EVOL (solid black line) and $\MESAstar$ (thick grey line). 
  Time has been set to zero for both tracks at the onset of the third dredge-up.
 \label{EVOL_MESA_CO}}
\end{figure}

Another important property of the He-shell flashes is the intershell
abundance as a result of the convective mixing and burning. Again, the
comparison of results from both codes shows good agreement, which is
expected since they both implement the same overshooting mixing assumptions
for the He-shell flash convection zone.
A consequence of the third dredge-up is the gradual increase of the
envelope C/O ratio as thermal pulses repeatedly occur. 
Since the $2\Msun$ models do not
experience hot-bottom burning, the evolution of this ratio is an
effective probe of the cumulative efficiency of the third dredge-up in
these simulations.\footnote{For massive AGB stars $\MESAstar$ shows the expected hot-bottom burning
behavior, including, for example, the avoidance of the C-star phase
for a $5\Msun$, $Z=0.01$ stellar model track despite efficient third
dredge-up.} The top panel of Figure \ref{EVOL_MESA_CO} 
shows the surface C/O ratio evolution according to EVOL (dashed-red line) 
and $\MESAstar$ (solid black line). They are in good agreement, e.g.\ in terms of the time
period over which the third dredge-up occurs and the amount by which
C/O increases. The mass loss history over the same time period, 
shown in the bottom panel of Figure \ref{EVOL_MESA_CO}, is similar by
design.

\subsubsection{Interior structure of Slowly Pulsating B Stars and Beta Cepheids} 

\label{s.pulsate}
The advent of space-based asteroseismology for main sequence B stars  with the {\it Corot} \citep{deg09} and {\it Kepler} \citep{gil10} satellites is probing 
 the slowly pulsating B stars (SPBs, $M\approx 3-8\Msun$) and the more massive ($M\approx 7-20\Msun$)  $\beta$ Cepheids \citep{deg09,deg10}. These stars are all undergoing main sequence H burning and are  unstably pulsating  due to the $\kappa$ mechanism 
from the Fe-group opacity bump at $T\approx 2\times 10^5 \ {\rm K}$ \citep{dzi93}. The observed modes have 
finite amplitudes deep in the stellar core, demanding a full interior model for mode frequency (and stability) prediction \citep{dzi93,pam04}. 

\begin{figure}[H]
\begin{center}
\includegraphics[width=0.6\textwidth]{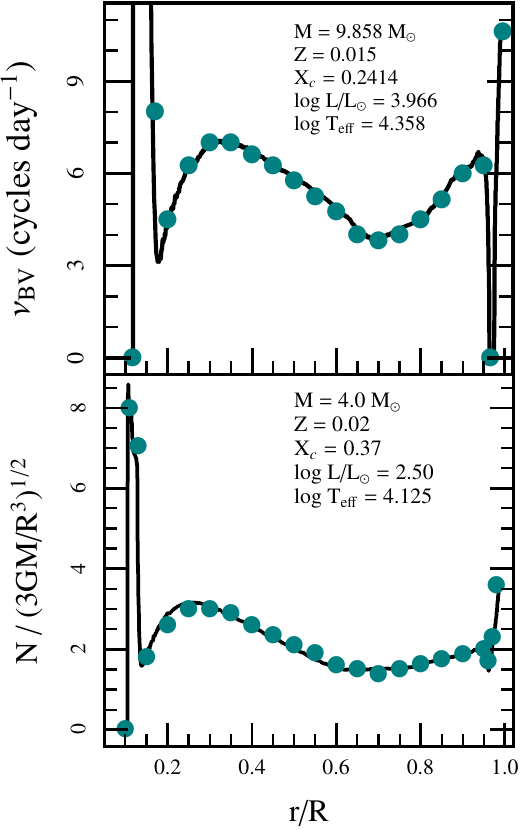}
\end{center}
\caption{Comparison of $\MESAstar$ predictions of the Brunt-V\"ais\"al\"a frequency, $N$,  to 
two cases from the literature; in both cases, the $\MESAstar$ model is shown as a solid line while the literature values are plotted as filled green circles. Comparisons are made at fixed $X_c$ for H burning stars.  The bottom panel shows a $4\Msun$ star from \citet{dzi93}, and the top panel shows a $M=9.858\Msun$ star from \citet{pam04}.  In keeping with the way the numbers are presented in these papers, the vertical axes are different in the two panels with the bottom one in dimensionless units of $N/(3GM/R^3)^{1/2}$  and the top in cycles per day. 
 \label{fig:bruntcompare}}
\end{figure} 
\clearpage

These papers provide a few specific models that allow a direct comparison to the 
$\MESAstar$ prediction of the  Brunt-V\"ais\"al\"a frequency
\begin{equation}
N^2=g\left({1\over \Gamma_1} {d\ln P \over dr}-{d\ln \rho\over dr}\right)=g\left(-{g\over c_s^2}-{d\ln \rho\over dr}\right),
\end{equation}
where $c_s^2=\Gamma_1P/\rho$ is the adiabatic sound speed, and we used hydrostatic balance, $dP/dr=-\rho g$. 
Numerically, these are obtained by interpolating the sound speed at the cell boundary, whereas $d\ln \rho/ dr$ is estimated by numerical differencing and then smoothed. This method naturally captures the  extra restoring force from composition gradients, especially relevant in these evolving stars that leave a He rich radiative region above the retreating convective core during the main sequence. 

Our first comparison is to \citet{dzi93}'s  $M=4\Msun$ main sequence  star with $Z=0.02$ at a time when the hydrogen abundance in the convective core is $X_c=0.37$.  With no overshoot from 
 the convective core, \citet{dzi93} found $\log L/\Lsun=2.51$ and $\log T_{\rm eff} = 4.142$ whereas $\MESAstar$ gives $\log L/\Lsun=2.50$ and $\log T_{\rm eff} = 4.125$. The top panel in Figure \ref{fig:bruntcompare} compares the $\MESAstar$ results (solid line) to the values  (green circles) from Figure 3 of \citet{dzi93}. The agreement is remarkable as an integral test  of $\MESAstar$. The bottom panel of Figure \ref{fig:bruntcompare} is a comparison to the more massive $M=9.858\Msun$ main sequence star with $Z=0.015$ from Figure 5 of \citet{pam04} at an age ($15.7$ Myr)  when $X_c=0.2414$ with  $\log L/\Lsun=3.969$ and $\log T_{\rm eff} = 4.3553$.  $\MESAstar$ gave $\log L/\Lsun=3.966$, $\log T_{\rm eff} = 4.358$ and an age of $16.4$ Myr  at the same value of $X_c$.  These comparisons highlight the readiness of $\MESAstar$ for adiabatic asteroseismological studies of main sequence stars.

\subsection{High Mass Stellar Structure and Evolution\label{s.vv_highmass}}

To explore $\MESAstar$'s results in this mass range, models of $15\Msun$,
$20\Msun$, and $25\Msun$ of solar metallicity and $1000\Msun$ of zero metallicity
were evolved from the Hayashi track to the onset of core-collapse.
Nuclear reactions are treated with the 21 isotope reaction network,
inspired by the 19 isotope network in \citet{wea78}, that is capable of
efficiently generating accurate nuclear energy generation rates from
hydrogen burning through silicon burning (see \S \ref{net}). This
network includes linkages for PP-I, steady-state CNO cycles, a
standard $\alpha$-chain, heavy ion reactions, and aspects of
photodisintegration into $^{54}$Fe.  Atmospheres are treated as a
$\tau$=2/3 Eddington gray surface as described in \S\ref{atm}.  Mass
loss for the solar metallicity stars uses the combined results of
\citet{gle09,vin01,nug00,nie90}, as described in \S \ref{s.masschanges}.
These massive star models are non-rotating, use no semi-convection,
employ a mixing length parameter of $\alpha_{MLT}$ = 1.6, and adopt $f$=0.01
for exponential diffusive overshoot (see \S\ref{overshoot}) for convective
regions that are either burning hydrogen or are not burning.

\begin{figure}[H]
\begin{center}
\includegraphics[width=0.6\textwidth]{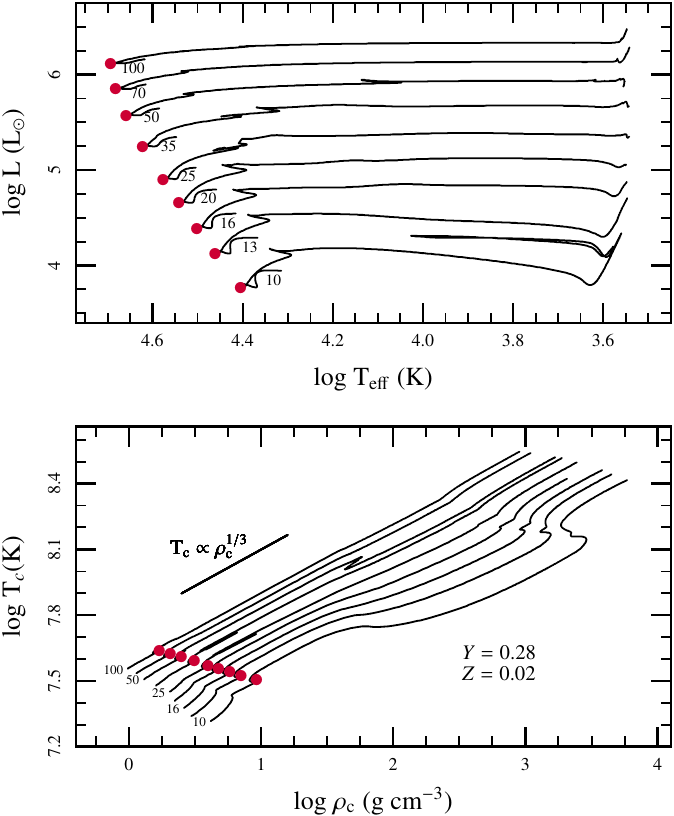}
\end{center}
\caption{Top: H-R diagram for  $10-100 \Msun$ models from the PMS  to the end of core Helium burning for $Z=0.02$ but with zero mass loss. Bottom: trajectories of the central conditions in the $T-\rho$ plane  over this same evolutionary period. 
\label{highduo}}
\end{figure}

Most of this section consists of comparisons to results from other stellar evolution 
codes. However, for consistency (and completeness), we show in Figure \ref{highduo} the 
H-R diagram and central condition evolution of  $10-100\Msun$ stars from the PMS to the 
end of core Helium-burning. Though these are stars with $Z=0.02$, we 
turned off mass loss during this calculation so that the plot would be easier to read 
and of some pedagogical use. The tendency of $T_c$ to scale with $\rho_c^{1/3}$ (also a 
constant radiation entropy) during these stages of evolution is expected from hydrostatic 
balance with only a mildly changing mean molecular weight. The rest of the calculations in 
this section included mass-loss as described above. 

\subsubsection{$25\Msun$ Model Comparisons\label{s.25msun}}

\begin{figure}[H]
\plotone{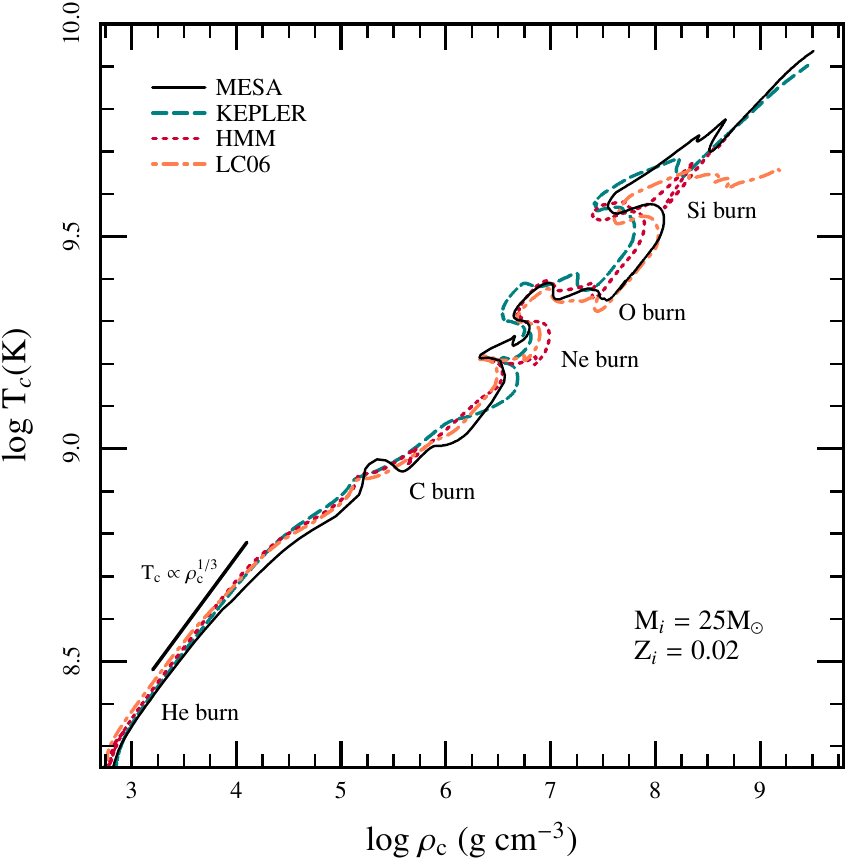}
\caption{
Evolution of the central temperature and central density in 
solar metallicity $M_i=25 \Msun$ models from different stellar evolution codes.
The locations of core helium, carbon, neon, oxygen, and silicon burning are labeled, 
as is the relation $T_c\propto \rho_c^{1/3}$. 
\label{25M:HRTRho}}
\end{figure}

Figure \ref{25M:HRTRho} shows the $T_c-\rho_c$ evolution 
in $M_i=25\Msun$ solar metallicity models from
$\MESAstar$, Kepler (private communication - Alex Heger), \citet{hmm},
and FRANEC \citep{lim06} from helium burning until iron-core collapse.
The curves fall below the $T_c\propto \rho_c^{1/3}$ scaling relation as 
the mean molecular weight increases due to the subsequent burning stages. 
The curves are also punctuated with
non-monotonic behavior when nuclear fuels are first ignited in shells.  Figure \ref{25M:HRTRho}
shows that $\MESAstar$ produces core evolutionary tracks
consistent with other pre-supernova efforts. The bump in the
$\MESAstar$ curve around carbon burning is due to the development of
central convection whereas the other codes do not \citep[although see
Figure 2 of][]{lsc}.  The development of a convective core during
carbon burning depends on the carbon abundance left over from core
helium burning \citep{lsc}.

\begin{figure}[H]
\begin{center}
\includegraphics[width=0.9\textwidth]{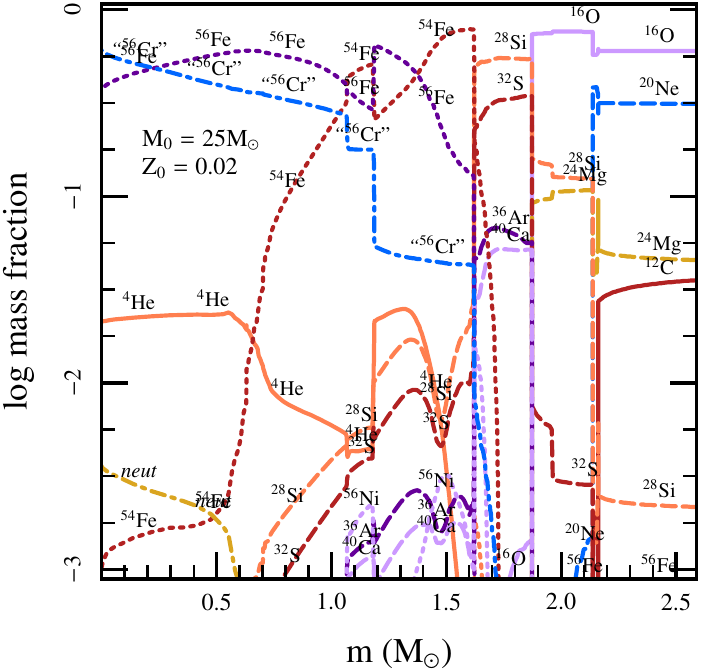}
\end{center}
\caption{
Mass fraction profiles of the inner 2.5$\Msun$ of the solar
metallicity $M_i= 25\Msun$ model at the onset of core collapse.  The reaction network includes links
between $^{54}$Fe, $^{56}$Cr, neutrons, and protons to model aspects
of photodisintegration and neutronization.
\label{25M:Abund}}
\end{figure}

\begin{figure}[H]
\plotone{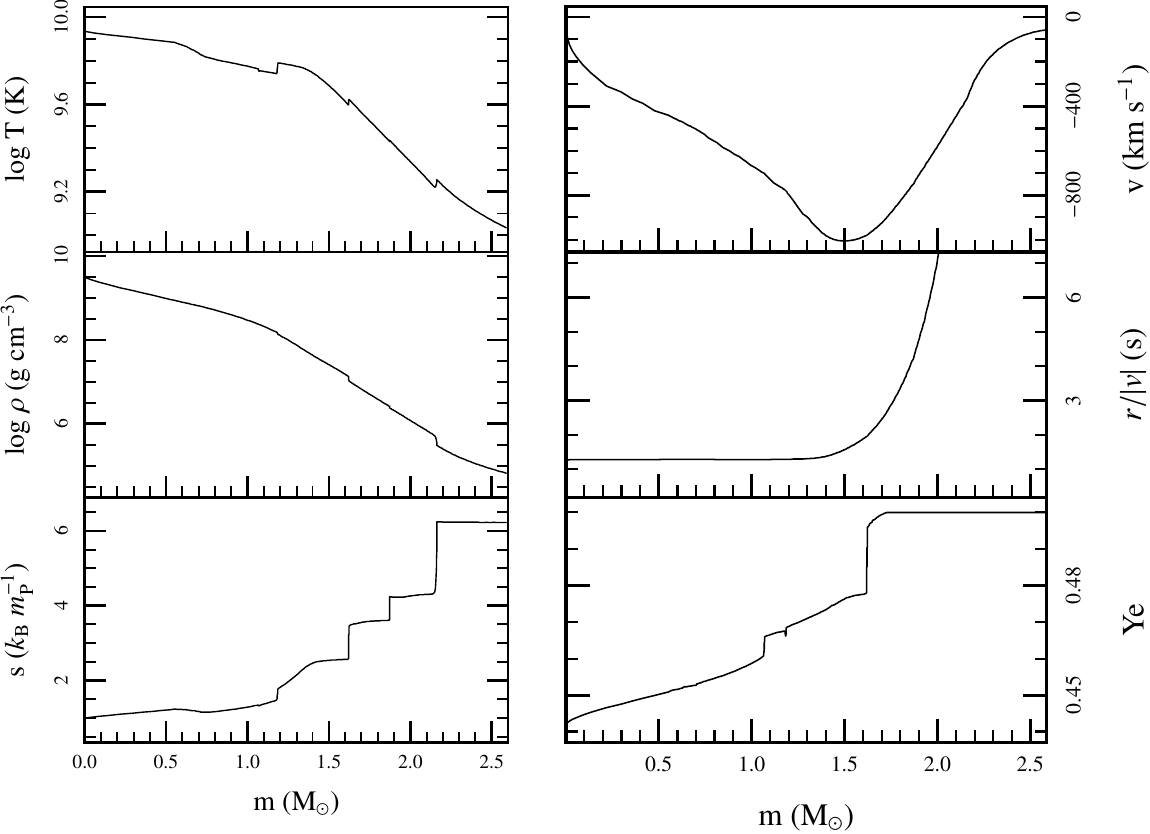}
\caption{
Profiles of $T$ (top left), $\rho$ (middle left),
dimensionless entropy (bottom left), material speed (top right),
infall timescale (middle right), and electron fraction $Y_e=\overline{\rm Z}/\overline{\rm A}$ (bottom right)  over the inner $2.5\Msun$ of the
$M_i=25\Msun$ star at the end of the pre-supernova evolution. 
\label{25M:profiles}}
\end{figure}

The mass fraction profiles of the inner $2.5\Msun$ of this $M_i=25\Msun$
model are shown in Figure \ref{25M:Abund} at the onset of core collapse. 
At the time of these plots, the infall speed has reached $\approx 1000\  {\rm km \ s^{-1}}$ just inside the iron core (at $m=1.5\Msun$) and the electron fraction, $Y_e$, has dropped below $\approx 0.48$. The oxygen shell lies at 
$1.88 \le m/\Msun \le 2.5 $, the silicon shell between $1.61 \le m/\Msun \le 1.88$,
and the iron core at $m \le 1.61 \Msun$.  Figure \ref{25M:profiles} shows $T$, $\rho$, $S$, the radial velocity, the infall timescale, and
$Y_e$ of this inner 2.5$\Msun$. Note the entropy
decrements at the oxygen, silicon and iron core boundaries.

\begin{figure}[H]
\plotone{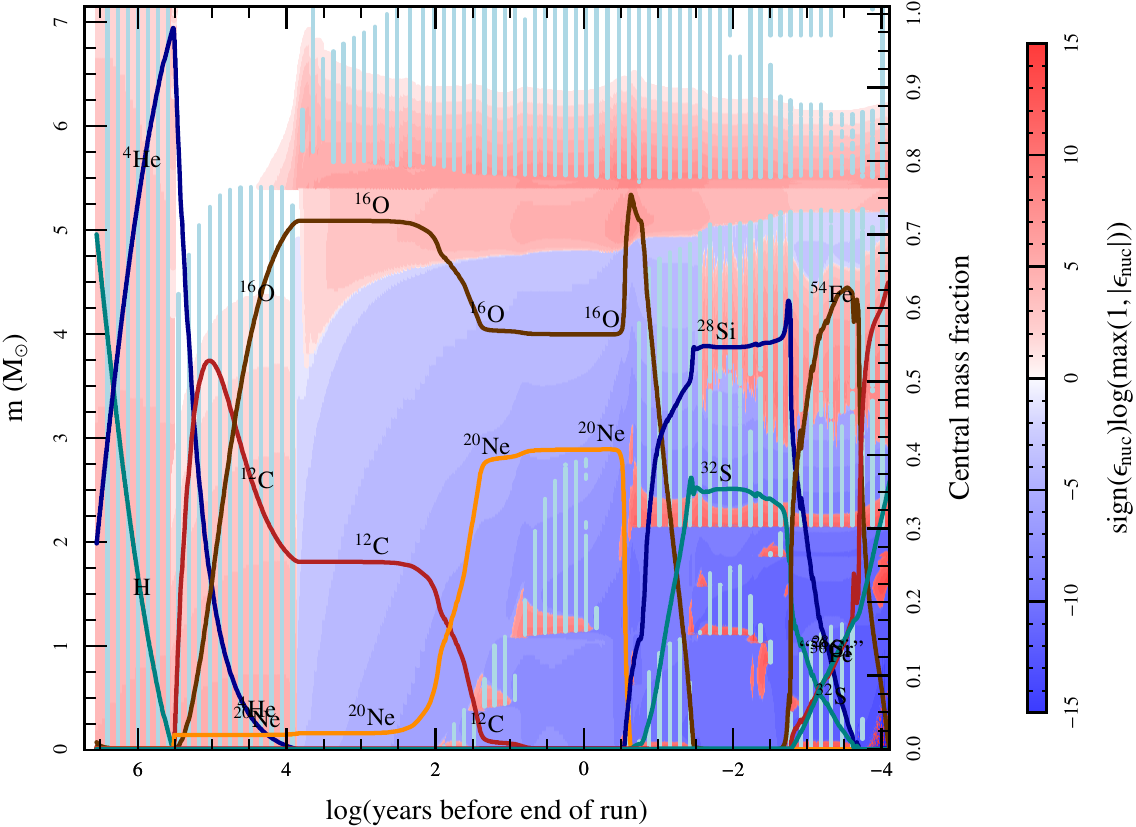}
\caption{
Kippenhahn diagram showing the full time evolution of the inner 7 $\Msun$ 
of the $M_i=25 \Msun$ 
evolutionary sequence from the main sequence to the onset of core
collapse. Mass coordinate and abundance mass fraction are labeled on
the left and right y-axes, respectively. The shaded bar on the right
indicates the net energy generation: red for positive values and blue
for negative values. The vertical lines indicate convection.
\label{25M:Kippenhahn}}
\end{figure}

Figure \ref{25M:Kippenhahn} summarizes the history of the inner
7$\Msun$ of this $M_i=25\Msun$ model as a function of interior mass (left
y-axis). Evolution is measured by the logarithm of time (in years)
remaining until the death of the star as a supernova (x-axis), which
reveals the late burning stages.  Levels of red and blue shading
indicate the magnitude of the net energy generation (nuclear energy
generation minus neutrino losses), with red reflecting positive values
and blue indicating negative ones. The vertical lines indicate regions
that are fully convective.  Note the appearance of a convective
envelope characteristic of a red supergiant late during helium
burning.  Abundance profiles of key isotopes during the major burning
stages are shown (right y-axis).  The hydrogen core shrinks towards
the end of hydrogen burning, and the helium core grows as helium is
depleted.  The total mass shrinks to about $M=12\Msun$ due to mass
loss.

\subsubsection{Comparison of 15, 20, and $25\Msun$ Models\label{s.highmasscompare}}

Now that we have shown that the $M_i=25\Msun$ $\MESAstar$ models compare 
well to previous efforts at the qualitative level, we will make more detailed 
comparisons to other available results. 
Table \ref{life25} compares the core burning lifetimes of solar
metallicity stars with $M_i=15,20$ and $25\Msun$, from $\MESAstar$,
\citet{hmm}, \citet{whw}, and \citet{lsc}.  We define a core burning
lifetime to begin when the central mass fraction of fuel has dropped
by 0.003 from its maximum value (or onset of central convection) and
to terminate when the central mass fraction has dropped below
$10^{-4}$ (or the end of central convection). Different authors adopt
different lifetime definitions, which likely contribute to some of the
scatter.  The hydrogen burning lifetimes for the 15$\Msun$, 20$\Msun$, 
and 25$\Msun$ models from the different authors are within 10\%
percent of each other, with the \citet{lsc} models generally having
the shortest lifetimes and the \citet{whw} models having the longest
lifetimes.  There is more spread in the helium burning lifetimes, with
$\MESAstar$ models showing shorter lifetimes and \cite{whw} models
having the longest lifetimes.  The carbon burning lifetimes 
show agreement within 20\% for the 15$\Msun$ model,
but differ by factors of $\sim$3 for the $20\Msun$ and $25\Msun$
models.  The neon, oxygen, and silicon burning lifetimes show
agreement within 20\% between some models, but factor of $\sim$5
differences in others.  It is beyond the scope of this paper to put the
different lifetime definitions on the same footing, and explore the
reasons for these differences. Nevertheless, Table \ref{life25}
suggests $\MESAstar$ produces lifetimes consistent with the range of 
lifetimes from other works.

Table \ref{mass25} compares  pre-supernova core masses of solar
metallicity stars with 
$M_i=15,20$ and $25\Msun$ models from $\MESAstar$, \citet{hmm}, \citet{rhw},
\citet{hlw00}, and \citet{lsc}.  $\MESAstar$ core masses are
defined as the mass interior to the location where the element mass
fraction is 0.5. 
The definitions used by various authors may differ,
contributing to scatter in the results.  However, most of the scatter
is probably due to the different mass loss prescriptions used by
different authors, resulting in different total masses.  The helium
yields differ by about 10\%, with the \citet{hlw00} models producing
less helium.  There is more diversity in the C+O+Ne bulk yields, up to
a factor of 2 for the $25\Msun$ model, with the \citet{rhw} models
producing the most and the \citet{hlw00} models producing the least.
Strikingly, the Fe core masses show less variations, with the
\citet{hmm} models producing the heaviest cores.  Table \ref{mass25}
suggests $\MESAstar$ produces bulk yields compatible with previous efforts.

\begin{deluxetable}{lccccc}
\tabletypesize{\footnotesize}
\tablecolumns{6}
\tablewidth{0pc}
\tablecaption{Massive Star Core Burning Lifetime Comparison\label{life25}}
\tablehead{ \colhead{Core Burning} & \multicolumn{4}{c}{Lifetime (years)} & \colhead{} \\
            \colhead{Element} & \colhead{HMM} & \colhead{WHW} & 
\colhead{LSC} & \colhead{$\MESA$} & \colhead{ } }
\startdata
\cutinhead{$M_i=15\Msun$}
H  & 1.13 &   1.11  &   1.07  &   1.14  & $\times 10^7$ \\
He & 1.34 &   1.97  &   1.4   &   1.25  & $\times 10^6$ \\
C  & 3.92 &   2.03  &   2.6   &   4.23  & $\times 10^3$ \\
Ne & 3.08 &   0.732 &   2.00  &   3.61  &  \\
O  & 2.43 &   2.58  &   2.43  &   4.10  &  \\
Si & 2.14 &   5.01  &   2.14  &   0.810 & $\times  10^{-2}$ \\

\cutinhead{$M_i=20\Msun$}
H  &  7.95 &   8.13  &  7.48 &  8.01  & $\times 10^6$ \\    
He &  8.75 &   11.7  &  9.3  &  8.10  & $\times 10^5$ \\    
C  &  9.56 &   9.76  & 14.5  & 13.5   & $\times 10^3$ \\    
Ne &  0.193&   0.599 & 1.46  &  0.916 &  \\                 
O  &  0.476&   1.25  & 0.72  &  0.751 &  \\                 
Si &  9.52 &   31.5  & 3.50  &  3.32  & $\times  10^{-3}$ \\

\cutinhead{$M_i=25\Msun$}
H  & 6.55   & 6.706   & 5.936   & 6.38  & $\times 10^6$\\
He & 6.85   & 8.395   & 6.85    &  6.30 & $\times 10^5$\\
C  & 3.17   & 5.222   & 9.72    &  9.07 & $\times 10^2$\\
Ne & 0.882  & 0.891   & 0.77    &  0.202&     \\
O  & 0.318  & 0.402   & 0.33    &  0.402&    \\
Si & 3.34   & 2.01    & 3.41    &  3.10 & $\times 10^{-3}$
\enddata
\tablerefs{
HMM--\citet{hmm}; WHW--\citet{whw}; LSC--\citet{lsc}; $\MESA$--this paper}
\end{deluxetable}

\begin{deluxetable}{lccccc}
\tablecolumns{6}
\tablewidth{0pc}
\tablecaption{Pre-Supernovae Core Mass Comparisons\label{mass25}}
\tablehead{\colhead{Mass ($\Msun$)} & \colhead{HMM} & \colhead{RHW} & 
\colhead{HLW} & \colhead{LSC} & \colhead{$\MESA$} }
\startdata
\cutinhead{$M_i=15\Msun$}
Total   & 13.232 &  12.612 & 13.55 &  15    &  12.81 \\
    He  &  4.168 &   4.163 &  3.82 &  4.10  &  4.37  \\
C+O+Ne  &  2.302 &   2.819 &  1.77 &  2.39  &   2.27 \\
``Fe"   &  1.514 &   1.452 &  1.33 &  1.429 &  1.510 \\

\cutinhead{$M_i=20\Msun$}
Total  & 15.69 & 14.74 & 16.31 & 20    & 15.50 \\
He     &  6.21 &  6.13 &  5.68 &  5.94 &  6.33 \\
C+O+Ne &  3.84 &  4.51 &  2.31 &  3.44 &  3.77 \\
``Fe"  &  1.75 &  1.46 &  1.64 &  1.52 &  1.58 \\

\cutinhead{$M_i=25\Msun$}
Total   & 16.002 & 13.079 & 18.72  & 25     & 15.28 \\
He      &  8.434 &  8.317 &  7.86  &  8.01  &  8.41 \\
C+O+Ne  &  5.834 &  6.498 &  3.11  &  4.90  &  5.49 \\
``Fe"   &  1.985 &  1.619 &  1.36  &  1.527 &  1.62 \\
\enddata
\tablerefs{
HMM--\citet{hmm};  RHW--\citet{rhw}; HLW--\citet{hlw00}; 
LSC--\citet{lsc}; $\MESA$--this paper }
\end{deluxetable}

\subsubsection{$1000\Msun$ metal-free star capabilities} 

\begin{figure}[H]
\begin{center}
\includegraphics[width=0.48\textwidth]{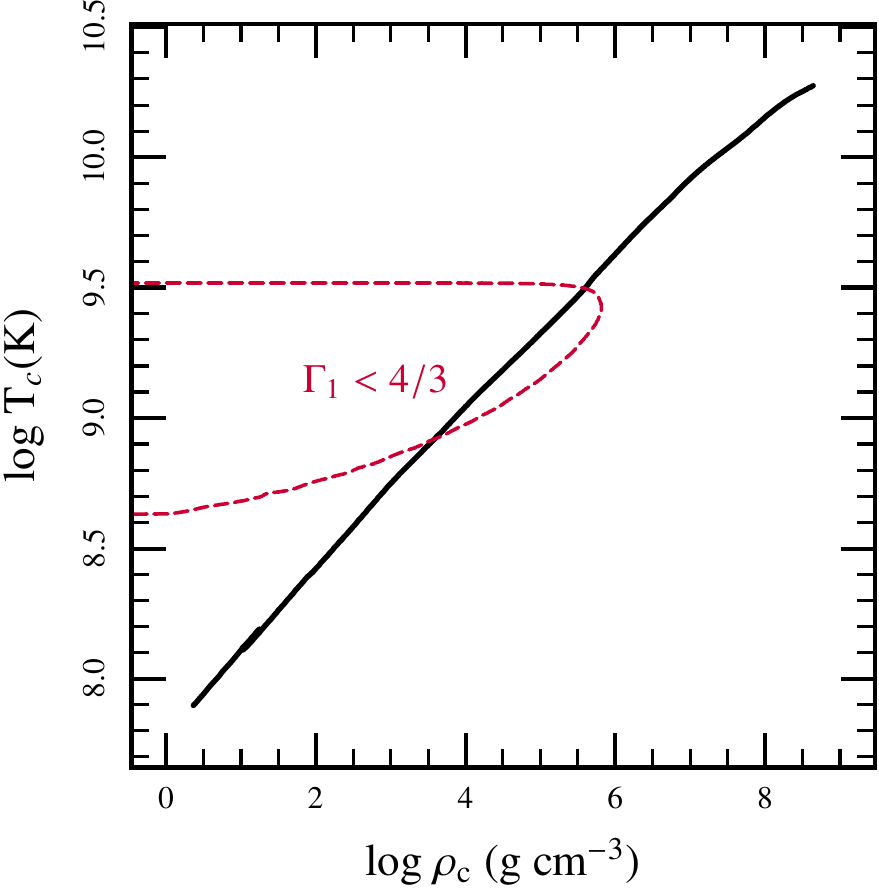}
\includegraphics[width=0.3\textwidth]{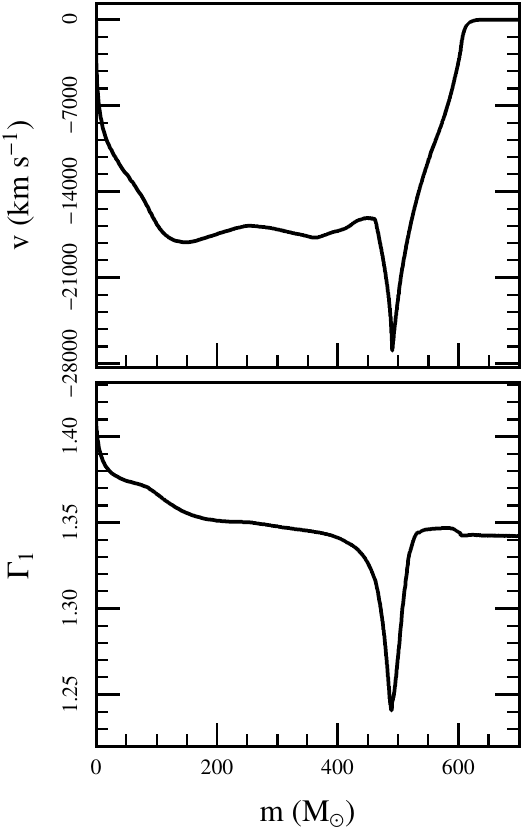}
\end{center}
\caption{
Time history (left panel) of $T_c$ and $\rho_c$ in a 
$1000\Msun$, zero metallicity, zero mass loss model. Also shown
are the boundaries within which $\Gamma_1 <$4/3. Material speed
and $\Gamma_1$ profiles (right panel) for the final model.
\label{fig.1000msun_t_rho}}
\end{figure}

We close this section with a demonstration of $\MESAstar$'s
capabilities by describing the unlikely scenario of a purely metal-free stellar evolution of a $M_i=1000\Msun$ star. The $T_c-\rho_c$ trajectory 
for a $1000\Msun$, zero metallicity,
zero mass loss model is shown in the left panel of Figure
\ref{fig.1000msun_t_rho}. The starting time point is in the lower left
corner and the final model, at the onset of core-collapse,  is in the upper right 
at very high values of $T_c$ and $\rho_c$. Fluid elements in the 
region to the left of the red-dashed line have 
$\Gamma_1 <4/3$. When the center enters this region, the
central portions of the star become dynamically unstable and begin
to contract.  However, the entire star does not collapse because the
infalling regions become denser and hotter, causing the central region
to leave the $\Gamma_1 <4/3$ region and the infall to slow.  Now
another part of the star moves into the $\Gamma_1 <$ 4/3 region
and begins to infall at high velocity. The net result is that the 
region where $\Gamma_1 <$ 4/3 starts at the center and moves outward.
The right panel of Figure \ref{fig.1000msun_t_rho} shows the material
speed and $\Gamma_1$ profiles for the final model, where the infalling
region is now at $m \approx 480 \Msun$.

\begin{figure}[H]
\plotone{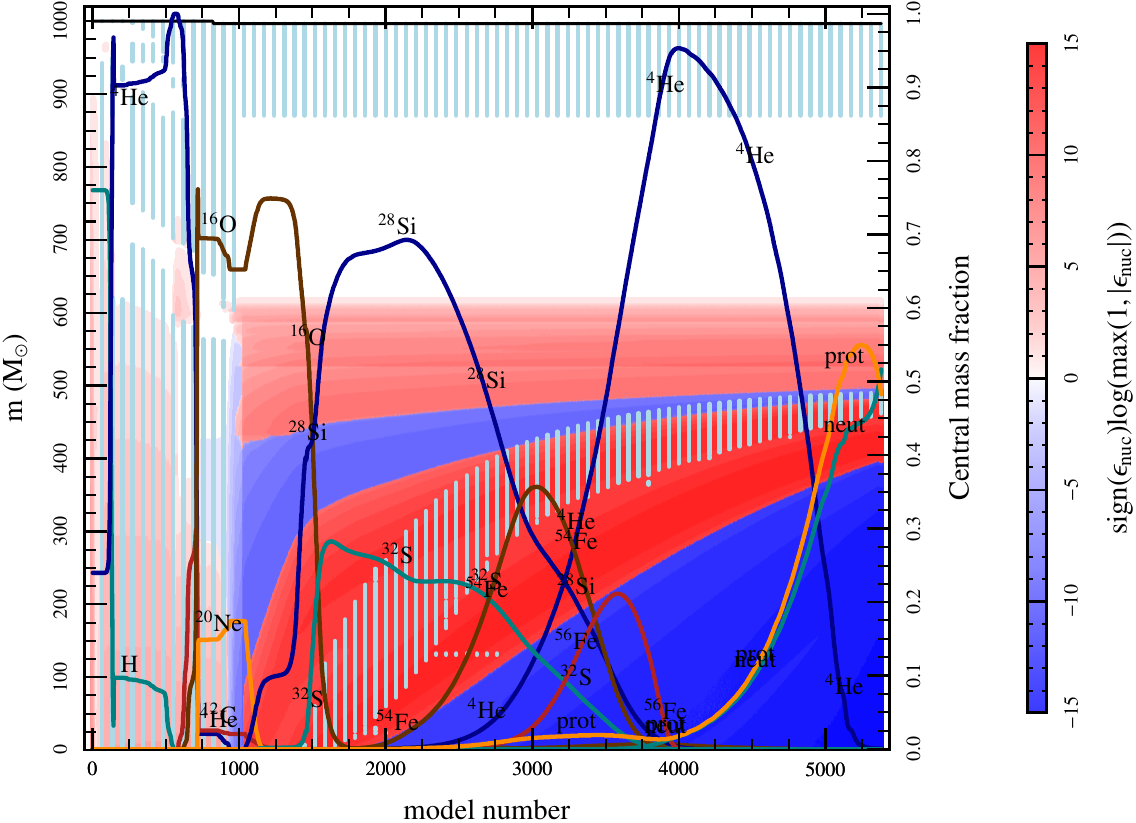}
\caption{
Kippenhahn diagram showing the evolution of the $1000\Msun$ model.
The format is the same as Figure \ref{25M:Kippenhahn}. 
\label{fig.1000msun_kippenhahn_burn_mix}}
\end{figure}

The global history of the $1000\Msun$ model as a function of time is
shown in the left panel of Figure \ref{fig.1000msun_kippenhahn_burn_mix}.  
 A convective envelope appears during late helium burning.  Abundance
profiles of key isotopes during the major burning stages are shown
(right y-axis).  Note the short carbon burning era. At late times the
core photodisintegrates to $^4$He instead of creating $^{56}$Ni because
of the lower central densities encountered in these supermassive
progenitors. This also partially causes the large endothermic central
regions of the star.

\subsection{Stellar Evolution with Mass Transfer } 
\label{s.accr} 

$\MESAstar$  can be used to examine how a star responds to mass loss or 
accretion (see \S \ref{s.masschanges}). This opens up a large variety of possible applications, including accretion onto white 
dwarfs for classical novae and thermonuclear supernovae, mass transfer in tight stellar binaries, and learning the response of a star to sudden mass loss. We show two examples where $\MESAstar$'s results 
can be compared to previous work. The first is a  mass-transfer scenario relevant to $P_{\rm orb}<2$ hour cataclysmic variables, and the second is the response of a neutron star to accretion of pure He.

\subsubsection{Mass Transfer in a Binary } 

To illustrate $\MESAstar$'s ability to calculate the impact of mass loss on a star, 
we model the evolution of a  compact binary consisting of a Roche Lobe filling low-mass ZAMS ($M<0.2M_\odot$) 
model and an accreting white dwarf with  $M_{\rm WD}=0.6\Msun$. These short orbital period ($P_{\rm orb}< 2$ hr) 
cataclysmic variables are the end points of these mass transferring systems \citep{pat98, kolb99} 
and are now being discovered in large numbers in the SDSS database \citep[more than 100 studied by][]{gan09}. 

\begin{figure}[H]
\plotone{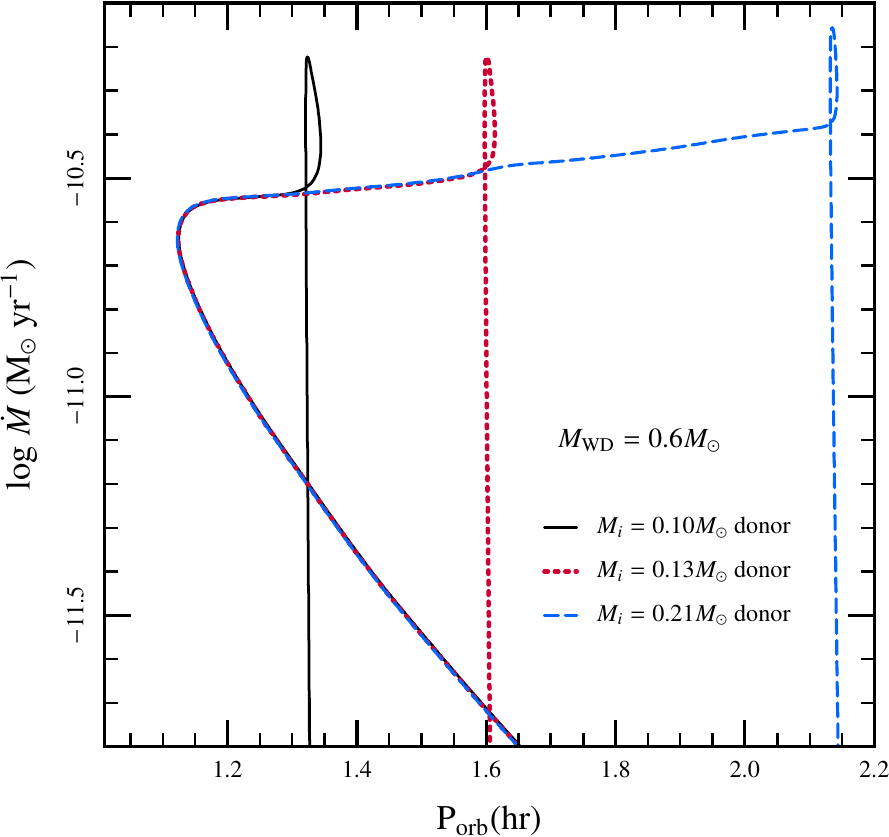}
\caption{Mass transfer rate for cataclysmic variables with low mass main sequence donor stars of varying initial masses $M_i$. Each line shows the $\dot M$ history for
different initial mass donors, all accreting onto a $M_{\rm WD}=0.6M_\odot$ white dwarf. After a period of initial adjustment to the mass transfer, each track tends to the same trajectory, showing the 
orbital period minimum at $P_{\rm orb}=67.4$ minutes. 
\label{rloverflow}}
\end{figure}

We model the parameters of the binary system and the Roche lobe overflow triggered mass transfer rate $\dot M$ as 
in \citet{mad06}. So as to compare to the previous work of \citet{kolb99}, we presume angular momentum losses  from gravitational wave emission and keep the  accreting WD mass fixed at its initial value, $M_{\rm WD}=0.6M_\odot$. The 
evolution of the donor star is carried out by $\MESAstar$, using the $\tau=100$ atmosphere tables from $\atm$. 
The evolution shown in Figure \ref{rloverflow} is followed for over 6 Gyr until the donor has been reduced to a brown dwarf remnant of $M\approx 0.03\Msun$ (see Table \ref{kolbbaraffe}).  During that time, the binary period drops to a minimum value of 67.4 minutes and then increases, independent of the initial donor mass. This plot is very similar to Figure 1 of \citet{kolb99}. We also show in Table \ref{kolbbaraffe} the evolution in time of the main properties of the donor star and mass transfer rate of the $M_i=0.21M_\odot$ model. Again, this agrees with the results in Table 2 of  \citet{kolb99}. The prime differences can be attributed to a slightly different $R(M)$ relation.

\begin{deluxetable}{rrrrrrr}
\tabletypesize{\footnotesize}
\tablecolumns{7}
\tablewidth{0pc}
\tablecaption{Mass Transfer History for $M_i=0.21M_\odot$ and $M_{WD}=0.6M_\odot$ \label{kolbbaraffe}}
\tablehead{
\colhead{Time (Gyr)} &
\colhead{$P_{\rm orb} ({\rm hr})$}   & 
\colhead{$M/M_\odot$}   &
\colhead{$T_{\rm eff}$ (K)} &
\colhead{$\log (L/L_\odot)$} &
\colhead{$R/R_\odot$} &
\colhead{$\log \dot M$} 
           }
\startdata
 0.00 &  2.1319&   0.2100   &  3278  & -2.2688 &  0.2279  & -10.24   \\
 0.25 &  2.0962  & 0.1987   &  3262  & -2.3041 &  0.2209  & -10.39    \\  
 0.50  & 2.0367   &0.1887   &  3242  & -2.3467 &  0.2129  & -10.40   \\   
 0.75   & 1.9770   &0.1787  &   3217  & -2.3939 &  0.2049  & -10.41  \\    
 1.00   & 1.9181   &0.1693  &   3191  & -2.4414 &  0.1971 &  -10.42 \\     
 1.25   & 1.8560   &0.1599  &   3167 &  -2.4904 &  0.1891  & -10.44  \\    
 1.50   & 1.7938   &0.1510   &  3142  & -2.5404  & 0.1814  & -10.45  \\    
 1.75   & 1.7299   &0.1421 &    3113  & -2.5952  & 0.1735   &-10.46  \\    
 2.00   & 1.6684 & 0.1336 &    3074  & -2.6563  & 0.1659 &  -10.47  \\    
 2.25   &1.6097 &  0.1252  &   3018  & -2.7277  & 0.1585  & -10.48 \\     
 2.50   &1.5475 & 0.1170  &   2965   &-2.8004 &  0.1510 &  -10.50  \\    
 2.75   & 1.4829 &  0.1093  &   2916  & -2.8737 & 0.1435 &  -10.51  \\    
 3.00  & 1.4151 &  0.1016   &  2854 &  -2.9590  & 0.1358 &  -10.52   \\   
 3.25  & 1.3484 &  0.0942  &   2774  & -3.0577  & 0.1283   &-10.53 \\     
 3.50   &1.2804 &  0.0868 &    2666  & -3.1796  & 0.1207 &  -10.54  \\    
 3.75  & 1.2172 &  0.0796  &   2525  & -3.3273 &  0.1135 &  -10.54  \\    
 4.00  & 1.1663 &  0.0726  &   2355 &  -3.4987  & 0.1071  & -10.55  \\    
 4.25 &  1.1345 &  0.0659  &   2161 &  -3.6915 &  0.1019  & -10.58  \\    
 4.50  & 1.1227&   0.0596  &   1963  & -3.8930  & 0.0980 &  -10.63  \\    
 4.75 & 1.1291 &  0.0541   &  1771  & -4.0949  & 0.0953  & -10.70  \\   
 5.00  & 1.1487 &  0.0495 &    1595  & -4.2924 &  0.0937 &  -10.78   \\   
 5.25  & 1.1752 &  0.0457  &   1445 &  -4.4725  & 0.0927  & -10.86  \\    
 5.50  & 1.2051 &  0.0426  &   1314  & -4.6436 &  0.0922  & -10.94  \\    
 5.75  & 1.2379  & 0.0399 &    1201  & -4.8016  & 0.0919 &  -11.02  \\    
 6.00  & 1.2706 &  0.0377  &   1110  & -4.9398  & 0.0918  & -11.09 \\     
 6.25  & 1.3035 &  0.0358 &    1028  & -5.0720 &  0.0918 &  -11.16   \\   
 6.50 &  1.3343 &  0.0343  &    969  & -5.1749  & 0.0919  & -11.22   \\  
 6.75  & 1.3636 &  0.0328  &    921  & -5.2630 &  0.0920  & -11.29\\     
 7.00 &  1.3902 &  0.0316 &     879  & -5.3425 &  0.0921&   -11.34  \\   
\enddata
\end{deluxetable}

\subsubsection{Rapid Helium Accretion onto a Neutron Star} 
\label{s.nsaccr} 

   The outer envelope of an accreting neutron star is modeled in $\MESAstar$ by using non-zero boundary conditions
$M_c$ and $L_c$ (see discussion in \S \ref{s.stareqns}) at a finite radius $R_c$. This allows for a time dependent 
calculation of the thermonuclear instability that yields Type I X-ray bursts \citep{stroh06}
for those accretion rates where the burning is thermally unstable ($\dot M\le 10^{-8}M_\odot \ {\rm yr^{-1}}$).
Such calculations have been performed with the KEPLER code \citep{woo04,cyb10} and prove very valuable in direct comparisons to observed Type I X-ray burst recurrence times and light curves, especially for the H-rich accreting ``clocked burster" GS 1826-24 \citep{heg07}. We  focus here on pure He accretion, 
relevant to neutron stars in ultra-compact binaries, such as 4U 1820-30 \citep{cum03}. 

For these simulations we set $M_c=1.4M_\odot$, $R_c=10$ km, $L_c=3.6 \times 10^{34} \ {\rm ergs \ s^{-1}}$, and $g=2.39\times 10^{14}\ {\rm cm \ s^{-2}}$ (correcting for the gravitational redshift). 
The initial model consisted of $3\times 10^{25} {\rm g}$ of pure $^{56}$Fe and accreted pure He at 
$\dot M=3\times 10^{-9}M_\odot \ {\rm yr^{-1}}$.  We require a slightly higher value of core luminosity $L_c/\dot M\approx 0.19 \ {\rm keV \ nucleon^{-1}}$
to reach the same ignition column depth ($5\times 10^8 \ {\rm g \ cm^{-2}}$) as  \citet{wei06}.  We used 31 species
in the nuclear reaction network,   including the $^{12}$C bypass reaction chain $^{12}$C(p,$\gamma$)$^{13}$N($\alpha$,p)$^{16}$O and elements ($^{23}$Na, $^{27}$Al, $^{31}$P, $^{35}$Cl, and $^{39}$K) 
that can appear as intermediates  in $(\alpha,{\rm p})({\rm p},\gamma)$ reactions and serve as 
the proton source for the $^{12}$C bypass \citep{wei06}. 

\begin{figure}[H]
\begin{center}
\includegraphics[width=0.8\textwidth]{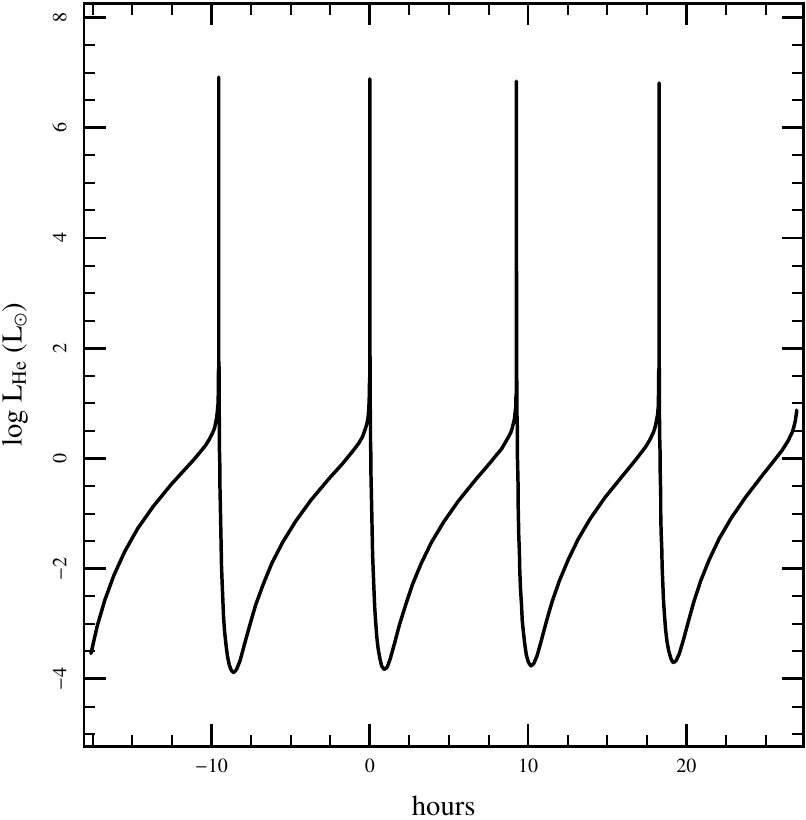}
\end{center}
\caption{The helium-burning luminosity, $L_{\rm He}$, as a function of time for a neutron star of mass $M_c=1.4M_\odot$ and radius $R_c=10$ km accreting pure helium at $\dot M=3\times 10^{-9}M_\odot \ {\rm yr^{-1}}$.  The Type I X-ray bursts occur every 9.56 hours. 
\label{lehens}}
\end{figure}

Figure \ref{lehens} shows a snapshot of the  time history of the helium burning luminosity, $L_{\rm He}$, which is periodic at the Type I burst recurrence time of 9.56 hours. This luminosity, as well as $L$, very quickly exceeds $L_{\rm Edd}$, in which case we allow for mass loss via a wind \citep{pac86}.
We arbitrarily set our time coordinate to zero at the time of maximum luminosity, $L$, in the second burst after the start of accretion. The peak for $L_{\rm He}$ is at $t=-0.0269$ s, and $L>L_{\rm Edd}$ for the time interval $-0.0047<t<1.2169$ seconds..\footnote{A movie of this flash (made with \texttt{PGstar}, see \S \ref{pgstar}) is at \url{http://mesa.sourceforge.net/pdfs/nshe.mov}.}

\begin{figure}[H]
\plotone{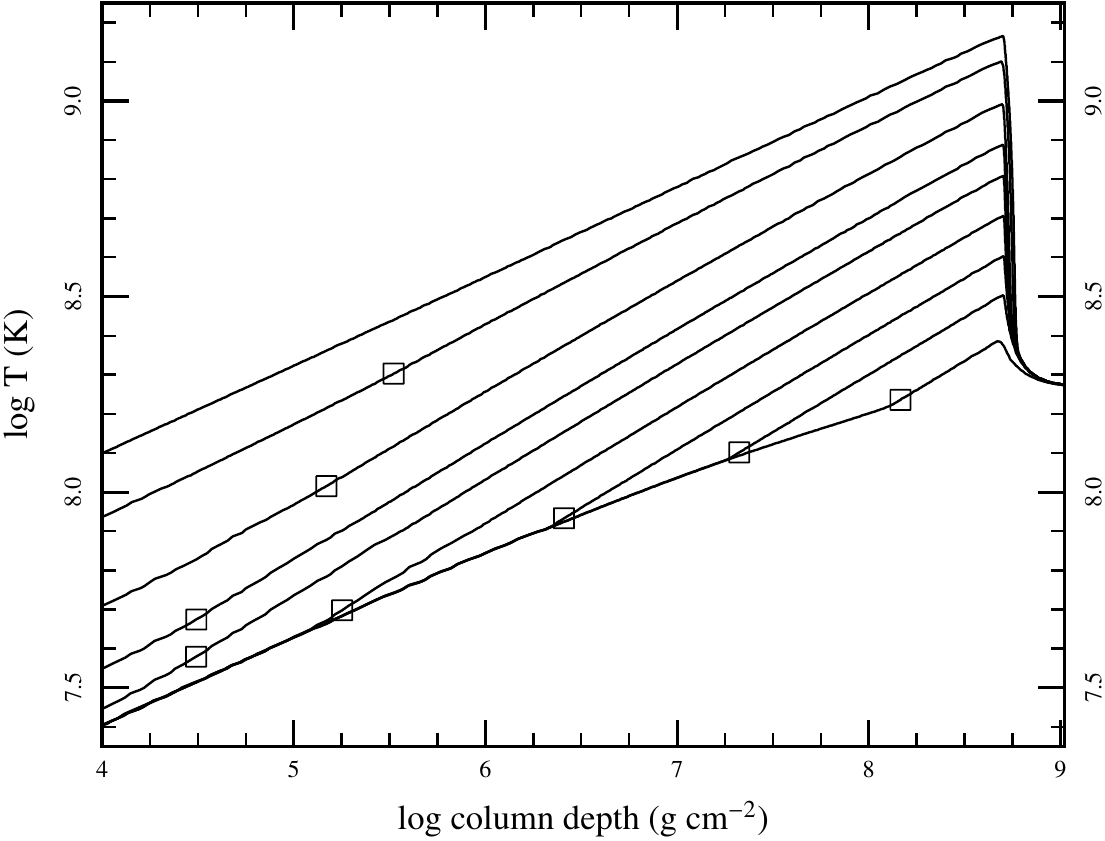}
\caption{The evolving temperature profile during the convective burning phase of a Type I burst, as a function of column depth, $P/g$. Starting from the bottom, each successive solid line is the temperature profile at a later time.  The open squares marks the location of the top of the convective zone. The top curve is at $t=-0.00716$. 
\label{tprofns}}
\end{figure}

\begin{figure}[H]
\begin{center}
\includegraphics[width=0.6\textwidth]{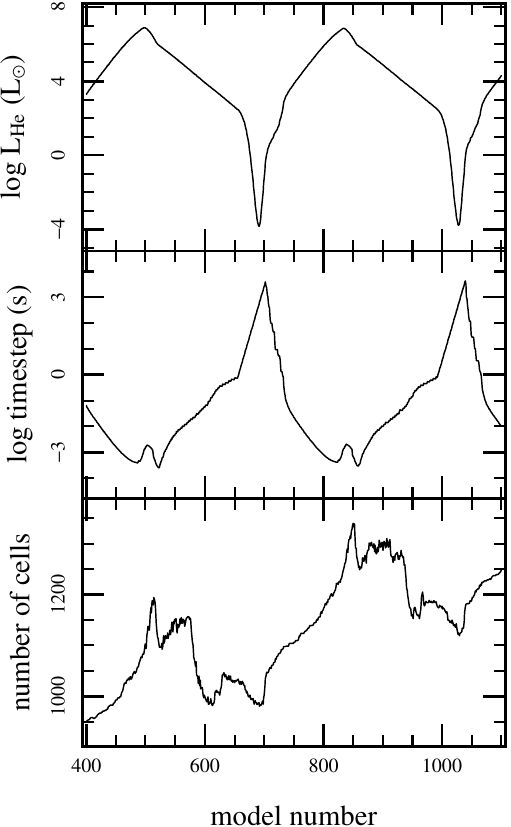}
\end{center}
\caption{The He burning luminosity,  timestep, and number of cells as a function of model number for the $\MESAstar$  simulation of an accreting neutron star. The timestep ranges from a millisecond to an hour, whereas the number of cells only grows by $\approx 25$\% during the burst and shows a secular trend upward as partially burned material accumulates. \label{dtnzL}}
\end{figure}

\begin{figure}[H]
\plotone{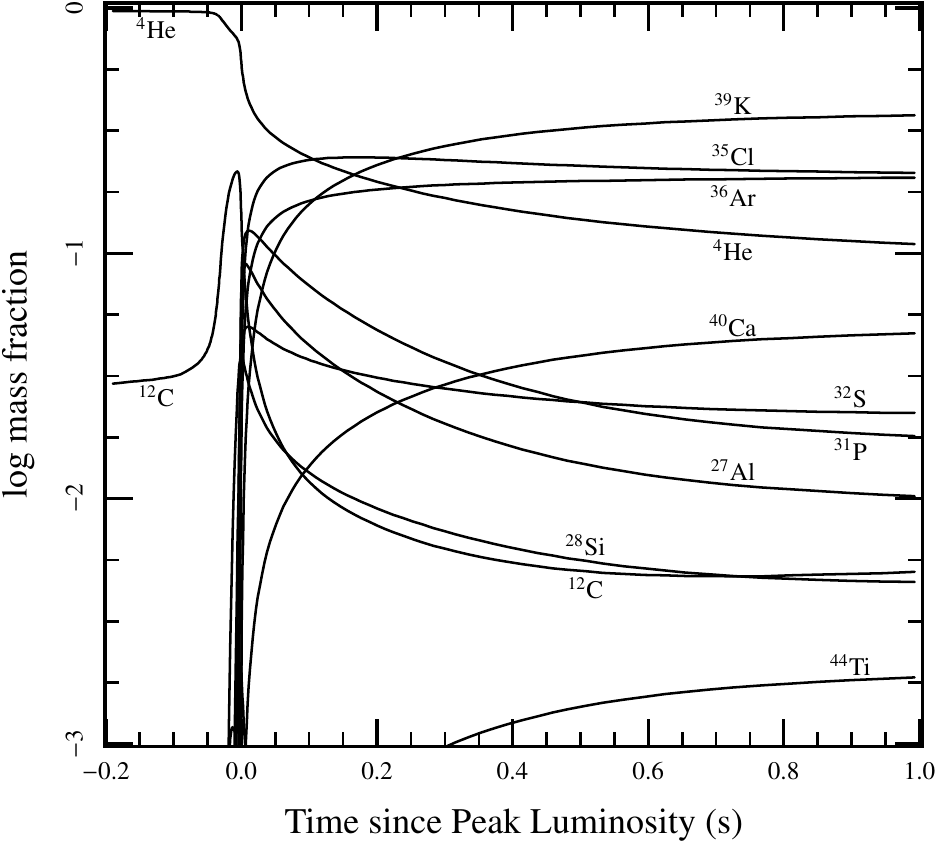}
\caption{Abundances of the dominant isotopes at the base of the convective zone as a function of time during the Type I burst. The temperature at the base of the convective zone at $t=1$ second is
$\log T=9.15$. 
\label{nsabund}}
\end{figure}

 Figure \ref{tprofns} shows the evolving temperature profile during the convective burning runaway, where time increases upwards. Though not for the same ignition depth, 
this plot is very similar to  Figure 2 of \citet{wei06}, including the evolution of the location of the top of the convective zone (open squares). \citet{wei06} discussed in detail the onset of heat transport in the outer, thin, radiative layer that allows for the retreat of the top of the convective zone. This $\MESAstar$ result is the first numerical confirmation of this transition for a pure helium accretor and
 demonstrates our ability to obtain excellent time and mass resolution as shown in Figure \ref{dtnzL}.  By using nonzero center
boundary conditions  so that the $dq$ variables (see \S \ref{s.masschanges}) cover only the relatively small envelope mass,
we reach a mass resolution of $\approx 1.5\times 10^{-20}M_\odot$.  The timestep adjustment algorithms (\S \ref{s.timestep}) 
provide a smooth change from timesteps of almost an hour between bursts down to millisecond steps at peak luminosity (see middle panel in 
Figure \ref{dtnzL}). The secular increase in the number of cells is to track the accumulation of the pile of ashes from each burst. 
The evolution of abundances at the base of the convective zone is shown in Figure \ref{nsabund}  and exhibits the presence of the isotopes $^{35}$Cl and $^{39}$K.

We have performed simulations at lower accretion rates but these become dynamical events where the temperature rises on a local dynamical timescale and are beyond the present scope of $\MESAstar$. 
While multi-dimensional hydrodynamical codes \citep[e.g.,][]{zin09} may be needed to follow the details of such an event, $\MESAstar$ can be used  for studying the longer timescale, hydrostatic evolution leading up to the point where hydrodynamic effects become dominant.

\section{Summary and conclusion\label{summary}}
Modules for Experiments in Stellar Astrophysics ($\MESA$) provides 
open source, portable, robust, efficient, thread-safe
libraries for stellar astrophysics and stellar evolution. It provides 
tools for a broad community of astrophysicists to explore a wide range 
of stellar masses and metallicities. State-of-the-art modules include the 
equation of state, opacity, nuclear reaction rates and networks, atmosphere
boundary conditions, and element diffusion.
$\MESA$ features a modern code architecture and run-time environment.

$\MESAstar$ solves the fully coupled structure and composition
equations simultaneously and is capable of calculating full evolutionary
tracks without user intervention. It implements adaptive mesh refinement,
sophisticated timestep adjustment, mass loss and accretion, and 
parallelism based on OpenMP.

$\MESA$ is subjected to an ongoing testing and verification process.
Current capabilities include evolutionary tracks of very low mass
stellar objects and gas giant planets, intermediate mass stars,
pulsating stars, accreting compact objects, and massive stars from 
the pre-main sequence to late times. Future versions of $\MESA$ will 
include the addition of a variety of new physics modules, features driven 
by the $\MESA$ user community, and architectural refinements.

\acknowledgments
We thank Edward Brown and Mike Zingale for carefully  reading the manuscript and providing 
cogent criticisms and clarifications. $\MESA$ has benefited from personal communications with 
many people including, but not limited to, the following: 
France Allard, 
Leandro Althaus,
Dave Arnett,
Phil Arras, 
Isabelle Baraffe, 
Arnold Boothroyd,
Adam Burgasser,
Adam Burrows,
Jeff Cash, 
Brian Chaboyer,
Philip Chang,
Alessandro Chieffi,
Joergen Christensen-Dalsgaard,
Chris Deloye,
Pavel Denisenkov,
Peter Eggleton,
J.~J.\ Eldridge, 
Jason Ferguson,
Jonathan Fortney,
Michael Gehmeyr,
Evert Glebbeek, 
Ernst Hairer,
Francois Hebert,
Alexander Heger, 
Lynne Hillenbrand,
Raphael Hirschi,
Piet Hut,
Alan Irwin,
Thomas Janka,
Stephen Justham,
David Kaplan,
Max Katz,
Attay Kovetz,
Michael Lederer,
Marco Limongi,
Marcin Mackiewicz,
Georgios Magkotsios,
Lars Mattsson,
Dan Meiron, 
Michael Montgomery,
Ehsan Moravveji,
Lorne Nelson,
Marco Pignatari, 
Marc Pinsonneault,
Philip Pinto, 
Phillip Podsiadlowski,
Onno Pols, 
Alexander Potekhin, 
Saul Rappaport,
Yousef Saad,
Didier Saumon,
Helmut Schlattl,
Aldo Serenelli,
Ken Shen,
Steinn Sigurdsson, 
Dave Spiegel,
Richard Stancliffe,
Sumner Starrfield, 
Justin Steinfadt,
Peter Teuben,
Jonathan Tomshine, 
Dean Townsley,
Don VandenBerg,
Roni Waldman,
Yan Wang,
Nevin Weinberg,
and Ofer Yaron.
This work was supported by the National Science Foundation under 
grants PHY 05-51164 and AST 07-07633. AD received support from a 
CITA National Fellowship. FH is supported by an NSERC Discovery Grant.
FXT acknowledges supported from the National Science Foundation, 
grants AST 08-07567 and AST 08-06720, and NASA, grant NNX09AD106.

\begin{appendix}
\section{{\sc Manifesto}\label{manifesto}}
$\MESA$  was developed through the concerted efforts of the lead author over a six year period 
with the engagement and deep involvement of many theoretical and computational astrophysicists. 
The public availability of $\MESA$ will serve education, scientific research, and outreach. This appendix 
describes the scientific motivation for $\MESA$, the philosophy and rules of use for $\MESA$, and 
the path forward on stewardship of $\MESA$ and advanced development of future research and education 
tools.  We make $\MESA$ openly available with the hope that it will grow into a community resource. 
We therefore consider it important to explain the guiding principles for using and contributing to MESA. 
Our goal is to assure the greatest usefulness for the largest number of research and educational projects.
   
\subsection{Motivation for a new tool} 
      
Stellar evolution calculations (i.e.\ stellar evolution tracks and
detailed information about the evolution of internal and global
properties) are a basic tool that enable a broad range of 
research in astrophysics. Areas that critically depend on
high-fidelity and modern stellar evolution include asteroseismology,
nuclear astrophysics, galactic chemical evolution and population synthesis, compact objects,
supernovae, stellar populations, stellar hydrodynamics, and stellar
activity. New observational capabilities are emerging in these fields that place a
high demand on exploration of stellar dependencies on metallicity and age. So, even though 
one dimensional stellar evolution is a mature discipline, we continue to ask new questions of 
stars. The emergence of demand requires the construction of a
general, modern stellar evolution code that combines the following advantages:

\begin{itemize}
\item {\bf Openess:} should be open to any researcher, both to advance the pace of
  scientific discovery, but also to share the load of updating physics, fine-tuning,
  and further development.
\item{\bf Modularity:} should provide independent, reusable modules. 
\item {\bf Wide Applicability:} should be capable of calculating the evolution of
  stars in a wide range of environments, including low- and massive
  stars, binaries, accreting, mass-losing stars, early and advanced
  phases of evolution etc. This will enable multi-problem,
  multi-object physics validation.
\item {\bf Modern techniques:} should employ modern numerical approaches,
  including high-order interpolation schemes, advanced AMR,
  simultaneous operator solution; should support well defined
  interfaces for related applications, e.g.  atmospheres, wind
  simulations, nucleosynthesis simulations, and  hydrodynamics.  
\item {\bf Microphysics:} should allow for up-to-date, wide-ranging,
  flexible and modular micro-physics.
\item {\bf Performance:} should parallelize on present and future
  shared-memory, multi-core/thread and possibly hybrid architectures
  so that performance continues to grow within the new computational
  paradigm.
\end{itemize}

A tool that combines the above features is a significant research and education resource for stellar
astrophysics. We acknowledge that some important aspects of stars are truly three-dimensional, such as convection,
rotation, and magnetism. Those applications remain in the realm of research frontiers with evolving 
understanding and insights, quite often profound. However, much remains to be gained scientifically 
(and pedagogically) by accurate one-dimensional  calculations, and this is the present focus of $\MESA$. 

\subsection{ $\MESA$ philosophy}

The $\MESA$ code library project is open. It explicitly
invites participation from anybody (researchers, students, interested
amateurs). Participation in $\MESA$ can take a wide range of forms, from
just using a $\MESA$ release for a science project, to testing and
debugging (i.e.\ report bugs, find fixes and submit them for
inclusion into the next release) as well as taking on responsibility
for the continued stewardship of certain aspects (modules) of the
code. The participation of experienced stellar evolution experts is
very welcome. 

Users are encouraged to add to the capabilities of $\MESA$, which will remain a community resource.
However, use of $\MESA$ requires adherence to the ``$\MESA$ code of conduct'': 
\begin{itemize}
\item That all publications and presentations (research, educational, or outreach) deriving from the use of $\MESA$ acknowledge the Paxton et al. (2010) publication and $\MESA$ website.
\item That user modifications and additions are given back to the community.
\item That users alert the $\MESA$ Council (see below) about their publications, either pre-release or at the time of publication.
\item That users make available in a timely fashion (e.g., online at the $\MESA$ website) 
all information needed for others to recreate their $\MESA$ results -- ``open know how''  to match ``open source.''
\item That users agree to help others learn $\MESA$, giving back as the project progresses. 
\end{itemize}
Users are requested to identify themselves by name, email contact, and home location.

\subsection{Establishment of the $\MESA$ council }

The $\MESA$ project began as an initiative to construct a reliable computational tool for stellar structure and evolution that takes full advantage of modern processor architectures, algorithms and community engagement. The release of\ $\MESA$ has forced some explicit thinking of what structure is needed so as to  achieve the mission of stewarding $\MESA$ in its use for scientific research, education and outreach, while also enabling the development of new tools and ideas. The $\MESA$ operating principles are simple: be open in your scientific discussions, give credit to all contributors, and be prepared to give back to the community of users. We hope that this creates an environment where the young are encouraged to become engaged in a career-enhancing manner. 

 We have  established  the $\MESA$ Council that consists of those engaged in working towards the shared missions outlined here: 
 \begin{itemize}
 \item {\bf Steward $\MESA$} 
There are many ways this will be done: supporting the contributors, maintaining the web access and web page updates, seeking enabling funding, 
holding yearly working groups that allow for continued engagement, documenting $\MESA$ development in the refereed literature, and sustaining
advanced development.

\item{\bf Interface with the User Community} 
This starts with answering questions from users, developing a way to accept new code in an integrated fashion, maintain a user registry, and 
identify new $\MESA$ Council members from those most active and engaged in the intelligent use of $\MESA$.

\item{\bf Enable Scientific Research and Education with $\MESA$}
Promote $\MESA$ and its goals, e.g., through scientific contributions at relevant conferences. Identify science opportunities that match 
$\MESA$ capabilities and facilitate and encourage appropriate collaborative activities. Track the science carried out by the community with $\MESA$.

\end{itemize}

\section{Code testing and verification\label{test}}
An important part of the ongoing development of a large, complex software project,
such as $\MESA$, is regular, systematic testing. Testing is necessary to ensure
that $\MESA$ continues to function as expected and that the addition of new
features does not have unintended consequences for existing features.

$\MESA$ is tested at the module level each time it is compiled from the install 
scripts. These tests check that each module produces results that are consistent with 
expectations. The next level is the $\MESAstar$ test suite, which consists of various 
evolutionary cases that are intended to cover a broad range of applications, including 
Roche lobe overflow, the He core flash in a low mass star, the evolution of sub-stellar 
mass objects, advanced nuclear burning in massive stars, accreting white dwarfs and 
neutron star envelopes, and more are being added all the time. The test cases 
come in both short and long varieties. 
Run in serial, the full set of short tests completes in less than one hour on modern hardware. 
The long tests might each take one 
or several hours to complete. Many of the evolutionary sequences presented in \S  
\ref{vandv} are included in the test suite.  Short tests include the very low mass models 
evolved to 10 Gyr (Figure \ref{VLMS}) and the $0.8\Msun, Z=10^{-4}$ track (Figure \ref{SCC:plot})
while longer tests include the Solar model calibration (Figure \ref{sound}), 
the ``hands off'' $1\Msun$ pre-main sequence to white dwarf calculation (Figure \ref{1MHRTRho}),
and Si-burning in a $15\Msun, Z=0.02$ model (Tables \ref{life25} and \ref{mass25}).

The test suite is readily extended in order to ensure regular testing of certain aspects 
of $\MESA$ that are not covered by the existing set but are important for a particular 
avenue of research.  A template is provided to encourage the creation of new test cases.
\end{appendix}

\end{document}